\documentclass[10pt]{iopart}
\usepackage[utf8]{inputenc}
\usepackage{graphicx}
\usepackage{siunitx}
\usepackage[table]{xcolor}
\usepackage{url}
\usepackage{hyperref}

\usepackage[title]{appendix}
\usepackage{multirow}
\usepackage{makecell}
\usepackage[export]{adjustbox}
\usepackage{placeins}
\usepackage{orcidlink}

\usepackage[singlelinecheck=false]{caption}

\usepackage{booktabs}

\usepackage[perpage]{footmisc}

\usepackage{geometry}
\usepackage{enumitem}
\setitemize{noitemsep,topsep=0pt,parsep=0pt,partopsep=0pt}

\usepackage[
backend=biber,
style=ieee,
sorting=none,
isbn=false, doi=true, url=false,
maxbibnames=3,minbibnames=3
]{biblatex}
\AtEveryBibitem{\clearlist{language}}
\AtEveryBibitem{\clearlist{publisher}}
\AtEveryCitekey{\clearfield{eprint}}
\newcommand{\implies}{\Rightarrow}

\begin{document}


\title[Validation of edge turbulence codes against TCV-X21]{Validation of edge turbulence codes against the TCV-X21 diverted L-mode reference case}

\author{
D. S. Oliveira$^1$\footnote{D. S. Oliveira and T. Body are co-first authors of this work.}\footnote{Corresponding authors} \orcidlink{0000-0002-6888-2594},
T. Body$^2$\addtocounter{footnote}{-1}\footnotemark[1]\addtocounter{footnote}{1}\footnotemark[2] \orcidlink{0000-0003-1904-6300},
D. Galassi$^1$ \orcidlink{0000-0003-3388-4538},
C. Theiler$^1$ \orcidlink{0000-0003-3926-1374},
E. Laribi$^{3,4}$,
P. Tamain$^3$,
A. Stegmeir$^2$ \orcidlink{0000-0001-8629-5062},
M. Giacomin$^1$ \orcidlink{0000-0003-2821-2008},
W. Zholobenko$^2$ \orcidlink{0000-0002-2624-0251},
P. Ricci$^1$ \orcidlink{0000-0003-3117-2238},
H. Bufferand$^3$ \orcidlink{0000-0002-9731-5855},
J. A. Boedo$^6$,
G. Ciraolo$^4$,
C. Colandrea$^1$,
D. Coster$^2$ \orcidlink{0000-0002-2470-9706},
H. de Oliveira$^1$ \orcidlink{0000-0002-6942-5491},
G. Fourestey$^1$,
S. Gorno$^1$ \orcidlink{0000-0003-0524-7283},
F. Imbeaux$^3$ \orcidlink{0000-0001-7461-314X},
F. Jenko$^2$ \orcidlink{0000-0001-6686-1469},
V. Naulin$^5$,
N. Offeddu$^1$ \orcidlink{0000-0002-0534-1993},
H. Reimerdes$^1$ \orcidlink{0000-0002-9726-1519},
E. Serre$^4$ \orcidlink{0000-0002-3174-7727},
C. K. Tsui$^6$ \orcidlink{0000-0002-7346-8312},
N. Varini$^1$,
N. Vianello$^7$ \orcidlink{0000-0003-4401-5346},
M. Wiesenberger$^5$ \orcidlink{0000-0002-5921-0163},
C. Wüthrich$^1$ \orcidlink{0000-0001-7548-2452}
and the TCV Team\footnote{See author list of S. Coda \textit{et al.} 2019 Nucl. Fusion 59 112023}}
\address{$^1$ Ecole Polytechnique Fédérale de Lausanne (EPFL), Swiss Plasma Center (SPC), CH-1015 Lausanne, Switzerland}
\address{$^2$ Max Planck Institute for Plasma Physics, Garching, Germany}
\address{$^3$ CEA, IRFM, F-13108 Saint-Paul-lez-Durance, France}
\address{$^4$ Aix-Marseille Univ., CNRS, Centrale Marseille, M2P2 Marseille, France}
\address{$^5$ Department of Physics, Technical University of Denmark, DK-2800 Kgs. Lyngby, Denmark}
\address{$^6$ Center for Energy Research (CER), University of California--San Diego (UCSD), La Jolla, California 92093, USA}
\address{$^7$ Consorzio RFX (CNR, ENEA, INFN, Universit\`a di Padova, Acciaierie Venete SpA), C.so Stati Uniti 4,35127 Padova, Italy}

\ead{\mailto{diego.oliveira@epfl.ch},\mailto{thomas.body@ipp.mpg.de}, \mailto{christian.theiler@epfl.ch}, \mailto{spc.data@epfl.ch} (long term data access)}

\begin{abstract}
Self-consistent full-size turbulent-transport simulations of the divertor and scrape-off-layer of existing tokamaks have recently become feasible. This enables the direct comparison of turbulence simulations against experimental measurements. In this work, we perform a series of diverted Ohmic L-mode discharges on the TCV tokamak, building a first-of-a-kind dataset for the validation of edge turbulence models. This dataset, referred to as \path{TCV-X21}, contains measurements from 5 diagnostic systems from the outboard midplane to the divertor targets -- giving a total of 45 one- and two-dimensional comparison observables in two toroidal magnetic field directions. The experimental dataset is used to validate three flux-driven 3D fluid-turbulence models -- GBS, GRILLIX and TOKAM3X. With each model, we perform simulations of the TCV-X21 scenario, individually tuning the particle and power source rates to achieve a reasonable match of the upstream separatrix value of density and electron temperature. We find that the simulations match the experimental profiles for most observables at the outboard midplane -- both in terms of profile shape and absolute magnitude -- while a comparatively poorer agreement is found towards the divertor targets. The match between simulation and experiment is seen to be sensitive to the value of the resistivity, the heat conductivities, the power injection rate and the choice of sheath boundary conditions. Additionally, despite targeting a sheath-limited regime, the discrepancy between simulations and experiment also suggests that the neutral dynamics should be included. The results of this validation show that turbulence models are able to perform simulations of existing devices and achieve reasonable agreement with experimental measurements. Where disagreement is found, the validation helps to identify how the models can be improved. By publicly releasing the experimental dataset and validation analysis, this work should help to guide and accelerate the development of predictive turbulence simulations of the edge and scrape-off-layer.
\end{abstract}

\noindent{\it Divertor, Simulation, Turbulence, Validation\/}

\submitto{\NF}

\maketitle

\ioptwocol

\textit{A summary of nomenclature is provided in Appendix \ref{sec:nomenclature and units}.}

\section{Introduction}\label{sec:introduction}

Magnetic-confinement-fusion devices cannot provide perfect confinement of the plasma in the core. Turbulent fluctuations and collisions lead to heat and particles being transported across field-lines, eventually reaching the solid walls of the device. The peak heat flux reaching plasma-facing components must be kept below engineering limits to prevent damage to the vessel. One key element to address the power exhaust problem is the diverted magnetic field geometry. In this configuration, an X-point is introduced in the tokamak boundary via shaping coils, diverting the boundary plasma to heat-resistant targets. Divertor geometries provide several benefits in comparison to simpler limited geometries. By increasing the separation of the confined region and the plasma-facing components, the divertor geometry improves the screening of impurities generated through plasma-wall interactions or intentionally injected to radiatively cool the plasma \cite{pitcherExperimentalDivertorPhysics1997}. They also increase the volume for scrape-off layer (SOL) plasma cooling and the connection length over which cross-field transport can broaden the heat flux channel \cite{pitcherExperimentalDivertorPhysics1997}, as well as helping to reach improved-confinement and detached regimes \cite{Wagner1982,krasheninnikovDivertorPlasmaDetachment2016a,Leonard2018Detachment}.

Divertor plasmas are, however, challenging to model and predict, due to the interplay of turbulence, drifts, plasma gradients, coherent filaments and interactions with neutrals and the walls \cite{ROGNLIEN1999,pitcherExperimentalDivertorPhysics1997}. Most divertor modelling is performed with transport codes such as SOLPS-ITER \cite{wiesen_new_2015} or SOLEDGE2D \cite{bufferandApplicationsSOLEDGE2DCode2011}, which treat the cross-field transport as an effective heat and particle diffusion, rather than directly modelling the small-scale convection due to turbulence. This reduces the computational cost, which in turn allows transport models to be run for large machines and over long time-scales. However, one limitation of transport modelling is that the diffusive transport coefficients are not self-consistently determined, and instead must be either heuristically fitted to experimental data, computed via reduced models \cite{Baschetti2021} or determined via a coupled turbulence code \cite{NishimuraCoupling2002,zhangSelfconsistentSimulationTransport2019a}. This can provide a reasonable match to the mean plasma profiles of existing devices, but cannot describe the time-dependent behaviour of the plasma. Furthermore, particularly in the SOL, the plasma can form coherent structures called filaments, or `blobs', which transport heat and particles ballistically rather than due to local gradients \cite{naulinTurbulentTransportPlasma2007}, breaking the assumption of diffusive transport. Turbulent self-organisation also leads to non-linear behaviour, which complicates direct extrapolations from current to future devices.

Therefore, to model the time-dependent dynamics and make predictive simulations of the divertor and SOL, it is necessary to simulate the turbulent nature of the transport\footnote{Here, reduced models such as Ref.\cite{Baschetti2021} could enable predictive mean-profile modelling via quasi-linear turbulence models. Since such models have a greatly reduced computational cost, they are expected to form part of a multi-fidelity `hierarchy of models' for predictive modelling.}. Direct numerical turbulence simulations require more sophisticated physical models than transport codes, and orders-of-magnitude more computational resources due to the 3D multi-scale nature of turbulence. Nevertheless, advances in numerical methods and larger, faster supercomputers mean that full-size turbulence simulations of the boundary region of existing experimental devices like COMPASS \cite{stegmeirGlobalTurbulenceSimulations2019, tataliImpactCollisionalityTurbulence2021, galassiTokamakEdgePlasma2019}, ISTTOK \cite{dudsonEdgeTurbulenceISTTOK2021}, MAST \cite{rivaThreedimensionalPlasmaEdge2019}, TCV \cite{rivaShapingEffectsScrapeoff2020}, Alcator C-mod \cite{changGyrokineticProjectionDivertor2017, Zholobenko2019, Halpern_2015}, AUG \cite{zholobenkoElectricFieldTurbulence2021a} and DIII-D \cite{changGyrokineticProjectionDivertor2017} are now achievable, allowing direct comparison between turbulence simulations and experimental results.

Full-size turbulence simulations of existing machines allow us to \textit{validate} our turbulence codes, which is an important step towards the development of predictive simulations for future devices such as ITER. Validation (in combination with verification) is a common technique in software testing. In the fusion community, a set of best practices for model validation was proposed by Terry et al, 2008 \cite{terryValidationFusionResearch2008} and Greenwald, 2010 \cite{greenwaldVerificationValidationMagnetic2010} -- outlining a rigorous validation methodology based on the quantitative comparison of multiple measurements at different `primacy hierarchies'. Importantly, validation here is not a binary result, but rather a tool for checking the fidelity of the simulations and guiding targeted development of the models -- with repeated validation suggested as part of a model testing and development cycle \cite{greenwaldVerificationValidationMagnetic2010}. This methodology has already been used to test boundary turbulence simulations in basic plasma physics devices \cite{ricciMethodologyTurbulenceCode2011, ricciApproachingInvestigationPlasma2015a,Galassi2021} and limited tokamak plasmas \cite{rivaShapingEffectsScrapeoff2020}.\\

In this paper, we extend these previous works to the validation of edge turbulence codes in diverted tokamak geometry, with the goal of guiding the development of the models and assessing how close simulations are to reproducing realistic plasma behaviour. For this purpose, a diverted, Ohmic L-mode scenario has been developed on the Tokamak à Configuration Variable (TCV) \cite{Coda2019}, performed in both toroidal magnetic field directions. Thanks to the large suite of edge/SOL diagnostics available on TCV, an extensive experimental dataset has been collected, allowing for a stringent assessment of the simulation-experiment agreement. We refer to this scenario and dataset as the \path{TCV-X21} validation reference case. This is used to test three 3D boundary turbulence codes -- namely GBS, developed at the Swiss Plasma Center at EPFL, Lausanne \cite{ricciSimulationPlasmaTurbulence2012, halpernGBSCodeTokamak2016}, GRILLIX \cite{stegmeirGlobalTurbulenceSimulations2019, zholobenkoElectricFieldTurbulence2021a}, developed at the Max Planck Institute for Plasma Physics, Garching and TOKAM3X \cite{Tamain2016, Nespoli_2019}, developed at the CEA, Cadarache in collaboration with Aix-Marseille University. The codes solve subsets of the drift-reduced Braginskii fluid equations \cite{Zeiler1997}, which require sufficient plasma collisionality such that each plasma species is close to a local thermodynamic equilibrium \cite{zeiler:habil99,Zeiler1997}. As such, the codes are not suitable for modelling the reactor core, and we focus our validation on the edge and open field-line region only.

By validating several codes against a common reference case, we can investigate how differences between the codes affect the results of divertor modelling, assess the importance of physical processes, and -- using the results of the validation -- guide the development of the codes. Validation against the \path{TCV-X21} case could also benefit other boundary turbulence codes -- such as XGC \cite{changGyrokineticProjectionDivertor2017}, COGENT \cite{dorfContinuumGyrokineticSimulations2021}, GENE-X \cite{michelsGENEXFullfGyrokinetic2021}, Gkeyll \cite{hakimContinuumElectromagneticGyrokinetic2020}, ORB5/PICLS \cite{boeslGyrokineticFullfParticleincell2019}, GYSELA \cite{cascheraImmersedBoundaryConditions2018}, FELTOR \cite{wiesenbergerReproducibilityAccuracyPerformance2019}, BOUT++ \cite{dudsonBOUTFrameworkParallel2009}, STORM \cite{rivaThreedimensionalPlasmaEdge2019}, Hermes \cite{Dudson2017Apr}, GDB \cite{zhuGDBGlobal3D2018}, HESEL \cite{thrysoePlasmaParticleSources2018} and SOLEDGE3X \cite{Bufferand2021} -- and could eventually be used as a common divertor reference case, similarly to the CYCLONE base case used for core modelling \cite{dimitsComparisonsPhysicsBasis2000}. To enable future testing against the \path{TCV-X21} case, we provide the experimental and simulation results in a Findable, Accessible, Interoperable, Reuseable (FAIR) data repository, along with additional documentation and data (such as the magnetic equilibrium) to help set up and post-process future validations. The data is available both as NetCDF files, and (for the experimental data only) in ITER Integrated Modelling and Analysis Suite (IMAS) format \cite{imbeauxDesignFirstApplications2015}. A dynamic repository is provided at \url{gitlab.mpcdf.mpg.de/tcv-x21/tcv-x21}, which will be updated with the results of future comparisons against the reference case (this is encouraged, to provide an evolving picture of the state-of-the-art of divertor modelling). We additionally provide a static repository for the version used in this paper at
\url{zenodo.org},
and a web-interface to the processing routines at \url{mybinder.org}.
Throughout this paper, wherever extended analyses are available through the repository, we indicate this via a file-path relative to the root of the \path{TCV-X21} repository.\\

The remainder of this paper is organised as follows: we first outline our validation methodology in Sec.\ref{sec:validation_methodology}, following the methodology in Ricci et al., 2015 \cite{ricciApproachingInvestigationPlasma2015a}. We then discuss the development of the experimental reference scenario and the collection of the experiment dataset in Sec.\ref{sec:experimental_scenario}. Next, we briefly introduce the three participating codes and discuss how the experimental reference scenario was simulated in Sec.\ref{sec:codes}. The results from the experiment and the simulations are compared both graphically and via a validation metric in Sec.\ref{sec:results}. We then discuss the results of the overall validation and the physics observed in Sec.\ref{sec:discussion}.

\section{Validation methodology}\label{sec:validation_methodology}

We start our validation by outlining the methodology. In this study, we perform both qualitative and quantitative validations in Sec.\ref{sec:results}, and consider the overall result in Sec.\ref{sec:discussion}. For our qualitative validation, we simply mean graphically comparing the simulated results and the experiment. This is helpful for evaluating the ability of the codes to make predictions of the dominant physical processes, of the shape and magnitude of the profiles, and for building an understanding of \textit{why} the simulations agree or disagree. Qualitative validation can, however, be imprecise or subjective -- which is why we also perform a quantitative validation.

The goal of quantitative validation is to provide a single numerical value of the \textit{level of agreement} between simulation and experiment. This is performed using a \textit{validation metric}, which is a type of summary statistic like the average-absolute-difference or Pearson correlation coefficient. Since a validation is more meaningful if more observables are compared, a complementary \textit{quality metric} is typically also given, which provides a measure for the number and precision of the observables used. For this study, we use the validation methodology presented in Ricci et al., 2015 \cite{ricciApproachingInvestigationPlasma2015a}, which is based on Terry et al., 2008 \cite{terryValidationFusionResearch2008}. We briefly review here the concepts and terms.\\

To quantify the level of agreement between simulation and experiment in a validation involving several observables, a `composite metric' is useful. In Ricci et al., 2015, a composite metric $\chi$ is computed from the individual levels-of-agreement for each observable $j$, denoted here $R(d_j)$, which are combined via a weighted average. Each observable is weighted according to its `primacy hierarchy' $H_j$ and its `sensitivity' $S_j$.

The individual level of agreement $R(d_j)$ is computed from the root-mean-square of the error-normalised experiment-simulation difference (roughly equivalent to the RMS Z-score) for some observable denoted `$j$'
\begin{equation}\label{eq:normalised_distance}
    d_j = \left[\frac{1}{N_j} \sum_{i=1}^{N_j} \frac{\left(e_{j,i} - s_{j,i}\right)^2}{\Delta e_{j,i}^2 + \Delta s_{j,i}^2} \right]^{1/2}
\end{equation}
where the experimental measurement has estimated values $e_{j,i}$ and uncertainties $\Delta e_{j,i}$ defined at some set of discrete data points $i=\left\{1, 2, ..., N_j\right\}$. The simulation result is assumed to be continuous and so is interpolated to the experimental measurement positions, giving computed values $s_{j,i}$ and uncertainties $\Delta s_{j,i}$. The uncertainties of the experimental results is evaluated for each diagnostic in Sec.\ref{sec:experimental_scenario}. A rigorous estimate of the simulation uncertainty is difficult to determine, so we simply set the simulation uncertainty to zero. We note here that this has the effect of increasing our $d_j$ values, which we discuss in Sec.\ref{sec:discussion}. For a discussion of sources of simulation uncertainty, see Ricci et al., 2015 \cite{ricciApproachingInvestigationPlasma2015a} and Chapter 5 of \textit{Computer Simulation Validation} \cite{Roy2019-ROYEAU}.

\begin{figure}[h]
    \centering
    \includegraphics[width=0.8\columnwidth]{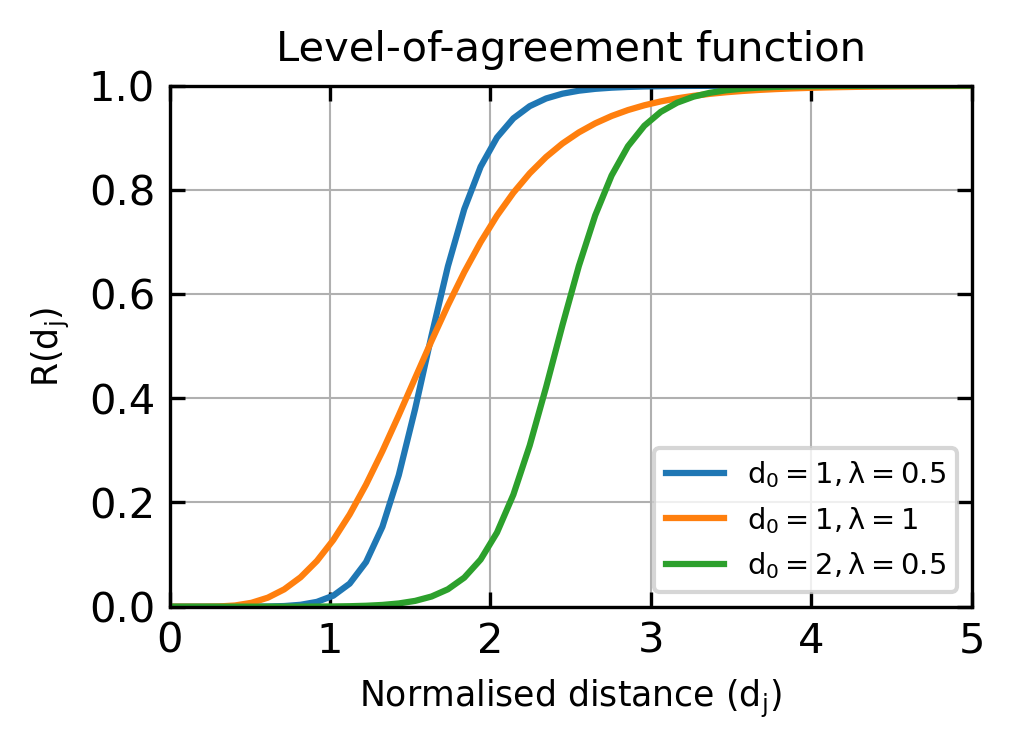}
    \caption{Converting from the normalised distance $d_j$ to a level of agreement $R(d_j)$, for different values of the agreement threshold $d_0$ and transition sharpness $\lambda$.}
    \label{fig:level_of_agreement_function}
\end{figure}

The level of agreement $R(d_j)$ is computed from $d_j$ via a smooth-step function
\begin{equation}\label{eq:individual level of agreement}
R(d_j) = \frac{\tanh\left[ \left( d_j - 1/d_j - d_0 \right)/\lambda \right]+1}{2}
\end{equation}
for $d_0$ the `agreement threshold' and $\lambda$ the `transition sharpness', which are usually set to $d_0 = 1$ and $\lambda=0.5$ \cite{ricciApproachingInvestigationPlasma2015a}. The shape of this function and the effect of the $d_0$ and $\lambda$ parameters are shown in Fig.\ref{fig:level_of_agreement_function}. For $d_j \ll d_0$ a value of $R(d_j)\approx0$ is returned, indicating quantitative agreement. For $d_j \gg d_0$, a value of $R(d_j) \approx 1$ indicates quantitative disagreement. For $d_j$ values in a region of width $\sim \lambda$ around $d_0$, an intermediate level of agreement is returned.

The level of agreement is combined with the primacy hierarchy and the sensitivity of the observable. The $H_j$ hierarchy weighting is computed as 
\begin{equation}\label{eq:hierarchy}
    H_j = \left[h_{Comp}\right]^{-1}_{j} = \left[ h_{Exp} + h_{Sim} - 1 \right]^{-1}_{j}
\end{equation}
where $h_{Exp}$ and $h_{Sim}$ are the primacy hierarchies of the observable for the experiment and simulation, and $h_{Comp}$ is the combined hierarchy of the comparison. Higher values of the primacy hierarchy indicate observables which require stronger assumptions, or calculation from a model combining multiple measurements (for $h_{Exp}$) or combinations of multiple directly simulated quantities (for $h_{Sim}$). An extended discussion of the primacy hierarchy can be found in Ref.\cite{ricciLangmuirProbebasedObservables2009}. By using the inverse of $h_{Comp}$ in Eq.\ref{eq:hierarchy}, we weight directly available observables more than indirect observables. The observables used in this validation together with their primacy hierarchies are given in Tab.\ref{tab:summary_of_measurements_and_hierarchies}.

The $S_j$ sensitivity weighting is computed as
\begin{equation}\label{eq:sensitivity}
S_j = \exp\left( - \frac{\sum_i \Delta e_{j,i} + \sum_i \Delta s_{j,i}}{\sum_i|e_{j,i}|+\sum_i|s_{j,i}|} \right)
\end{equation}
using the same notation as in Eq.\ref{eq:normalised_distance}. The sensitivity is a measure of the relative total uncertainty of the observable. It approaches $1$ for observables with very high precision, and $0$ for observables which have very high uncertainties.\\

Finally, we compute the composite metric via the weighted average
\begin{equation}\label{eq:ricci_chi}
    \chi = \frac{\sum_j R(d_j) H_j S_j}{\sum_j H_j S_j}    
\end{equation}
which gives values between $0$ (quantitative agreement) and $1$ (quantitative disagreement). This is combined with the overall `quality'
\begin{equation}\label{eq:quality}
Q=\sum_j H_j S_j
\end{equation}
which gives higher values for validations considering more directly-computed, high-precision measurements.

\section{Experimental Scenario} \label{sec:experimental_scenario}
\begin{figure}
    \centering
    \includegraphics[width=\columnwidth]{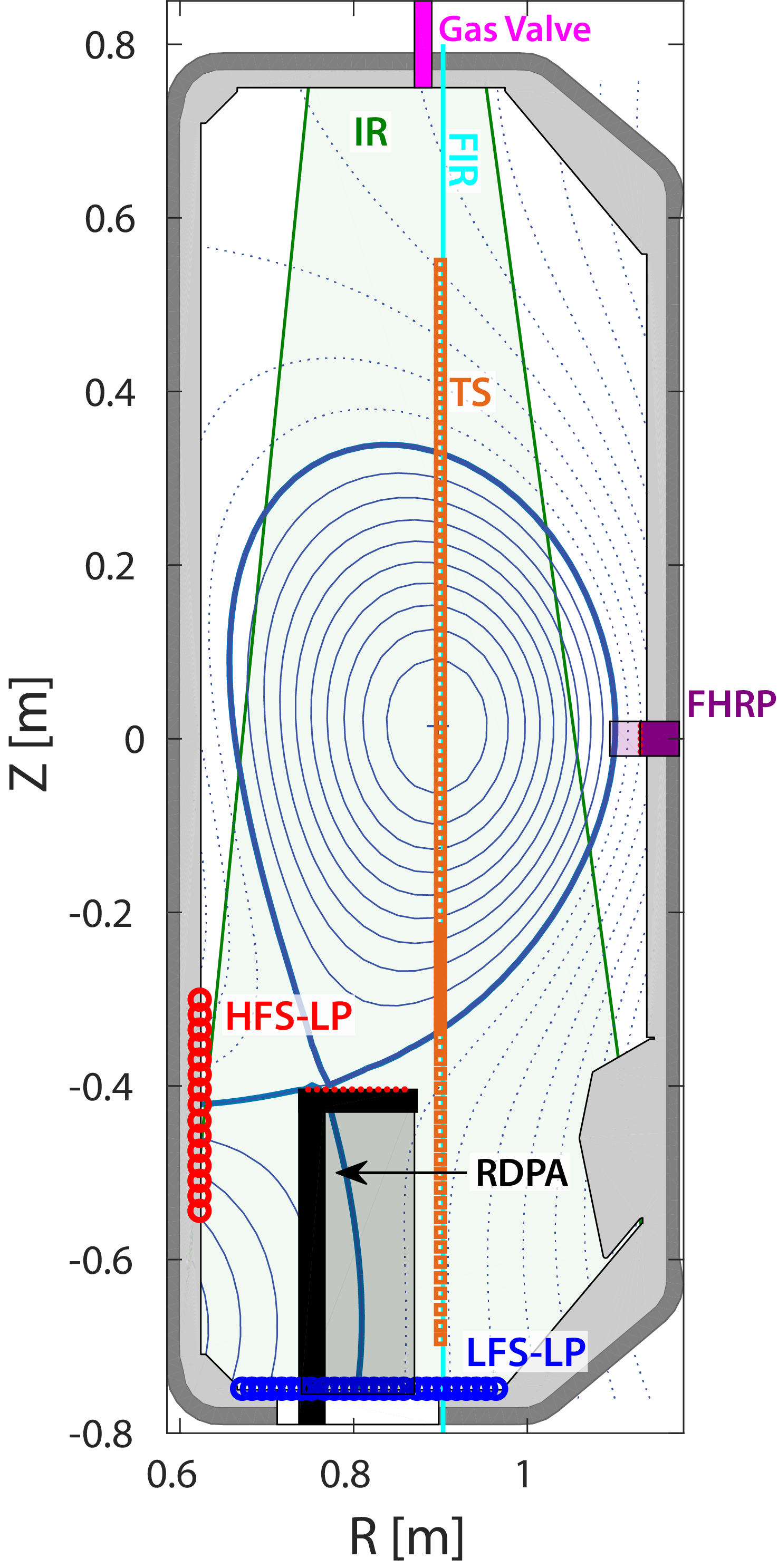}
    \caption{\textbf{TCV-X21 magnetic geometry and diagnostics.} Poloidal cross section showing the magnetic surfaces (dark blue lines) reconstructed by LIUQE and the diagnostics used to gather the dataset. We indicate the location of the wall-embedded Langmuir probes at the high-field-side and low-field-side divertor targets (HFS-LP and LFS-LP -- red and blue circles), the Reciprocating Divertor Probe Array (RDPA -- black L-shaped structure) and its swept area (black watermark box), the Thomson Scattering system (TS -- orange squares), the Fast Horizontal Reciprocating Probe (FHRP -- solid and watermark boxes in purple), the far infrared interferometer (used to estimate the line-average density, FIR -- cyan), the field-of-view of the Infrared system (IR -- green), and the position of the top valve used to fuel the plasma (magenta).}
    \label{fig:pol_view_diag}
\end{figure}

In this work, we developed an experimental scenario in TCV for the validation of boundary turbulence codes. TCV is a medium size tokamak ($R_{axis}=0.88\si{\meter}$) with nominal vacuum toroidal magnetic field of $B_{\phi,axis}\simeq 1.45 \si{\tesla}$ equipped with 16 independently-powered shaping coils providing extreme plasma shaping capabilities \cite{Coda2019}. The experimental scenario, referred to as the \path{TCV-X21} reference scenario, is a Lower-Single-Null L-mode Ohmic plasma. A poloidal cross-section showing the magnetic flux surfaces of this scenario, obtained from the LIUQE magnetic reconstruction code \cite{moretTokamakEquilibriumReconstruction2015}, can be seen in Fig.\ref{fig:pol_view_diag} together with the set of diagnostics used to collect the experimental dataset.

The discharges were performed in Deuterium and at a reduced toroidal field of $B_{\phi,axis}\simeq0.95\si{\tesla}$. This has the benefit of increasing the characteristic perpendicular scale length of the turbulence, which is given in terms of the sound drift scale
\begin{equation}\label{eq:drift_scale}
\rho_s = \frac{\sqrt{T_e m_i}}{eB}
\end{equation}
where $m_i$ is the mass of the Deuterium ions. To locally resolve the turbulence drive due to ballooning, drift-wave and ITG instabilities, a numerical grid resolution in the direction perpendicular to the magnetic field on the order of the local drift scale is required (the exact resolution requirement depends on the instability and the numerical scheme). By reducing the toroidal magnetic field by a factor of $1.5$, we can resolve the drift scale (or some multiple of it) with a $1.5\times$ lower poloidal and radial resolution, reducing the number of simulation grid points and therefore the computational cost of the simulations. To avoid MHD modes and Ohmic H-mode transitions in the forward-field case, we also reduced the plasma current to $I_p\simeq 165\si{\kilo\ampere}$, giving an edge safety factor of $q_{95} \approx 3.2$.

\begin{figure}[h]
    \centering
   \includegraphics[width=\columnwidth]{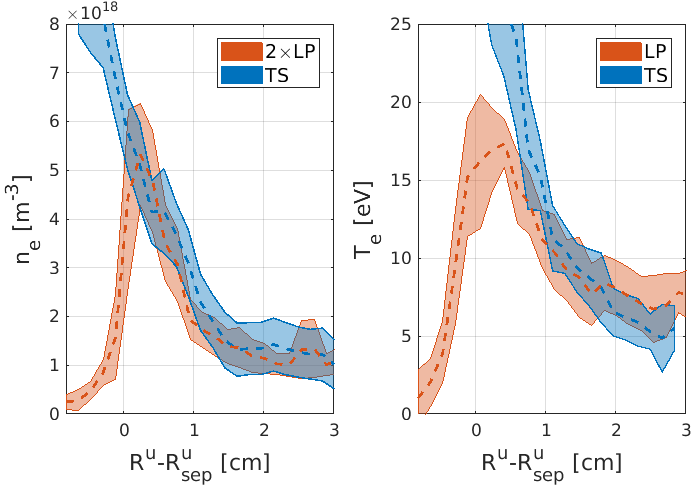}
    \caption{\textbf{Comparison of upstream and target profiles in reversed-field.} No significant reduction of $T_e$ in the LFS target (right) is observed. $n_e$ shows the expected factor 2 drop, characteristic of the sheath-limited regime.}
    \label{fig:sheath_lim}
\end{figure}

Since neutrals were not included in the simulations, the effect of their dynamics in the divertor volume was minimised in the \path{TCV-X21} scenario by using a low electron line-average density (determined from the FIR chord shown in Fig.\ref{fig:pol_view_diag}) of $\langle n_{e}\rangle\sim2.5\times10^{19} \si{\per\cubic\meter}$, corresponding to a Greenwald fraction of $\approx 0.25$. The discharges were fuelled with $D_2$ from the top valve, indicated in Fig.\ref{fig:pol_view_diag}. Fig.\ref{fig:sheath_lim} shows no significant difference between the SOL profiles of electron temperature and density in the divertor entrance and the LFS target profiles, suggesting a sheath-limited regime and, therefore, negligible temperature parallel gradients due to recycling in the divertor region \cite{stangeby2000plasma}. The substantial difference of the electron temperature near the separatrix in Fig.\ref{fig:sheath_lim} is because the divertor entrance is connected to the hot confined plasma, while the LFS target is disconnected from the confined plasma.

To investigate the effect of drifts, experiments were performed in both the ``forward'' and ``reversed'' toroidal field directions. In the forward (`favourable') field case, the ion $\nabla B$ drift points downwards from the plasma core towards the X-point, whereas in the reversed (`unfavourable') field case it points upwards, away from the X-point.

\subsection{Diagnostics \& Observables}\label{sec:diagnostics}

\begin{table*}
\centering
\caption{Available observables and their respective comparison hierarchies.}
\label{tab:summary_of_measurements_and_hierarchies}
\begin{tabular}{c c c c c c}
\toprule
\multirow{2}{*}{Diagnostic} & \multirow{2}{*}{Observable} & \multicolumn{3}{c}{Hierarchy}\\ 
        & & Exp & Sim & $H_j$ \\
\midrule
\multirow{4}{*}{\makecell{Wall Langmuir Probes (LP)\\ at the low-field-side and\\ high-field-side targets}}
        & $n$, $T_e$, $V_{pl}$ & 2 & 1 & 1/2 \\ 
        & \makecell{$J_{sat}$, $\sigma\left(J_{sat}\right)$,\\ $\mathrm{skew}\left(J_{sat}\right)$, $\mathrm{kurt}\left(J_{sat}\right)$} & 1 & 2 & 1/2 \\ 
        & $V_{fl}$, $\sigma\left(V_{fl}\right)$ & 1 & 2 & 1/2 \\ 
        & $J_\parallel$, $\sigma\left(J_\parallel\right)$ & 1 & 2 & 1/2 \\
        \midrule
        \makecell{Infrared camera (IR) \\ for low-field-side target}  & $q_\parallel$, $\lambda_q$ & 2 & 2 & 1/3\\
        \midrule
        \multirow{5}{*}{\makecell{Reciprocating divertor \\probe array (RDPA)\\for divertor volume}}
        & $n$, $T_e$, $V_{pl}$ & 2 & 1 & 1/2 \\ 
        & $M_\parallel$ & 2 & 2 & 1/3 \\ 
        & \makecell{$J_{sat}$, $\sigma\left(J_{sat}\right)$,\\ $\mathrm{skew}\left(J_{sat}\right)$, $\mathrm{kurt}\left(J_{sat}\right)$} & 1 & 2 & 1/2 \\ 
        & $V_{fl}$, $\sigma\left(V_{fl}\right)$ & 1 & 2 & 1/2 \\
        \midrule
        \makecell{Thomson scattering (TS) \\ for divertor entrance} & $n$, $T_e$ & 2 & 1 & 1/2 \\
        \hline
        \multirow{4}{*}{\makecell{Fast horizontally-\\reciprocating probe (FHRP)\\for outboard midplane}}
        & $n$, $T_e$, $V_{pl}$ & 2 & 1 & 1/2 \\ 
        & $M_\parallel$ & 2 & 2 & 1/3 \\
        & \makecell{$J_{sat}$, $\sigma\left(J_{sat}\right)$,\\ $\mathrm{skew}\left(J_{sat}\right)$, $\mathrm{kurt}\left(J_{sat}\right)$} & 1 & 2 & 1/2 \\ 
        & $V_{fl}$, $\sigma\left(V_{fl}\right)$ & 1 & 2 & 1/2 \\
        \bottomrule
\end{tabular}
\end{table*}

In the following section, we introduce the diagnostics and the basic analyses used to compute the experimental profiles and uncertainties. Tab.\ref{tab:summary_of_measurements_and_hierarchies} shows all observables, divided between the different regions of the SOL and the respective diagnostic used to determine them. For all profiles, we use as the radial coordinate $R^u - R^u_{sep}$, which is the distance between the measurement location and the separatrix, mapped along flux surfaces to the outboard midplane. The flux surfaces in the private-flux region (PFR) are not connected to the outboard midplane (OMP), and so the upstream mapping is carried out using the corresponding surface with the same poloidal flux $\psi$ in the confined region \footnote{A more involved method for computing the flux-surface label in the PFR is presented in Ref.\cite{Maurizio2018}. Our method is simpler to compute, while the method in Ref.\cite{Maurizio2018} may be preferable for detailed analysis of the PFR.}. The separatrix, flux surfaces and poloidal flux are obtained from the LIUQE magnetic reconstruction \cite{moretTokamakEquilibriumReconstruction2015}. This approach of using $R^u-R_{sep}^u$ removes small differences in the plasma positioning and magnetic geometry when combining data from repeated discharges, between different time-stamps, or in the comparison with the simulation profiles.

The magnetic reconstruction reveals a rapid oscillation of the strike-point, with a period of $\sim\SI{35}{\milli\second}$. This movement has a peak-to-peak value of $\Delta (R^u - R^u_{sep} )<\sim 2.5\si{\milli\meter}$ in forward-field and $\Delta(R^u - R^u_{sep})<\sim 1.5\si{\milli\meter}$ in reversed-field, affecting spatially fixed measurements which average over a long time interval. This is the case for the parallel heat flux estimated by the infrared vertical camera (Sec.\ref{sec:ir}) and density, electron temperature and plasma potential estimated by the wall-embedded Langmuir Probes in swept-bias mode (Sec.\ref{sec:langmuir}). In these cases, the rapid oscillation adds a broadening of the order of the peak-to-peak movement to the profiles.

The reconstructed magnetic equilibrium has an associated uncertainty \cite{Mele2021,fischerEstimationUncertaintiesProfiles2020}. For the experimental measurements, small variations of the separatrix position contribute, to some extent, to the reproducibility uncertainty of the experimental profiles (introduced in the next paragraph). For the simulations, quantifying the effect of the uncertainty in the magnetic geometry would require a sensitivity scan. This analysis is not performed in this work due to computational cost, and as such for the simulations we neglect the uncertainty in the magnetic reconstruction.

In general, we categorise the experimental uncertainty into three sources. The first is $\Delta e_{fit}$, the uncertainty related to fitting experimental data to a model. The second is $\Delta e_{dia}$, due to inherent characteristics of the diagnostics, e.g. uncertainties in the effective ion collection surface of Langmuir probes. Finally, the third is $\Delta e_{rep}$, the uncertainty related to the reproducibility of the observable assessed by comparing repeated discharges, typically performed in separate experimental sessions. Then, the total experimental uncertainty is evaluated as $\Delta e_{tot}=\sqrt{\Delta e_{fit}^2+\Delta e_{dia}^2+\Delta e_{rep}^2}$. Depending on the diagnostic and operation mode, not all the sources of uncertainty defined above are present.

\subsection{Wall-embedded Langmuir probes} \label{sec:langmuir}

In TCV, both targets are covered by wall-embedded, dome-shaped Langmuir probes (LP) as shown in Fig.\ref{fig:pol_view_diag}. The probes are operated in four different modes: swept-bias, ion saturation current, floating potential, and ground current mode. The details of the basic probe analysis can be consulted in Refs.\cite{Fevrier2018,DeOliveira2019}. Mean profiles of the electron density $n_e$, the electron temperature $T_e$, the floating potential $V_{fl}$, the plasma potential (obtained as $V_{pl}=V_{fl}+3T_e$), the ion saturation current density $J_{sat}$ parallel to the magnetic field, and the parallel current density $J_{||}$ are obtained from the swept-bias mode. Since we additionally have separate discharges where we use the Langmuir probes to measure $V_{fl}$, $J_{sat}$, and $J_{||}$ as a function of time, the time-averaged profiles for these quantities obtained from swept-bias mode are only used to determine the uncertainty $\Delta e_{rep}$ associated with the reproducibility. The quantities obtained from the non-swept-bias modes are evaluated over time windows of $1\si{\milli\second}$. In ion saturation current mode, a constant negative bias of $-100\si{\volt}$ is applied to the probes, resulting in a direct measurement of $J_{sat}$, which is used to estimate the mean, $J_{sat}$, the fluctuations (standard deviation), $\sigma(J_{sat})$, the skewness, $\mathrm{skew}(J_{sat})$, and the Pearson kurtosis, $\mathrm{kurt}(J_{sat})$. Direct measurements of the $V_{fl}$ and $J_{||}$ time histories are performed using, respectively, the floating potential mode (measuring the potential of the probe when floating with respect to the plasma) and the ground current measurement mode (current measured when the probe is biased to the vessel potential), allowing the estimate of mean ($V_{fl}$ and $J_{||}$) and fluctuation ($\sigma(V_{fl})$ and $\sigma(J_{||})$) profiles. For an improved spatial resolution, both strike-point positions were swept during the discharges.

For most of the quantities, $\Delta e_{rep}$ is estimated using different shots, the only exceptions being $\sigma(V_{fl}) $ and $\sigma(J_{||})$, where the single discharge available is divided into $100\si{\milli\second}$ intervals and $\Delta e_{rep}$ is estimated using the profiles resulting from the comparison of these sub-intervals. $\Delta e_{dia}$ is estimated by assuming an uncertainty of $\pm 0.1\si{\milli\meter}$ in the height of the probes (the wall LPs have $4\si{\milli\meter}$ of diameter and the domed-shape head protrudes from the tile shadow by $1\si{\milli\meter}$ \cite{Fevrier2018}). This source of error affects the quantities $n_e$, $J_{sat}$, $\sigma(J_{sat})$, $J_{||}$, and $\sigma(J_{||})$. The last source of uncertainty is $\Delta e_{fit}$, which affects only the swept-bias mode and is estimated as the $95\%$ confidence interval of the IV four-parameter fit.

\subsection{Infrared Cameras} \label{sec:ir}

The vertical Infrared thermography system (IR) covering the TCV outer target (see Fig.\ref{fig:pol_view_diag}) uses a camera operating with a frame rate of $160\si{\hertz}$ and a spatial resolution is $2.5\si{\milli\meter}$ \cite{Maurizio2018}. We estimate the heat flux at the LFS target for every camera frame and then average the results to obtain the averaged parallel heat flux $q_{||}$ as a function of $R^u-R_{sep}^u$. We also use the standard parametrisation of the heat flux profiles \cite{Eich2011} to determine the SOL power fall-off length $\lambda_q$ and spreading factor $S$.

The only source of uncertainty for $q_{||}$ is $\Delta e_{rep}$ which is estimated comparing profiles from different time frames. This accounts for uncertainties related to the strike point oscillation mentioned in Sec.\ref{sec:diagnostics} and the spacial calibration of the IR. 

\subsection{Reciprocating Divertor Probe Array}

The Reciprocating Divertor Probe Array (RDPA) installed at the bottom of TCV (see Fig.\ref{fig:pol_view_diag}) provides 2D measurements of a variety of quantities by combining a fast, vertical linear motion and a radial array of 12 rooftop Mach probes \cite{Oliveira2021}. A typical RPDA plunge takes approximately $350\si{\milli\second}$ and its Mach probes are operated in three different modes: sweep-bias, $J_{sat}$, and $V_{fl}$. The quantities obtained in this way are time-averages of $n_e$, $T_e$, $V_{fl}$, $V_{pl}$, $M_{||}$ and time histories of $J_{sat}$ and $V_{fl}$. The time histories are used to estimate mean and fluctuation profiles of $J_{sat}$ and $V_{fl}$, as well as the skewness and kurtosis of $J_{sat}$.

The vertical positions $Z$ of the RDPA dataset are translated to the coordinate system $Z-Z_X$, where $Z_X$ is the vertical X-point position determined by LIUQE. Analogously to the radial profiles, this approach removes small differences of the plasma positioning when combining data from repeated discharges or when comparing with the simulation data. 

The two main sources of uncertainty in swept-bias and $J_{sat}$ mode are $\Delta e_{dia}$, which is estimated considering an uncertainty of $10\%$ in the probe area, and $\Delta e_{rep}$, which is estimated comparing different shots. For $V_{fl}$ mode, the only source of uncertainty is $\Delta e_{rep}$, since the value of $V_{fl}$ does not depend on the probe area. This is also the case for $T_e$ and $V_{pl}$ estimated from swept-bias mode.

\subsection{Thomson Scattering}

The Thomson scattering system (TS) installed on TCV consists of 109 observation positions (chords) covering the region between $Z=-69\si{\centi\meter}$ to $Z=+55\si{\centi\meter}$ at a radial location of $R=0.9\si{\meter}$ (see Fig.\ref{fig:pol_view_diag}). This TS system can provide spatial profiles of the $n_e$ and $T_e$ covering a range of $T_e$ from $6\si{\electronvolt}$ to $20\si{\kilo\electronvolt}$ \cite{Hawke2017} (and even down to $1.4\si{\electronvolt}$ in the divertor \cite{Blanchard2019}). In our analysis, we use TS data measured near the separatrix in the divertor entrance to produce divertor-entrance profiles, seen in Fig.\ref{fig:sheath_lim}. The uncertainty sources are $\Delta e_{fit}$, estimated from the analysis procedure used to obtain $n_e$ and $T_e$, and $\Delta e_{rep}$, which is estimated from the comparison of the profiles obtained in different shots.

\subsection{Fast Horizontal Reciprocating Probe}

The horizontal reciprocating probe mounted at the outboard midplane (FHRP, at $Z=0$), shown in Fig.\ref{fig:pol_view_diag}, consists of a probe head with ten electrodes, which are used in different configurations and operation modes to provide measurements of time-averaged and fluctuation quantities \cite{Tsui2018}. The double probe configuration, operated in swept-bias mode, is used to determine $T_e$ and $n_e$. Similarly to the wall LPs, direct, time-resolved measurements of $J_{sat}$ and $V_{fl}$, performed at $2.5-5 \si{\mega\hertz}$, are used to estimate the mean and the fluctuations for both quantities, and the skewness and kurtosis for $J_{sat}$. Two electrodes are used to determine the parallel Mach number $M_{||}$. The sign convention of $M_\parallel$ for the FHRP is such that positive values refer to flows in the counter-clockwise direction if the torus is seen from top. For the magnetic helicity used in the present experiments (standard helicity), this corresponds to a parallel flow directed towards the LFS target. For $T_e$ and $n_e$, all three  sources of uncertainty are present and estimated following the same procedure employed for the wall LPs. For the quantities related to $J_{sat}$, the main sources of uncertainty are $\Delta e_{dia}$, estimated considering an uncertainty of $10\%$ in the probe area, and $\Delta e_{rep}$, which is estimated by comparing different discharges. For the fluctuations and mean value of $V_{fl}$, only $\Delta e_{rep}$ is relevant since this quantity is independent of the probe area.

The ion collection area of the FHRP electrodes is calculated here using the total probe surface area (and accounting for sheath expansion as detailed in \cite{Tsui2018}) rather than its projection along the magnetic field. This weak magnetic field assumption is made because the FHRP electrode dimensions (cylinders of length $1.5 \si{\milli\metre}$ and radius $0.75\si{\milli\metre}$) are comparable to the ion Larmor radius at the outboard midplane ($\rho_s=c_s/\Omega_i\approx 1.2\si{\milli\meter}$ for $T_e=T_i=20\si{\electronvolt}$ and $B_\phi=0.76T$ at $R=1.1\si{\meter}$). This is not the case for the wall LP and RDPA electrodes, which are larger than the local ion Larmor radius and therefore are treated as strongly magnetised.

\section{Simulation codes} \label{sec:codes}
\begin{figure*}
    \centering
    \includegraphics[width=\textwidth]{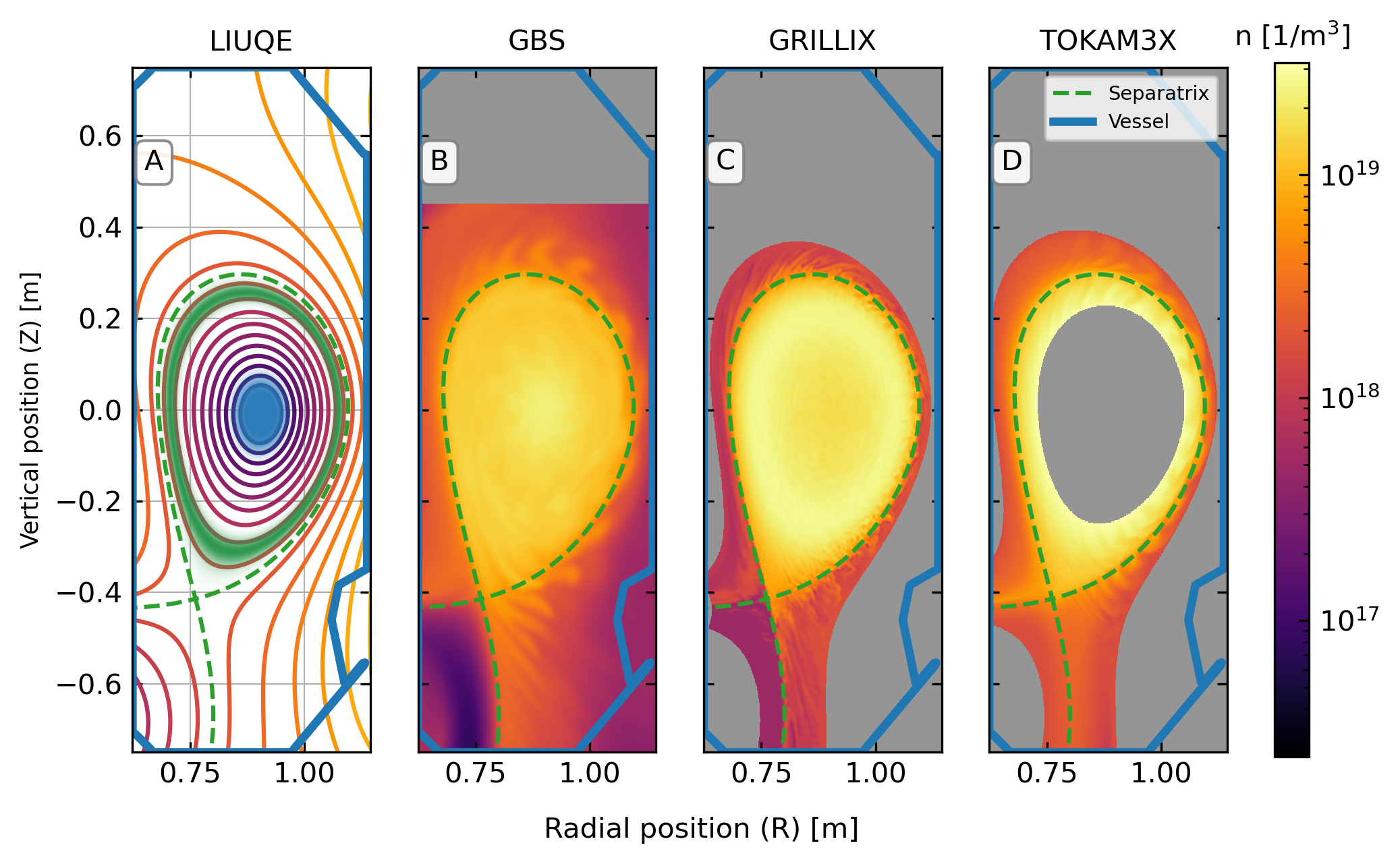}
    \caption{Poloidal profiles of the magnetic flux surfaces from LIUQE (\textit{A}) for TCV shot 65402 at $t=1.0\si{\second}$, with the GBS and GRILLIX density source position (\textit{green-shaded}) and temperature or power source position (\textit{blue-shaded}) superimposed. TOKAM3X applies both density and pressure sources at the inner core boundary, roughly corresponding to the \textit{green-shaded} position. Snapshots of the plasma density from GBS, GRILLIX and TOKAM3X simulations are shown in figures \textit{B, C,} and \textit{D} respectively. For all figures, the separatrix and vessel are indicated by \textit{green} and \textit{blue} lines respectively. Here, GBS and GRILLIX show the reversed-field case, while TOKAM3X shows the forward-field case.}
    \label{fig:simulation_snaps}
\end{figure*}

The \path{TCV-X21} scenario was simulated with the GBS, GRILLIX and TOKAM3X 3D two-fluid drift-reduced Braginskii turbulence codes. In this section we first provide a brief introduction of the models in each of the codes, and then discuss the choice of sources and parameters to model the TCV-X21 scenario. For the complete physical models (excluding boundary conditions) of the respective codes, the reader is directed to Giacomin and Ricci, 2020 \cite{giacominInvestigationTurbulentTransport2020} for GBS, appendix A of Zholobenko et al, 2021 \cite{zholobenkoElectricFieldTurbulence2021a} for GRILLIX, and Tatali et al, 2021 \cite{tataliImpactCollisionalityTurbulence2021} for TOKAM3X. For a discussion of the sheath boundary conditions used in this study, see Sec.\ref{subsec:contrasting_the_codes} and appendix \ref{sec:boundary_conditions}. The codes have all previously been \textit{verified} via the Method of Manufactured Solutions \cite{Paruta2018Nov,stegmeirGlobalTurbulenceSimulations2019,Tamain2016}, to ensure that the model equations have been correctly numerically implemented. TOKAM3X has additionally been verified via the \textit{a posteriori} iPOPE method \cite{cartier-michaudPosterioriErrorEstimate2020}.

The codes all solve versions of the drift-reduced Braginskii fluid equations \cite{Zeiler1997} -- giving the evolution of a plasma density $n$ (under the assumption of quasi-neutrality), separate electron and ion temperatures $T_e$ and $T_i$, the parallel ion velocity $u_\parallel$, the parallel electron velocity $v_\parallel$ or parallel current density $J_\parallel = e n (u_\parallel - v_\parallel)$ and the electrostatic potential $V_{pl}$. Additionally, GRILLIX evolves the parallel component of the electromagnetic vector potential $A_\parallel$. In this work, the codes neglect the neutral dynamics. To approximate the particle source due to neutral ionisation, a simple confined-region source (discussed in Sec.\ref{subsec:reference_equilibrium}) is used.

Rigorously, the fluid theory does not allow for modelling low-collisionality plasma regions \cite{zeiler:habil99,Zeiler1997}. The collisionality in the plasma core is too low for fluid models to be formally valid, and as such we do not expect to have a good agreement with the experiment in this region. Nevertheless, both GBS and GRILLIX include the plasma core in their simulations (see Fig.\ref{fig:simulation_snaps}). This circumvents the need to apply boundary conditions at the core, which do not have a clear physical analogue \cite{giacominInvestigationTurbulentTransport2020}. Additionally, the ion drift approximation breaks down at the entrance to the magnetic presheath \cite{loizuBoundaryConditionsPlasma2012} and as such the codes aim to mimic the effects of the plasma sheath via `sheath boundary conditions', rather than directly modelling the sheath.

\subsection{Contrasting the codes}\label{subsec:contrasting_the_codes}

There are a number of significant differences between the codes, despite the fact that they all are based on drift-reduced Braginskii models. We consider a few of the most significant differences here, to help interpret the differences found between the simulated results. Considering first the models, the codes apply different assumptions to simplify the numerical implementation of the model. In this work, both GBS and TOKAM3X use the Boussinesq approximation (although with different flavors, for details see Ref.\cite{giacominTurbulenceFlowsPlasma2020} for GBS and Ref.\cite{tataliImpactCollisionalityTurbulence2021} for TOKAM3X) and treat the electrostatic limit of the dynamics, while GRILLIX does not apply the Boussinesq approximation and includes the effects of electromagnetic induction. Additionally, GRILLIX and TOKAM3X include terms for electron-ion heat exchange, in contrast to GBS. Since the start of this project, new versions of the codes with extended models have been developed -- these are discussed in Sec.\ref{subsec:towards_a_match}.

The codes employ different sheath boundary conditions at the magnetic presheath entrance near the divertor targets. The details of the boundary conditions used for each code are given in Appendix A, with the key differences summarised here. For GBS, the parallel ion velocity $u_\parallel$ is set to $c_s$. In GRILLIX and TOKAM3X, corrections for the $E\times B$ drift (both codes) and curvature drift (TOKAM3X only) are included in the $u_\parallel$ boundary condition, and the drift-corrected ion velocity is set to $\geq c_s$ to allow supersonic transients to be freely advected across the boundaries (see Appendix.\ref{sec:boundary_conditions} and Ref.\cite{bufferandApplicationsSOLEDGE2DCode2011,stangebyIonVelocityBohm1995}). For the electron and ion temperatures, GBS enforces $\nabla_\parallel T =0$, while in GRILLIX and TOKAM3X sheath heat transmission factors are used. GBS and TOKAM3X determine coupled expressions for the current and plasma potential in the electrostatic limit, whereas GRILLIX assumes that internally-generated currents freely flow across the boundaries (via a $\nabla_\parallel J_\parallel = 0$ or `free-flowing' boundary condition) and sets the plasma potential such that $V_{fl} \to 0$. The effect of the boundary conditions is discussed in Sec.\ref{sec:discussion boundary conditions}.

The codes use different methods to discretise their model equations -- from a fourth-order non-field-aligned scheme in GBS \cite{giacomin2021}, to a domain-decomposed flux-aligned scheme in TOKAM3X \cite{tataliImpactCollisionalityTurbulence2021}, to a locally-field-aligned scheme in GRILLIX \cite{stegmeirGlobalTurbulenceSimulations2019}. In GBS and TOKAM3X, the boundary conditions are enforced at boundary grid points, while in GRILLIX an immersed boundary method is used \cite{stegmeirGlobalTurbulenceSimulations2019}. Since all the codes have been verified, the choice of discretisation will have no impact on the solution which the codes will converge to at arbitrarily high grid resolution. For a given grid resolution however, the discretisation error can vary depending on the choice of discretisation. Furthermore, the choice of discretisation affects the geometrical flexibility, computational efficiency and scalability of the codes.

\subsection{Physical parameters}\label{subsec:physical_params}

The physical and numerical parameters of the simulations can be varied to permit coarser spatial resolutions and a larger time-step, which reduces the computational cost of the simulations. The simulations set their resistivity in terms of the Braginskii value \cite{Braginskii1965}
\begin{equation}\label{eq:resistivity}
\eta_\parallel = \frac{0.51 m_e}{e \tau_{ee}} = 2.70 \left(\frac{T_e}{40 \si{\electronvolt}}\right)^{-3/2} \si{\micro\ohm\meter}
\end{equation}
where $m_e$, $e$ and $\tau_{ee}$ are the electron mass, elementary charge and electron collision time, and we have taken the Coulomb logarithm to be equal to a constant value of $13$ (the weak parametric dependence of the Coulomb logarithm is dropped). GRILLIX used the value of $\eta_\parallel$ as defined in Eq.\ref{eq:resistivity}, while GBS and TOKAM3X increased $\eta_\parallel$ by factors of 3 and 1.8 respectively, to permit the use of a larger time-step (see Ref.\cite{dudsonOhmLawReduced2021}) and avoid numerical instabilities.
The codes also set their electron and ion heat conductivities in terms of the Braginskii values \cite{Braginskii1965}. The electron heat conductivity is
\begin{equation}\label{eq:electron_conductivity}
\chi_{\parallel,e} = 3.16 \frac{n T_e \tau_{ee}}{m_e} = 23.9\left(\frac{T_e}{40\si{\electronvolt}}\right)^{5/2} \si{\mega\watt}/(\si{\electronvolt\meter})    
\end{equation}
and the ion heat conductivity is
\begin{equation}\label{eq:ion_conductivity}
\chi_{\parallel,i} = 3.9 \frac{n T_i \tau_{ii}}{m_i} = 0.69\left(\frac{T_i}{40\si{\electronvolt}}\right)^{5/2} \si{\mega\watt}/(\si{\electronvolt\meter})    
\end{equation}
which we have evaluated for Deuterium ions. GRILLIX used the heat conductivities as defined in Eqs.\ref{eq:electron_conductivity}-\ref{eq:ion_conductivity}, with a periodicity limiter (Eq.B.63 from the SOLPS-ITER manual \cite{wiesen_new_2015}) to limit the core heat flux. Conversely, both GBS and TOKAM3X used reduced heat conductivities, to avoid time-step limitations. TOKAM3X reduced the heat conductivities given by Eqs.\ref{eq:electron_conductivity}-\ref{eq:ion_conductivity} by a factor of 1.8. GBS used an effective heat conductivity of $\chi_{\parallel,e} = 1.29 (n/n_{ref})\si{\mega\watt}/(\si{\electronvolt\meter})$ for the electrons and $\chi_{\parallel,i} = 0.037 (n/n_{ref})\si{\mega\watt}/(\si{\electronvolt\meter})$ for the ions, with $n_{ref} = 6\times10^{18}\si{\per\cubic\meter}$ (corresponding to Eqs.\ref{eq:electron_conductivity}-\ref{eq:ion_conductivity} evaluated at $n=n_{ref}$ and $T_e = 41.3\si{\electronvolt}$ and then reduced by a factor of 20). Using the experimental values, we see that this gives a heat conductivity reduced by a factor of 20 at the OMP, and by a factor of 4.8 near the targets\footnote{Here, we see that including the strong temperature dependence reduces the heat conductivity in the SOL (which makes the simulations less expensive). In this work, it was not included in GBS due to the divergence of the Braginskii heat flux in the core, while a new version of GBS includes the temperature dependence and a heat flux limiter.}. The effect of using relaxed parameters is discussed in Sec.\ref{sec:discussion reduced params}.

\subsection{Equilibrium, resolution and sources}\label{subsec:reference_equilibrium}

We select a single `reference equilibrium' -- TCV shot 65402 at time $t=1.0\si{\second}$ -- which is representative of the experimental discharges. The magnetic field structure of the reference equilibrium is computed by LIUQE and is provided as an input to the codes for the simulations in both toroidal field directions. By using the magnetic field from LIUQE and the electron temperature from Thomson scattering, we can approximately determine the drift scale (Eq.\ref{eq:drift_scale}) as a function of position. We find that the confined region should have $\rho_s \geq 1\si{\milli\meter}$, while the open field-line region has $\rho_s$ as small as $0.3 \si{\milli\meter}$. Therefore, a perpendicular resolution of the order of a few \si{\milli\meter} should resolve most of the `primary' turbulence drive in the confined region, which ballistically drives SOL turbulence, and partially resolve `secondary' instabilities, which locally drive turbulence in the open field-line region. In this work, GBS used a perpendicular resolution of $\approx 2 \si{\milli\meter}$, GRILLIX used a perpendicular resolution of $\approx 1 \si{\milli\meter}$ and TOKAM3X used a resolution approximately equivalent to 1\si{\milli\meter} radially and $4\si{\milli\meter}$ poloidally at the outboard midplane. In the toroidal direction, GBS used 128 planes for $2\pi$ of toroidal angle, GRILLIX used 16 planes for $2\pi$ of toroidal angle and TOKAM3X used 32 planes for $\pi$ of toroidal angle (half-torus). Due to the different discretisation methods the resolution requirements may vary dramatically between the codes, although this is difficult to quantify without resolution scans. Due to computational cost, different resolution were tested with GRILLIX only, with the results discussed in Sec.\ref{subsec:disc_quant_and_sensitivity}.

The simulations are flux-driven, with freely evolving profiles determined by the balance of source terms, transport mechanisms and sinks at the device walls. The sources for this study are selected to provide simple approximations of Ohmic heating and neutral ionisation. The temperature source is selected to be close to the magnetic axis, since this is the position where Ohmic heating is primarily expected \cite{WessonTokamaks}. The density source is placed just inside the confined region, as shown by the green shaded region in Fig.\ref{fig:simulation_snaps}A. Ionisation in the divertor is not taken into account. For GBS and GRILLIX, the source positions are indicated in Fig.\ref{fig:simulation_snaps} and given in \path{TCVX21/grillix_post/components/sources_m.py}. For TOKAM3X, pressure sources (i.e. combined density and temperature sources) are located in the vicinity of the core limiting flux surface, which corresponds approximately to the same position as the GBS and GRILLIX density source. Treating the neutral dynamics via a simple confined-region source is a strong approximation in this work -- this is discussed in Sec.\ref{subsec:towards_a_match}.

In addition to the confined-region sources, small additional sources are added in the open field-line region to prevent numerical instabilities which occur at very low temperatures or densities. For GBS, additional particle sources are added in the private-flux region and at the inner boundary where flux surfaces become tangent to the wall. These sources are intended to prevent the density from dropping below $~10^{16}\si{\per\cubic\meter}$. For GRILLIX, point-wise adaptive sources are used to prevent the density from dropping below $5\times10^{17}\si{\per\cubic\meter}$ and the electron and ion temperatures from dropping below $2\si{\electronvolt}$. For TOKAM3X, additional sources were not required in this work.

\subsection{Constraints and free parameters}\label{sec:free_parameters}

Since the simulations self-consistently evolve the plasma profiles as well as the turbulence, they have only a few free parameters which can be tuned to match the experiment. The most significant free parameters are the power and particle source rates. In all codes, the density source rate is adjusted such that the separatrix value of the simulated density profile approximately matches the separatrix value measured by Thomson Scattering.
In GBS and GRILLIX, both density and temperature sources act as sources for power (since adding particles at non-zero temperature requires energy). The total power added to the plasma is
\begin{equation}\label{eq:power_integral}
P = \frac{3}{2}\int n \left(S_{T_e} + S_{T_i}\right) + (T_e + T_i) S_n \textrm{d}^3V
\end{equation}
where $n$, $T_e$, $T_i$, $S_n$, $S_{T_e}$ and $S_{T_i}$ are all functions of $R$, $Z$ and $\phi$, and the ionisation energy, stored in the plasma as potential energy, is not included here.
In GBS, the electron temperature source rate is adjusted such that the separatrix $T_e$ value at the outboard midplane matches the value measured by Thomson Scattering, and the ion temperature source rate is set to $25\%$ of $S_{T_e}$.
In GRILLIX, $S_{T_i} = 0$ and \linebreak$S_{T_e} = \frac{1}{n}\left(S_P - (T_e+T_i)S_n\right)$, which simplifies Eq.\ref{eq:power_integral} to $P = \frac{3}{2}\int S_P d^3V$. Therefore, a negative $T_e$ source is applied at the $n$ source position to maintain a constant power. The $S_P$ can be considered a power source for electrons, which is adjusted to match the Ohmic power, and the ions are heated via the equipartition term.
In TOKAM3X, sources are used for the pressure instead of for the temperatures, such that the total power is $P = \frac{3}{2} \int ( S_{p_e} + S_{p_i}) \textrm{d}^3V$. Equal pressure sources are used for the electrons and ions, and the electron pressure source is adjusted such that the separatrix $T_e$ value matches the value measured by Thomson Scattering. 

Therefore, in all simulations, the density value at the separatrix should approximately match the Thomson scattering separatrix value \textit{as a result of tuning}, while the rest of the profile is free to vary. Additionally, in GBS and TOKAM3X, the electron temperature value at the separatrix should match the Thomson scattering separatrix value (while the rest of the $T_e$ profile is free), while in GRILLIX the total injected power is set to approximately match the experimental Ohmic-heating power and the whole $T_e$ profile is free.

The order-of-magnitude of the resulting source rates can be compared to the experiment. For the power injection, the simulation source rate was equivalent to $170\si{\kilo\watt}$ for GBS, $150\si{\kilo\watt}$ for GRILLIX and $30\si{\kilo\watt}$ for TOKAM3X -- compared to a total Ohmic-heating power of $150\si{\kilo\watt}$, of which approximately $120\si{\kilo\watt}$ crossed the separatrix (estimated from a tomographic reconstruction of the radiated power measured with bolometry). For the particle source, the simulation source rate was equivalent to $\sim \SI{2e21}{\per\second}$ for GBS and $\SI{1.85e21}{\per\second}$ for GRILLIX and TOKAM3X -- compared to $\approx \SI{3e21}{\per\second}$ inferred from the total out-flux to the Langmuir probes, assuming perfect recycling. Therefore, the simulated and the expected experimental source rates come out at similar orders-of-magnitude. However, the power varied by more than a factor of 5 between the simulations, despite each simulation achieving upstream separatrix $T_e$ values which are similar to the experiment. From a simple two-point model, we expect that the upstream separatrix $T_e$ will be weakly dependent on the input power ($T_{e,upstream}\propto P_{sep}^{2/7}$, Eq.5.7 in Ref.\cite{stangeby2000plasma}) -- and as such it is possible to achieve roughly the same upstream $T_e$ value with a wide range of input powers. However, for the target $T_e$ value, a stronger $P_{sep}^{10/7}$ dependence (Eq.5.10 in Ref.\cite{stangeby2000plasma}) is expected.

\subsection{Simulations and post-processing}

Each of the three codes performed simulations of the \path{TCV-X21} scenario for a physical time of at least $2 \si{\milli\second}$, allowing the sources, cross-field turbulent transport, plasma profiles and sinks to approach a dynamic equilibrium (or \textit{saturated}) state. Statistical moments were calculated over the last $1\si{\milli\second}$ of each simulation, sampled at approximately $1\si{\micro\second}$ intervals. In Sec.\ref{sec:results} we compare results from both field directions for GBS and GRILLIX, while TOKAM3X performed a simulation only in the forward-field direction. During the setup of the simulation, an issue in the aspect ratio resulted in the $R$ coordinate of the TOKAM3X being effectively shifted inwards by $-25\si{\centi\meter}$, which was corrected for in post-processing.
In addition, the normalisation parameters of the TOKAM3X simulations were adjusted in post-processing to improve the match of the outboard midplane separatrix values of $n$ and $T_e$. This renormalisation can be performed consistently since the equations are implemented in a dimensionless form which retains the parametric dependencies\footnote{Setting the Coulomb logarithm equal to a constant is one exception.}. Renormalisation also changes the effective value of other physical parameters such as the resistivity, heat conductivity and source rates (the values given in Sec.\ref{subsec:physical_params} and Sec.\ref{sec:free_parameters} are computed after renormalisation).
A worked example showing how the observables in Tab.\ref{tab:summary_of_measurements_and_hierarchies} are calculated is given in \path{TCV-X21/notebooks/simulation_postprocessing.ipynb}.

\section{Validation}\label{sec:results}
In this section, we present the overall result of the validation and show individual profiles from the experiment and simulations. We start by giving the overall quantitative result, to quickly indicate which observables agree particularly well or poorly. We then show the profiles obtained at the outboard midplane and divertor entrance (Sec.\ref{subsec:omp profiles}), which are found to give reasonable agreement. This is contrasted to the divertor target profiles (Sec.\ref{subsec:divertor target profiles}), where a reduced level of agreement is found. Finally, we show the divertor volume profiles from the RDPA (Sec.\ref{subsec:divertor volume measurements}). Note that due to space limitations it is not possible to show figures for all observables. Figures for all observables (and additionally the simulated ion temperature) can be found in \path{TCV-X21/3.results}.

We limit our validation analysis to the range $R^u - R^u_{sep} < 2.5\si{\centi\metre}$ for all diagnostics, and to $R^u - R^u_{sep} > -0.9\si{\centi\metre}$ for the wall Langmuir probes. This removes regions where the signal acquired by the probes is very low, which can prevent the correct fitting of the IV curve by the four-parameter model. Additionally, this range ensures that the comparison points are on the grid of all simulations (since the codes use different radial grid extents, indicated in Fig.\ref{fig:simulation_snaps}). We also note that the RDPA 2D profiles are limited to $Z - Z_X > -0.32 \si{\meter}$, with points close to the targets cropped to avoid possible effects of the pre-sheath entrance. The data over an extended range is available in \path{TCV-X21/1.experimental_data}.\\

\begin{table*}
    \centering
    \resizebox{1.0\textwidth}{!}{
    \begin{tabular}{llcccccccccc}
\toprule
 & & \multicolumn{2}{c}{GBS($+$)} & \multicolumn{2}{c}{GBS($-$)} & \multicolumn{2}{c}{GRILLIX($+$)} & \multicolumn{2}{c}{GRILLIX($-$)} & \multicolumn{2}{c}{TOKAM3X($+$)} \\ 
Diagnostic & observable & $d_j$ & $S$ & $d_j$ & $S$ & $d_j$ & $S$ & $d_j$ & $S$ & $d_j$ & $S$ \\ 
\midrule
\multirow{11}{*}{\makecell{Fast\\horizontally-\\reciprocating\\probe (FHRP)\\for outboard\\midplane}}
& $n$                                      & \cellcolor[rgb]{ 0.796,  0.914,  0.510}\textcolor[rgb]{ 0.000,  0.000,  0.000}{1.35  } & \cellcolor[rgb]{ 1.000,  1.000,  1.000}\textcolor[rgb]{ 0.000,  0.000,  0.000}{0.841 } & \cellcolor[rgb]{ 0.655,  0.853,  0.418}\textcolor[rgb]{ 0.000,  0.000,  0.000}{0.712 } & \cellcolor[rgb]{ 1.000,  1.000,  1.000}\textcolor[rgb]{ 0.000,  0.000,  0.000}{0.899 } & \cellcolor[rgb]{ 0.646,  0.849,  0.415}\textcolor[rgb]{ 0.000,  0.000,  0.000}{0.698 } & \cellcolor[rgb]{ 1.000,  1.000,  1.000}\textcolor[rgb]{ 0.000,  0.000,  0.000}{0.865 } & \cellcolor[rgb]{ 0.869,  0.945,  0.569}\textcolor[rgb]{ 0.000,  0.000,  0.000}{1.73  } & \cellcolor[rgb]{ 1.000,  1.000,  1.000}\textcolor[rgb]{ 0.000,  0.000,  0.000}{0.91  } & \cellcolor[rgb]{ 1.000,  0.988,  0.729}\textcolor[rgb]{ 0.000,  0.000,  0.000}{2.6   } & \cellcolor[rgb]{ 1.000,  1.000,  1.000}\textcolor[rgb]{ 0.000,  0.000,  0.000}{0.893 }\\
& $T_e$                                    & \cellcolor[rgb]{ 0.636,  0.845,  0.414}\textcolor[rgb]{ 0.000,  0.000,  0.000}{0.66  } & \cellcolor[rgb]{ 1.000,  1.000,  1.000}\textcolor[rgb]{ 0.000,  0.000,  0.000}{0.765 } & \cellcolor[rgb]{ 0.577,  0.819,  0.408}\textcolor[rgb]{ 0.000,  0.000,  0.000}{0.426 } & \cellcolor[rgb]{ 1.000,  1.000,  1.000}\textcolor[rgb]{ 0.000,  0.000,  0.000}{0.825 } & \cellcolor[rgb]{ 0.686,  0.866,  0.439}\textcolor[rgb]{ 0.000,  0.000,  0.000}{0.866 } & \cellcolor[rgb]{ 1.000,  1.000,  1.000}\textcolor[rgb]{ 0.000,  0.000,  0.000}{0.733 } & \cellcolor[rgb]{ 0.765,  0.900,  0.489}\textcolor[rgb]{ 0.000,  0.000,  0.000}{1.21  } & \cellcolor[rgb]{ 1.000,  1.000,  1.000}\textcolor[rgb]{ 0.000,  0.000,  0.000}{0.776 } & \cellcolor[rgb]{ 0.694,  0.870,  0.444}\textcolor[rgb]{ 0.000,  0.000,  0.000}{0.916 } & \cellcolor[rgb]{ 1.000,  1.000,  1.000}\textcolor[rgb]{ 0.000,  0.000,  0.000}{0.756 }\\
& $V_{pl}$                                 & \cellcolor[rgb]{ 0.587,  0.823,  0.409}\textcolor[rgb]{ 0.000,  0.000,  0.000}{0.482 } & \cellcolor[rgb]{ 1.000,  1.000,  1.000}\textcolor[rgb]{ 0.000,  0.000,  0.000}{0.774 } & \cellcolor[rgb]{ 0.950,  0.979,  0.681}\textcolor[rgb]{ 0.000,  0.000,  0.000}{2.21  } & \cellcolor[rgb]{ 1.000,  1.000,  1.000}\textcolor[rgb]{ 0.000,  0.000,  0.000}{0.78  } & \cellcolor[rgb]{ 0.597,  0.827,  0.410}\textcolor[rgb]{ 0.000,  0.000,  0.000}{0.52  } & \cellcolor[rgb]{ 1.000,  1.000,  1.000}\textcolor[rgb]{ 0.000,  0.000,  0.000}{0.738 } & \cellcolor[rgb]{ 0.780,  0.907,  0.499}\textcolor[rgb]{ 0.000,  0.000,  0.000}{1.3   } & \cellcolor[rgb]{ 1.000,  1.000,  1.000}\textcolor[rgb]{ 0.000,  0.000,  0.000}{0.773 } & \cellcolor[rgb]{ 0.671,  0.859,  0.428}\textcolor[rgb]{ 0.000,  0.000,  0.000}{0.801 } & \cellcolor[rgb]{ 1.000,  1.000,  1.000}\textcolor[rgb]{ 0.000,  0.000,  0.000}{0.75  }\\
& $J_{sat}$                                & \cellcolor[rgb]{ 0.780,  0.907,  0.499}\textcolor[rgb]{ 0.000,  0.000,  0.000}{1.27  } & \cellcolor[rgb]{ 1.000,  1.000,  1.000}\textcolor[rgb]{ 0.000,  0.000,  0.000}{0.89  } & \cellcolor[rgb]{ 0.636,  0.845,  0.414}\textcolor[rgb]{ 0.000,  0.000,  0.000}{0.663 } & \cellcolor[rgb]{ 1.000,  1.000,  1.000}\textcolor[rgb]{ 0.000,  0.000,  0.000}{0.893 } & \cellcolor[rgb]{ 0.773,  0.903,  0.494}\textcolor[rgb]{ 0.000,  0.000,  0.000}{1.24  } & \cellcolor[rgb]{ 1.000,  1.000,  1.000}\textcolor[rgb]{ 0.000,  0.000,  0.000}{0.9   } & \cellcolor[rgb]{ 0.804,  0.917,  0.515}\textcolor[rgb]{ 0.000,  0.000,  0.000}{1.4   } & \cellcolor[rgb]{ 1.000,  1.000,  1.000}\textcolor[rgb]{ 0.000,  0.000,  0.000}{0.882 } & \cellcolor[rgb]{ 0.993,  0.725,  0.416}\textcolor[rgb]{ 0.000,  0.000,  0.000}{4.09  } & \cellcolor[rgb]{ 1.000,  1.000,  1.000}\textcolor[rgb]{ 0.000,  0.000,  0.000}{0.918 }\\
& $\sigma\left(J_{sat}\right)$             & \cellcolor[rgb]{ 0.975,  0.557,  0.323}\textcolor[rgb]{ 0.000,  0.000,  0.000}{4.73  } & \cellcolor[rgb]{ 1.000,  1.000,  1.000}\textcolor[rgb]{ 0.000,  0.000,  0.000}{0.889 } & \cellcolor[rgb]{ 0.857,  0.940,  0.553}\textcolor[rgb]{ 0.000,  0.000,  0.000}{1.64  } & \cellcolor[rgb]{ 1.000,  1.000,  1.000}\textcolor[rgb]{ 0.000,  0.000,  0.000}{0.939 } & \cellcolor[rgb]{ 0.898,  0.957,  0.609}\textcolor[rgb]{ 0.000,  0.000,  0.000}{1.9   } & \cellcolor[rgb]{ 1.000,  1.000,  1.000}\textcolor[rgb]{ 0.000,  0.000,  0.000}{0.927 } & \cellcolor[rgb]{ 0.898,  0.957,  0.609}\textcolor[rgb]{ 0.000,  0.000,  0.000}{1.9   } & \cellcolor[rgb]{ 1.000,  1.000,  1.000}\textcolor[rgb]{ 0.000,  0.000,  0.000}{0.934 } & \cellcolor[rgb]{ 0.997,  0.898,  0.577}\textcolor[rgb]{ 0.000,  0.000,  0.000}{3.25  } & \cellcolor[rgb]{ 1.000,  1.000,  1.000}\textcolor[rgb]{ 0.000,  0.000,  0.000}{0.933 }\\
& $\mathrm{skew}\left(J_{sat}\right)$      & \cellcolor[rgb]{ 0.991,  0.996,  0.737}\textcolor[rgb]{ 0.000,  0.000,  0.000}{2.44  } & \cellcolor[rgb]{ 1.000,  1.000,  1.000}\textcolor[rgb]{ 0.000,  0.000,  0.000}{0.81  } & \cellcolor[rgb]{ 0.886,  0.952,  0.593}\textcolor[rgb]{ 0.000,  0.000,  0.000}{1.81  } & \cellcolor[rgb]{ 1.000,  1.000,  1.000}\textcolor[rgb]{ 0.000,  0.000,  0.000}{0.912 } & \cellcolor[rgb]{ 0.994,  0.794,  0.474}\textcolor[rgb]{ 0.000,  0.000,  0.000}{3.78  } & \cellcolor[rgb]{ 1.000,  1.000,  1.000}\textcolor[rgb]{ 0.000,  0.000,  0.000}{0.898 } & \cellcolor[rgb]{ 0.964,  0.477,  0.286}\textcolor[rgb]{ 0.000,  0.000,  0.000}{12.1  } & \cellcolor[rgb]{ 1.000,  1.000,  1.000}\textcolor[rgb]{ 0.000,  0.000,  0.000}{0.942 } & \cellcolor[rgb]{ 0.892,  0.954,  0.601}\textcolor[rgb]{ 0.000,  0.000,  0.000}{1.85  } & \cellcolor[rgb]{ 1.000,  1.000,  1.000}\textcolor[rgb]{ 0.000,  0.000,  0.000}{0.847 }\\
& $\mathrm{kurt}\left(J_{sat}\right)$      & \cellcolor[rgb]{ 0.999,  0.959,  0.681}\textcolor[rgb]{ 0.000,  0.000,  0.000}{2.8   } & \cellcolor[rgb]{ 1.000,  1.000,  1.000}\textcolor[rgb]{ 0.000,  0.000,  0.000}{0.829 } & \cellcolor[rgb]{ 0.980,  0.991,  0.721}\textcolor[rgb]{ 0.000,  0.000,  0.000}{2.37  } & \cellcolor[rgb]{ 1.000,  1.000,  1.000}\textcolor[rgb]{ 0.000,  0.000,  0.000}{0.934 } & \cellcolor[rgb]{ 0.973,  0.547,  0.318}\textcolor[rgb]{ 0.000,  0.000,  0.000}{4.78  } & \cellcolor[rgb]{ 1.000,  1.000,  1.000}\textcolor[rgb]{ 0.000,  0.000,  0.000}{0.886 } & \cellcolor[rgb]{ 0.964,  0.477,  0.286}\textcolor[rgb]{ 0.000,  0.000,  0.000}{20.2  } & \cellcolor[rgb]{ 1.000,  1.000,  1.000}\textcolor[rgb]{ 0.000,  0.000,  0.000}{0.954 } & \cellcolor[rgb]{ 0.985,  0.994,  0.729}\textcolor[rgb]{ 0.000,  0.000,  0.000}{2.4   } & \cellcolor[rgb]{ 1.000,  1.000,  1.000}\textcolor[rgb]{ 0.000,  0.000,  0.000}{0.83  }\\
& $V_{fl}$                                 & \cellcolor[rgb]{ 0.983,  0.617,  0.350}\textcolor[rgb]{ 0.000,  0.000,  0.000}{4.52  } & \cellcolor[rgb]{ 1.000,  1.000,  1.000}\textcolor[rgb]{ 0.000,  0.000,  0.000}{0.833 } & \cellcolor[rgb]{ 0.964,  0.477,  0.286}\textcolor[rgb]{ 0.000,  0.000,  0.000}{6.38  } & \cellcolor[rgb]{ 1.000,  1.000,  1.000}\textcolor[rgb]{ 0.000,  0.000,  0.000}{0.901 } & \cellcolor[rgb]{ 0.939,  0.974,  0.665}\textcolor[rgb]{ 0.000,  0.000,  0.000}{2.15  } & \cellcolor[rgb]{ 1.000,  1.000,  1.000}\textcolor[rgb]{ 0.000,  0.000,  0.000}{0.749 } & \cellcolor[rgb]{ 0.965,  0.487,  0.290}\textcolor[rgb]{ 0.000,  0.000,  0.000}{4.99  } & \cellcolor[rgb]{ 1.000,  1.000,  1.000}\textcolor[rgb]{ 0.000,  0.000,  0.000}{0.824 } & \cellcolor[rgb]{ 0.857,  0.940,  0.553}\textcolor[rgb]{ 0.000,  0.000,  0.000}{1.65  } & \cellcolor[rgb]{ 1.000,  1.000,  1.000}\textcolor[rgb]{ 0.000,  0.000,  0.000}{0.696 }\\
& $\sigma\left(V_{fl}\right)$              & \cellcolor[rgb]{ 0.964,  0.477,  0.286}\textcolor[rgb]{ 0.000,  0.000,  0.000}{5.11  } & \cellcolor[rgb]{ 1.000,  1.000,  1.000}\textcolor[rgb]{ 0.000,  0.000,  0.000}{0.949 } & \cellcolor[rgb]{ 0.964,  0.477,  0.286}\textcolor[rgb]{ 0.000,  0.000,  0.000}{8.78  } & \cellcolor[rgb]{ 1.000,  1.000,  1.000}\textcolor[rgb]{ 0.000,  0.000,  0.000}{0.963 } & \cellcolor[rgb]{ 0.964,  0.477,  0.286}\textcolor[rgb]{ 0.000,  0.000,  0.000}{5.67  } & \cellcolor[rgb]{ 1.000,  1.000,  1.000}\textcolor[rgb]{ 0.000,  0.000,  0.000}{0.953 } & \cellcolor[rgb]{ 0.965,  0.487,  0.290}\textcolor[rgb]{ 0.000,  0.000,  0.000}{4.98  } & \cellcolor[rgb]{ 1.000,  1.000,  1.000}\textcolor[rgb]{ 0.000,  0.000,  0.000}{0.94  } & \cellcolor[rgb]{ 0.993,  0.702,  0.397}\textcolor[rgb]{ 0.000,  0.000,  0.000}{4.22  } & \cellcolor[rgb]{ 1.000,  1.000,  1.000}\textcolor[rgb]{ 0.000,  0.000,  0.000}{0.952 }\\
& $M_\parallel$                            & \cellcolor[rgb]{ 0.944,  0.977,  0.673}\textcolor[rgb]{ 0.000,  0.000,  0.000}{2.18  } & \cellcolor[rgb]{ 1.000,  1.000,  1.000}\textcolor[rgb]{ 0.000,  0.000,  0.000}{0.925 } & \cellcolor[rgb]{ 0.964,  0.477,  0.286}\textcolor[rgb]{ 0.000,  0.000,  0.000}{7.9   } & \cellcolor[rgb]{ 1.000,  1.000,  1.000}\textcolor[rgb]{ 0.000,  0.000,  0.000}{0.92  } & \cellcolor[rgb]{ 0.964,  0.477,  0.286}\textcolor[rgb]{ 0.000,  0.000,  0.000}{6.53  } & \cellcolor[rgb]{ 1.000,  1.000,  1.000}\textcolor[rgb]{ 0.000,  0.000,  0.000}{0.942 } & \cellcolor[rgb]{ 0.980,  0.597,  0.341}\textcolor[rgb]{ 0.000,  0.000,  0.000}{4.61  } & \cellcolor[rgb]{ 1.000,  1.000,  1.000}\textcolor[rgb]{ 0.000,  0.000,  0.000}{0.944 } & \cellcolor[rgb]{ 0.991,  0.996,  0.737}\textcolor[rgb]{ 0.000,  0.000,  0.000}{2.46  } & \cellcolor[rgb]{ 1.000,  1.000,  1.000}\textcolor[rgb]{ 0.000,  0.000,  0.000}{0.901 }\\
\cline{2-12}
& $\left(\chi; Q\right)$\textsubscript{FHRP} & \multicolumn{2}{c}{$ \textbf{(0.62; \ 4.02)} $ } & \multicolumn{2}{c}{$ \textbf{(0.61; \ 4.25)} $ } & \multicolumn{2}{c}{$ \textbf{(0.59; \ 4.06)} $ } & \multicolumn{2}{c}{$ \textbf{(0.69; \ 4.2)} $ } & \multicolumn{2}{c}{$ \textbf{(0.75; \ 4.01)} $ }\\
\midrule
\multirow{3}{*}{\makecell{Thomson scattering\\(TS) for divertor\\entrance}}
& $n$                                      & \cellcolor[rgb]{ 0.733,  0.887,  0.469}\textcolor[rgb]{ 0.000,  0.000,  0.000}{1.09  } & \cellcolor[rgb]{ 1.000,  1.000,  1.000}\textcolor[rgb]{ 0.000,  0.000,  0.000}{0.877 } & \cellcolor[rgb]{ 0.617,  0.836,  0.412}\textcolor[rgb]{ 0.000,  0.000,  0.000}{0.59  } & \cellcolor[rgb]{ 1.000,  1.000,  1.000}\textcolor[rgb]{ 0.000,  0.000,  0.000}{0.908 } & \cellcolor[rgb]{ 0.577,  0.819,  0.408}\textcolor[rgb]{ 0.000,  0.000,  0.000}{0.435 } & \cellcolor[rgb]{ 1.000,  1.000,  1.000}\textcolor[rgb]{ 0.000,  0.000,  0.000}{0.887 } & \cellcolor[rgb]{ 0.718,  0.880,  0.459}\textcolor[rgb]{ 0.000,  0.000,  0.000}{0.992 } & \cellcolor[rgb]{ 1.000,  1.000,  1.000}\textcolor[rgb]{ 0.000,  0.000,  0.000}{0.907 } & \cellcolor[rgb]{ 0.999,  0.983,  0.721}\textcolor[rgb]{ 0.000,  0.000,  0.000}{2.63  } & \cellcolor[rgb]{ 1.000,  1.000,  1.000}\textcolor[rgb]{ 0.000,  0.000,  0.000}{0.914 }\\
& $T_e$                                    & \cellcolor[rgb]{ 0.997,  0.893,  0.569}\textcolor[rgb]{ 0.000,  0.000,  0.000}{3.28  } & \cellcolor[rgb]{ 1.000,  1.000,  1.000}\textcolor[rgb]{ 0.000,  0.000,  0.000}{0.89  } & \cellcolor[rgb]{ 0.994,  0.763,  0.448}\textcolor[rgb]{ 0.000,  0.000,  0.000}{3.93  } & \cellcolor[rgb]{ 1.000,  1.000,  1.000}\textcolor[rgb]{ 0.000,  0.000,  0.000}{0.907 } & \cellcolor[rgb]{ 0.733,  0.887,  0.469}\textcolor[rgb]{ 0.000,  0.000,  0.000}{1.09  } & \cellcolor[rgb]{ 1.000,  1.000,  1.000}\textcolor[rgb]{ 0.000,  0.000,  0.000}{0.872 } & \cellcolor[rgb]{ 0.796,  0.914,  0.510}\textcolor[rgb]{ 0.000,  0.000,  0.000}{1.37  } & \cellcolor[rgb]{ 1.000,  1.000,  1.000}\textcolor[rgb]{ 0.000,  0.000,  0.000}{0.871 } & \cellcolor[rgb]{ 0.999,  0.969,  0.697}\textcolor[rgb]{ 0.000,  0.000,  0.000}{2.72  } & \cellcolor[rgb]{ 1.000,  1.000,  1.000}\textcolor[rgb]{ 0.000,  0.000,  0.000}{0.874 }\\
\cline{2-12}
& $\left(\chi; Q\right)$\textsubscript{TS} & \multicolumn{2}{c}{$ \textbf{(0.52; \ 0.883)} $ } & \multicolumn{2}{c}{$ \textbf{(0.5; \ 0.908)} $ } & \multicolumn{2}{c}{$ \textbf{(0.018; \ 0.88)} $ } & \multicolumn{2}{c}{$ \textbf{(0.1; \ 0.889)} $ } & \multicolumn{2}{c}{$ \textbf{(0.99; \ 0.894)} $ }\\
\midrule
\multirow{11}{*}{\makecell{Reciprocating\\divertor probe\\array (RDPA)\\for divertor\\volume}}
& $n$                                      & \cellcolor[rgb]{ 0.997,  0.907,  0.593}\textcolor[rgb]{ 0.000,  0.000,  0.000}{3.2   } & \cellcolor[rgb]{ 1.000,  1.000,  1.000}\textcolor[rgb]{ 0.000,  0.000,  0.000}{0.853 } & \cellcolor[rgb]{ 0.898,  0.957,  0.609}\textcolor[rgb]{ 0.000,  0.000,  0.000}{1.87  } & \cellcolor[rgb]{ 1.000,  1.000,  1.000}\textcolor[rgb]{ 0.000,  0.000,  0.000}{0.882 } & \cellcolor[rgb]{ 0.997,  0.907,  0.593}\textcolor[rgb]{ 0.000,  0.000,  0.000}{3.18  } & \cellcolor[rgb]{ 1.000,  1.000,  1.000}\textcolor[rgb]{ 0.000,  0.000,  0.000}{0.851 } & \cellcolor[rgb]{ 0.993,  0.725,  0.416}\textcolor[rgb]{ 0.000,  0.000,  0.000}{4.1   } & \cellcolor[rgb]{ 1.000,  1.000,  1.000}\textcolor[rgb]{ 0.000,  0.000,  0.000}{0.868 } & \cellcolor[rgb]{ 0.998,  0.945,  0.657}\textcolor[rgb]{ 0.000,  0.000,  0.000}{2.91  } & \cellcolor[rgb]{ 1.000,  1.000,  1.000}\textcolor[rgb]{ 0.000,  0.000,  0.000}{0.883 }\\
& $T_e$                                    & \cellcolor[rgb]{ 0.964,  0.477,  0.286}\textcolor[rgb]{ 0.000,  0.000,  0.000}{7.18  } & \cellcolor[rgb]{ 1.000,  1.000,  1.000}\textcolor[rgb]{ 0.000,  0.000,  0.000}{0.925 } & \cellcolor[rgb]{ 0.983,  0.617,  0.350}\textcolor[rgb]{ 0.000,  0.000,  0.000}{4.54  } & \cellcolor[rgb]{ 1.000,  1.000,  1.000}\textcolor[rgb]{ 0.000,  0.000,  0.000}{0.919 } & \cellcolor[rgb]{ 0.857,  0.940,  0.553}\textcolor[rgb]{ 0.000,  0.000,  0.000}{1.65  } & \cellcolor[rgb]{ 1.000,  1.000,  1.000}\textcolor[rgb]{ 0.000,  0.000,  0.000}{0.897 } & \cellcolor[rgb]{ 0.991,  0.996,  0.737}\textcolor[rgb]{ 0.000,  0.000,  0.000}{2.46  } & \cellcolor[rgb]{ 1.000,  1.000,  1.000}\textcolor[rgb]{ 0.000,  0.000,  0.000}{0.891 } & \cellcolor[rgb]{ 0.921,  0.967,  0.641}\textcolor[rgb]{ 0.000,  0.000,  0.000}{2.02  } & \cellcolor[rgb]{ 1.000,  1.000,  1.000}\textcolor[rgb]{ 0.000,  0.000,  0.000}{0.901 }\\
& $V_{pl}$                                 & \cellcolor[rgb]{ 0.964,  0.477,  0.286}\textcolor[rgb]{ 0.000,  0.000,  0.000}{5.6   } & \cellcolor[rgb]{ 1.000,  1.000,  1.000}\textcolor[rgb]{ 0.000,  0.000,  0.000}{0.926 } & \cellcolor[rgb]{ 0.969,  0.517,  0.304}\textcolor[rgb]{ 0.000,  0.000,  0.000}{4.86  } & \cellcolor[rgb]{ 1.000,  1.000,  1.000}\textcolor[rgb]{ 0.000,  0.000,  0.000}{0.903 } & \cellcolor[rgb]{ 0.892,  0.954,  0.601}\textcolor[rgb]{ 0.000,  0.000,  0.000}{1.84  } & \cellcolor[rgb]{ 1.000,  1.000,  1.000}\textcolor[rgb]{ 0.000,  0.000,  0.000}{0.88  } & \cellcolor[rgb]{ 0.995,  0.840,  0.513}\textcolor[rgb]{ 0.000,  0.000,  0.000}{3.56  } & \cellcolor[rgb]{ 1.000,  1.000,  1.000}\textcolor[rgb]{ 0.000,  0.000,  0.000}{0.886 } & \cellcolor[rgb]{ 0.874,  0.947,  0.577}\textcolor[rgb]{ 0.000,  0.000,  0.000}{1.74  } & \cellcolor[rgb]{ 1.000,  1.000,  1.000}\textcolor[rgb]{ 0.000,  0.000,  0.000}{0.883 }\\
& $J_{sat}$                                & \cellcolor[rgb]{ 0.997,  0.921,  0.617}\textcolor[rgb]{ 0.000,  0.000,  0.000}{3.09  } & \cellcolor[rgb]{ 1.000,  1.000,  1.000}\textcolor[rgb]{ 0.000,  0.000,  0.000}{0.856 } & \cellcolor[rgb]{ 0.968,  0.986,  0.705}\textcolor[rgb]{ 0.000,  0.000,  0.000}{2.3   } & \cellcolor[rgb]{ 1.000,  1.000,  1.000}\textcolor[rgb]{ 0.000,  0.000,  0.000}{0.869 } & \cellcolor[rgb]{ 0.982,  0.607,  0.346}\textcolor[rgb]{ 0.000,  0.000,  0.000}{4.54  } & \cellcolor[rgb]{ 1.000,  1.000,  1.000}\textcolor[rgb]{ 0.000,  0.000,  0.000}{0.875 } & \cellcolor[rgb]{ 0.964,  0.477,  0.286}\textcolor[rgb]{ 0.000,  0.000,  0.000}{8.98  } & \cellcolor[rgb]{ 1.000,  1.000,  1.000}\textcolor[rgb]{ 0.000,  0.000,  0.000}{0.872 } & \cellcolor[rgb]{ 0.964,  0.477,  0.286}\textcolor[rgb]{ 0.000,  0.000,  0.000}{25.7  } & \cellcolor[rgb]{ 1.000,  1.000,  1.000}\textcolor[rgb]{ 0.000,  0.000,  0.000}{0.902 }\\
& $\sigma\left(J_{sat}\right)$             & \cellcolor[rgb]{ 0.992,  0.686,  0.384}\textcolor[rgb]{ 0.000,  0.000,  0.000}{4.29  } & \cellcolor[rgb]{ 1.000,  1.000,  1.000}\textcolor[rgb]{ 0.000,  0.000,  0.000}{0.832 } & \cellcolor[rgb]{ 0.997,  0.893,  0.569}\textcolor[rgb]{ 0.000,  0.000,  0.000}{3.28  } & \cellcolor[rgb]{ 1.000,  1.000,  1.000}\textcolor[rgb]{ 0.000,  0.000,  0.000}{0.823 } & \cellcolor[rgb]{ 0.994,  0.771,  0.455}\textcolor[rgb]{ 0.000,  0.000,  0.000}{3.88  } & \cellcolor[rgb]{ 1.000,  1.000,  1.000}\textcolor[rgb]{ 0.000,  0.000,  0.000}{0.846 } & \cellcolor[rgb]{ 0.995,  0.817,  0.493}\textcolor[rgb]{ 0.000,  0.000,  0.000}{3.67  } & \cellcolor[rgb]{ 1.000,  1.000,  1.000}\textcolor[rgb]{ 0.000,  0.000,  0.000}{0.836 } & \cellcolor[rgb]{ 0.994,  0.755,  0.442}\textcolor[rgb]{ 0.000,  0.000,  0.000}{3.98  } & \cellcolor[rgb]{ 1.000,  1.000,  1.000}\textcolor[rgb]{ 0.000,  0.000,  0.000}{0.837 }\\
& $\mathrm{skew}\left(J_{sat}\right)$      & \cellcolor[rgb]{ 0.996,  0.871,  0.539}\textcolor[rgb]{ 0.000,  0.000,  0.000}{3.43  } & \cellcolor[rgb]{ 1.000,  1.000,  1.000}\textcolor[rgb]{ 0.000,  0.000,  0.000}{0.779 } & \cellcolor[rgb]{ 0.827,  0.927,  0.530}\textcolor[rgb]{ 0.000,  0.000,  0.000}{1.5   } & \cellcolor[rgb]{ 1.000,  1.000,  1.000}\textcolor[rgb]{ 0.000,  0.000,  0.000}{0.715 } & \cellcolor[rgb]{ 0.964,  0.477,  0.286}\textcolor[rgb]{ 0.000,  0.000,  0.000}{32.8  } & \cellcolor[rgb]{ 1.000,  1.000,  1.000}\textcolor[rgb]{ 0.000,  0.000,  0.000}{0.919 } & \cellcolor[rgb]{ 0.964,  0.477,  0.286}\textcolor[rgb]{ 0.000,  0.000,  0.000}{8.08  } & \cellcolor[rgb]{ 1.000,  1.000,  1.000}\textcolor[rgb]{ 0.000,  0.000,  0.000}{0.872 } & \cellcolor[rgb]{ 0.993,  0.725,  0.416}\textcolor[rgb]{ 0.000,  0.000,  0.000}{4.12  } & \cellcolor[rgb]{ 1.000,  1.000,  1.000}\textcolor[rgb]{ 0.000,  0.000,  0.000}{0.768 }\\
& $\mathrm{kurt}\left(J_{sat}\right)$      & \cellcolor[rgb]{ 0.939,  0.974,  0.665}\textcolor[rgb]{ 0.000,  0.000,  0.000}{2.13  } & \cellcolor[rgb]{ 1.000,  1.000,  1.000}\textcolor[rgb]{ 0.000,  0.000,  0.000}{0.883 } & \cellcolor[rgb]{ 0.863,  0.942,  0.561}\textcolor[rgb]{ 0.000,  0.000,  0.000}{1.67  } & \cellcolor[rgb]{ 1.000,  1.000,  1.000}\textcolor[rgb]{ 0.000,  0.000,  0.000}{0.901 } & \cellcolor[rgb]{ 0.964,  0.477,  0.286}\textcolor[rgb]{ 0.000,  0.000,  0.000}{415   } & \cellcolor[rgb]{ 1.000,  1.000,  1.000}\textcolor[rgb]{ 0.000,  0.000,  0.000}{0.989 } & \cellcolor[rgb]{ 0.964,  0.477,  0.286}\textcolor[rgb]{ 0.000,  0.000,  0.000}{94.0  } & \cellcolor[rgb]{ 1.000,  1.000,  1.000}\textcolor[rgb]{ 0.000,  0.000,  0.000}{0.97  } & \cellcolor[rgb]{ 0.950,  0.979,  0.681}\textcolor[rgb]{ 0.000,  0.000,  0.000}{2.18  } & \cellcolor[rgb]{ 1.000,  1.000,  1.000}\textcolor[rgb]{ 0.000,  0.000,  0.000}{0.885 }\\
& $V_{fl}$                                 & \cellcolor[rgb]{ 0.964,  0.477,  0.286}\textcolor[rgb]{ 0.000,  0.000,  0.000}{31.3  } & \cellcolor[rgb]{ 1.000,  1.000,  1.000}\textcolor[rgb]{ 0.000,  0.000,  0.000}{0.915 } & \cellcolor[rgb]{ 0.964,  0.477,  0.286}\textcolor[rgb]{ 0.000,  0.000,  0.000}{113   } & \cellcolor[rgb]{ 1.000,  1.000,  1.000}\textcolor[rgb]{ 0.000,  0.000,  0.000}{0.882 } & \cellcolor[rgb]{ 0.964,  0.477,  0.286}\textcolor[rgb]{ 0.000,  0.000,  0.000}{8.66  } & \cellcolor[rgb]{ 1.000,  1.000,  1.000}\textcolor[rgb]{ 0.000,  0.000,  0.000}{0.727 } & \cellcolor[rgb]{ 0.964,  0.477,  0.286}\textcolor[rgb]{ 0.000,  0.000,  0.000}{72.2  } & \cellcolor[rgb]{ 1.000,  1.000,  1.000}\textcolor[rgb]{ 0.000,  0.000,  0.000}{0.809 } & \cellcolor[rgb]{ 0.964,  0.477,  0.286}\textcolor[rgb]{ 0.000,  0.000,  0.000}{16.0  } & \cellcolor[rgb]{ 1.000,  1.000,  1.000}\textcolor[rgb]{ 0.000,  0.000,  0.000}{0.736 }\\
& $\sigma\left(V_{fl}\right)$              & \cellcolor[rgb]{ 0.964,  0.477,  0.286}\textcolor[rgb]{ 0.000,  0.000,  0.000}{59.6  } & \cellcolor[rgb]{ 1.000,  1.000,  1.000}\textcolor[rgb]{ 0.000,  0.000,  0.000}{0.911 } & \cellcolor[rgb]{ 0.964,  0.477,  0.286}\textcolor[rgb]{ 0.000,  0.000,  0.000}{6.42  } & \cellcolor[rgb]{ 1.000,  1.000,  1.000}\textcolor[rgb]{ 0.000,  0.000,  0.000}{0.897 } & \cellcolor[rgb]{ 0.964,  0.477,  0.286}\textcolor[rgb]{ 0.000,  0.000,  0.000}{48.2  } & \cellcolor[rgb]{ 1.000,  1.000,  1.000}\textcolor[rgb]{ 0.000,  0.000,  0.000}{0.902 } & \cellcolor[rgb]{ 0.964,  0.477,  0.286}\textcolor[rgb]{ 0.000,  0.000,  0.000}{9.25  } & \cellcolor[rgb]{ 1.000,  1.000,  1.000}\textcolor[rgb]{ 0.000,  0.000,  0.000}{0.866 } & \cellcolor[rgb]{ 0.964,  0.477,  0.286}\textcolor[rgb]{ 0.000,  0.000,  0.000}{48.4  } & \cellcolor[rgb]{ 1.000,  1.000,  1.000}\textcolor[rgb]{ 0.000,  0.000,  0.000}{0.899 }\\
& $M_\parallel$                            & \cellcolor[rgb]{ 0.964,  0.477,  0.286}\textcolor[rgb]{ 0.000,  0.000,  0.000}{11.3  } & \cellcolor[rgb]{ 1.000,  1.000,  1.000}\textcolor[rgb]{ 0.000,  0.000,  0.000}{0.912 } & \cellcolor[rgb]{ 0.964,  0.477,  0.286}\textcolor[rgb]{ 0.000,  0.000,  0.000}{14.7  } & \cellcolor[rgb]{ 1.000,  1.000,  1.000}\textcolor[rgb]{ 0.000,  0.000,  0.000}{0.91  } & \cellcolor[rgb]{ 0.964,  0.477,  0.286}\textcolor[rgb]{ 0.000,  0.000,  0.000}{16.8  } & \cellcolor[rgb]{ 1.000,  1.000,  1.000}\textcolor[rgb]{ 0.000,  0.000,  0.000}{0.926 } & \cellcolor[rgb]{ 0.964,  0.477,  0.286}\textcolor[rgb]{ 0.000,  0.000,  0.000}{25.2  } & \cellcolor[rgb]{ 1.000,  1.000,  1.000}\textcolor[rgb]{ 0.000,  0.000,  0.000}{0.943 } & \cellcolor[rgb]{ 0.964,  0.477,  0.286}\textcolor[rgb]{ 0.000,  0.000,  0.000}{21.6  } & \cellcolor[rgb]{ 1.000,  1.000,  1.000}\textcolor[rgb]{ 0.000,  0.000,  0.000}{0.948 }\\
\cline{2-12}
& $\left(\chi; Q\right)$\textsubscript{RDPA} & \multicolumn{2}{c}{$ \textbf{(0.99; \ 4.17)} $ } & \multicolumn{2}{c}{$ \textbf{(0.87; \ 4.12)} $ } & \multicolumn{2}{c}{$ \textbf{(0.93; \ 4.17)} $ } & \multicolumn{2}{c}{$ \textbf{(1.0; \ 4.17)} $ } & \multicolumn{2}{c}{$ \textbf{(0.95; \ 4.08)} $ }\\
\midrule
\multirow{2}{*}{\makecell{Infrared camera (IR)\\for low-field-side target}}
& $q_\parallel$                            & \cellcolor[rgb]{ 0.993,  0.702,  0.397}\textcolor[rgb]{ 0.000,  0.000,  0.000}{4.19  } & \cellcolor[rgb]{ 1.000,  1.000,  1.000}\textcolor[rgb]{ 0.000,  0.000,  0.000}{0.878 } & \cellcolor[rgb]{ 0.964,  0.477,  0.286}\textcolor[rgb]{ 0.000,  0.000,  0.000}{10.2  } & \cellcolor[rgb]{ 1.000,  1.000,  1.000}\textcolor[rgb]{ 0.000,  0.000,  0.000}{0.939 } & \cellcolor[rgb]{ 0.985,  0.627,  0.355}\textcolor[rgb]{ 0.000,  0.000,  0.000}{4.49  } & \cellcolor[rgb]{ 1.000,  1.000,  1.000}\textcolor[rgb]{ 0.000,  0.000,  0.000}{0.866 } & \cellcolor[rgb]{ 0.964,  0.477,  0.286}\textcolor[rgb]{ 0.000,  0.000,  0.000}{7.87  } & \cellcolor[rgb]{ 1.000,  1.000,  1.000}\textcolor[rgb]{ 0.000,  0.000,  0.000}{0.917 } & \cellcolor[rgb]{ 0.964,  0.477,  0.286}\textcolor[rgb]{ 0.000,  0.000,  0.000}{6.19  } & \cellcolor[rgb]{ 1.000,  1.000,  1.000}\textcolor[rgb]{ 0.000,  0.000,  0.000}{0.887 }\\
\cline{2-12}
& $\left(\chi; Q\right)$\textsubscript{LFS-IR} & \multicolumn{2}{c}{$ \textbf{(1.0; \ 0.293)} $ } & \multicolumn{2}{c}{$ \textbf{(1.0; \ 0.313)} $ } & \multicolumn{2}{c}{$ \textbf{(1.0; \ 0.289)} $ } & \multicolumn{2}{c}{$ \textbf{(1.0; \ 0.306)} $ } & \multicolumn{2}{c}{$ \textbf{(1.0; \ 0.296)} $ }\\
\midrule
\multirow{12}{*}{\makecell{Wall Langmuir\\probes for\\low-field-side\\target}}
& $n$                                      & \cellcolor[rgb]{ 0.886,  0.952,  0.593}\textcolor[rgb]{ 0.000,  0.000,  0.000}{1.81  } & \cellcolor[rgb]{ 1.000,  1.000,  1.000}\textcolor[rgb]{ 0.000,  0.000,  0.000}{0.861 } & \cellcolor[rgb]{ 0.993,  0.702,  0.397}\textcolor[rgb]{ 0.000,  0.000,  0.000}{4.2   } & \cellcolor[rgb]{ 1.000,  1.000,  1.000}\textcolor[rgb]{ 0.000,  0.000,  0.000}{0.89  } & \cellcolor[rgb]{ 0.886,  0.952,  0.593}\textcolor[rgb]{ 0.000,  0.000,  0.000}{1.81  } & \cellcolor[rgb]{ 1.000,  1.000,  1.000}\textcolor[rgb]{ 0.000,  0.000,  0.000}{0.859 } & \cellcolor[rgb]{ 0.974,  0.989,  0.713}\textcolor[rgb]{ 0.000,  0.000,  0.000}{2.35  } & \cellcolor[rgb]{ 1.000,  1.000,  1.000}\textcolor[rgb]{ 0.000,  0.000,  0.000}{0.862 } & \cellcolor[rgb]{ 0.997,  0.893,  0.569}\textcolor[rgb]{ 0.000,  0.000,  0.000}{3.28  } & \cellcolor[rgb]{ 1.000,  1.000,  1.000}\textcolor[rgb]{ 0.000,  0.000,  0.000}{0.902 }\\
& $T_e$                                    & \cellcolor[rgb]{ 0.964,  0.477,  0.286}\textcolor[rgb]{ 0.000,  0.000,  0.000}{6.01  } & \cellcolor[rgb]{ 1.000,  1.000,  1.000}\textcolor[rgb]{ 0.000,  0.000,  0.000}{0.937 } & \cellcolor[rgb]{ 0.995,  0.825,  0.500}\textcolor[rgb]{ 0.000,  0.000,  0.000}{3.63  } & \cellcolor[rgb]{ 1.000,  1.000,  1.000}\textcolor[rgb]{ 0.000,  0.000,  0.000}{0.911 } & \cellcolor[rgb]{ 0.874,  0.947,  0.577}\textcolor[rgb]{ 0.000,  0.000,  0.000}{1.76  } & \cellcolor[rgb]{ 1.000,  1.000,  1.000}\textcolor[rgb]{ 0.000,  0.000,  0.000}{0.907 } & \cellcolor[rgb]{ 0.909,  0.962,  0.625}\textcolor[rgb]{ 0.000,  0.000,  0.000}{1.94  } & \cellcolor[rgb]{ 1.000,  1.000,  1.000}\textcolor[rgb]{ 0.000,  0.000,  0.000}{0.868 } & \cellcolor[rgb]{ 0.898,  0.957,  0.609}\textcolor[rgb]{ 0.000,  0.000,  0.000}{1.88  } & \cellcolor[rgb]{ 1.000,  1.000,  1.000}\textcolor[rgb]{ 0.000,  0.000,  0.000}{0.908 }\\
& $V_{pl}$                                 & \cellcolor[rgb]{ 0.964,  0.477,  0.286}\textcolor[rgb]{ 0.000,  0.000,  0.000}{9.63  } & \cellcolor[rgb]{ 1.000,  1.000,  1.000}\textcolor[rgb]{ 0.000,  0.000,  0.000}{0.951 } & \cellcolor[rgb]{ 0.994,  0.794,  0.474}\textcolor[rgb]{ 0.000,  0.000,  0.000}{3.79  } & \cellcolor[rgb]{ 1.000,  1.000,  1.000}\textcolor[rgb]{ 0.000,  0.000,  0.000}{0.925 } & \cellcolor[rgb]{ 1.000,  0.993,  0.737}\textcolor[rgb]{ 0.000,  0.000,  0.000}{2.55  } & \cellcolor[rgb]{ 1.000,  1.000,  1.000}\textcolor[rgb]{ 0.000,  0.000,  0.000}{0.912 } & \cellcolor[rgb]{ 0.972,  0.537,  0.313}\textcolor[rgb]{ 0.000,  0.000,  0.000}{4.8   } & \cellcolor[rgb]{ 1.000,  1.000,  1.000}\textcolor[rgb]{ 0.000,  0.000,  0.000}{0.896 } & \cellcolor[rgb]{ 0.962,  0.984,  0.697}\textcolor[rgb]{ 0.000,  0.000,  0.000}{2.26  } & \cellcolor[rgb]{ 1.000,  1.000,  1.000}\textcolor[rgb]{ 0.000,  0.000,  0.000}{0.915 }\\
& $J_{sat}$                                & \cellcolor[rgb]{ 0.998,  0.945,  0.657}\textcolor[rgb]{ 0.000,  0.000,  0.000}{2.9   } & \cellcolor[rgb]{ 1.000,  1.000,  1.000}\textcolor[rgb]{ 0.000,  0.000,  0.000}{0.891 } & \cellcolor[rgb]{ 0.964,  0.477,  0.286}\textcolor[rgb]{ 0.000,  0.000,  0.000}{16.1  } & \cellcolor[rgb]{ 1.000,  1.000,  1.000}\textcolor[rgb]{ 0.000,  0.000,  0.000}{0.942 } & \cellcolor[rgb]{ 0.997,  0.902,  0.585}\textcolor[rgb]{ 0.000,  0.000,  0.000}{3.22  } & \cellcolor[rgb]{ 1.000,  1.000,  1.000}\textcolor[rgb]{ 0.000,  0.000,  0.000}{0.884 } & \cellcolor[rgb]{ 0.995,  0.832,  0.506}\textcolor[rgb]{ 0.000,  0.000,  0.000}{3.62  } & \cellcolor[rgb]{ 1.000,  1.000,  1.000}\textcolor[rgb]{ 0.000,  0.000,  0.000}{0.88  } & \cellcolor[rgb]{ 0.999,  0.964,  0.689}\textcolor[rgb]{ 0.000,  0.000,  0.000}{2.76  } & \cellcolor[rgb]{ 1.000,  1.000,  1.000}\textcolor[rgb]{ 0.000,  0.000,  0.000}{0.91  }\\
& $\sigma\left(J_{sat}\right)$             & \cellcolor[rgb]{ 0.967,  0.497,  0.295}\textcolor[rgb]{ 0.000,  0.000,  0.000}{4.93  } & \cellcolor[rgb]{ 1.000,  1.000,  1.000}\textcolor[rgb]{ 0.000,  0.000,  0.000}{0.859 } & \cellcolor[rgb]{ 0.996,  0.888,  0.561}\textcolor[rgb]{ 0.000,  0.000,  0.000}{3.34  } & \cellcolor[rgb]{ 1.000,  1.000,  1.000}\textcolor[rgb]{ 0.000,  0.000,  0.000}{0.894 } & \cellcolor[rgb]{ 0.964,  0.477,  0.286}\textcolor[rgb]{ 0.000,  0.000,  0.000}{5.52  } & \cellcolor[rgb]{ 1.000,  1.000,  1.000}\textcolor[rgb]{ 0.000,  0.000,  0.000}{0.854 } & \cellcolor[rgb]{ 0.964,  0.477,  0.286}\textcolor[rgb]{ 0.000,  0.000,  0.000}{5.3   } & \cellcolor[rgb]{ 1.000,  1.000,  1.000}\textcolor[rgb]{ 0.000,  0.000,  0.000}{0.872 } & \cellcolor[rgb]{ 0.964,  0.477,  0.286}\textcolor[rgb]{ 0.000,  0.000,  0.000}{5.38  } & \cellcolor[rgb]{ 1.000,  1.000,  1.000}\textcolor[rgb]{ 0.000,  0.000,  0.000}{0.85  }\\
& $\mathrm{skew}\left(J_{sat}\right)$      & \cellcolor[rgb]{ 0.998,  0.931,  0.633}\textcolor[rgb]{ 0.000,  0.000,  0.000}{3.02  } & \cellcolor[rgb]{ 1.000,  1.000,  1.000}\textcolor[rgb]{ 0.000,  0.000,  0.000}{0.849 } & \cellcolor[rgb]{ 0.964,  0.477,  0.286}\textcolor[rgb]{ 0.000,  0.000,  0.000}{9.03  } & \cellcolor[rgb]{ 1.000,  1.000,  1.000}\textcolor[rgb]{ 0.000,  0.000,  0.000}{0.922 } & \cellcolor[rgb]{ 0.964,  0.477,  0.286}\textcolor[rgb]{ 0.000,  0.000,  0.000}{71.3  } & \cellcolor[rgb]{ 1.000,  1.000,  1.000}\textcolor[rgb]{ 0.000,  0.000,  0.000}{0.957 } & \cellcolor[rgb]{ 0.964,  0.477,  0.286}\textcolor[rgb]{ 0.000,  0.000,  0.000}{45.5  } & \cellcolor[rgb]{ 1.000,  1.000,  1.000}\textcolor[rgb]{ 0.000,  0.000,  0.000}{0.943 } & \cellcolor[rgb]{ 0.964,  0.477,  0.286}\textcolor[rgb]{ 0.000,  0.000,  0.000}{5.4   } & \cellcolor[rgb]{ 1.000,  1.000,  1.000}\textcolor[rgb]{ 0.000,  0.000,  0.000}{0.808 }\\
& $\mathrm{kurt}\left(J_{sat}\right)$      & \cellcolor[rgb]{ 0.909,  0.962,  0.625}\textcolor[rgb]{ 0.000,  0.000,  0.000}{1.96  } & \cellcolor[rgb]{ 1.000,  1.000,  1.000}\textcolor[rgb]{ 0.000,  0.000,  0.000}{0.904 } & \cellcolor[rgb]{ 0.964,  0.477,  0.286}\textcolor[rgb]{ 0.000,  0.000,  0.000}{15.5  } & \cellcolor[rgb]{ 1.000,  1.000,  1.000}\textcolor[rgb]{ 0.000,  0.000,  0.000}{0.971 } & \cellcolor[rgb]{ 0.964,  0.477,  0.286}\textcolor[rgb]{ 0.000,  0.000,  0.000}{1290  } & \cellcolor[rgb]{ 1.000,  1.000,  1.000}\textcolor[rgb]{ 0.000,  0.000,  0.000}{0.994 } & \cellcolor[rgb]{ 0.964,  0.477,  0.286}\textcolor[rgb]{ 0.000,  0.000,  0.000}{68.1  } & \cellcolor[rgb]{ 1.000,  1.000,  1.000}\textcolor[rgb]{ 0.000,  0.000,  0.000}{0.982 } & \cellcolor[rgb]{ 0.999,  0.974,  0.705}\textcolor[rgb]{ 0.000,  0.000,  0.000}{2.7   } & \cellcolor[rgb]{ 1.000,  1.000,  1.000}\textcolor[rgb]{ 0.000,  0.000,  0.000}{0.895 }\\
& $J_\parallel$                            & \cellcolor[rgb]{ 0.998,  0.940,  0.649}\textcolor[rgb]{ 0.000,  0.000,  0.000}{2.93  } & \cellcolor[rgb]{ 1.000,  1.000,  1.000}\textcolor[rgb]{ 0.000,  0.000,  0.000}{0.765 } & \cellcolor[rgb]{ 0.964,  0.477,  0.286}\textcolor[rgb]{ 0.000,  0.000,  0.000}{7.53  } & \cellcolor[rgb]{ 1.000,  1.000,  1.000}\textcolor[rgb]{ 0.000,  0.000,  0.000}{0.863 } & \cellcolor[rgb]{ 0.964,  0.477,  0.286}\textcolor[rgb]{ 0.000,  0.000,  0.000}{7.78  } & \cellcolor[rgb]{ 1.000,  1.000,  1.000}\textcolor[rgb]{ 0.000,  0.000,  0.000}{0.841 } & \cellcolor[rgb]{ 0.993,  0.709,  0.403}\textcolor[rgb]{ 0.000,  0.000,  0.000}{4.16  } & \cellcolor[rgb]{ 1.000,  1.000,  1.000}\textcolor[rgb]{ 0.000,  0.000,  0.000}{0.85  } & \cellcolor[rgb]{ 0.992,  0.686,  0.384}\textcolor[rgb]{ 0.000,  0.000,  0.000}{4.29  } & \cellcolor[rgb]{ 1.000,  1.000,  1.000}\textcolor[rgb]{ 0.000,  0.000,  0.000}{0.735 }\\
& $\sigma\left(J_\parallel\right)$         & \cellcolor[rgb]{ 0.997,  0.898,  0.577}\textcolor[rgb]{ 0.000,  0.000,  0.000}{3.26  } & \cellcolor[rgb]{ 1.000,  1.000,  1.000}\textcolor[rgb]{ 0.000,  0.000,  0.000}{0.884 } & \cellcolor[rgb]{ 0.964,  0.477,  0.286}\textcolor[rgb]{ 0.000,  0.000,  0.000}{10.4  } & \cellcolor[rgb]{ 1.000,  1.000,  1.000}\textcolor[rgb]{ 0.000,  0.000,  0.000}{0.92  } & \cellcolor[rgb]{ 0.999,  0.969,  0.697}\textcolor[rgb]{ 0.000,  0.000,  0.000}{2.71  } & \cellcolor[rgb]{ 1.000,  1.000,  1.000}\textcolor[rgb]{ 0.000,  0.000,  0.000}{0.896 } & \cellcolor[rgb]{ 0.996,  0.855,  0.526}\textcolor[rgb]{ 0.000,  0.000,  0.000}{3.51  } & \cellcolor[rgb]{ 1.000,  1.000,  1.000}\textcolor[rgb]{ 0.000,  0.000,  0.000}{0.905 } & \cellcolor[rgb]{ 0.992,  0.694,  0.390}\textcolor[rgb]{ 0.000,  0.000,  0.000}{4.24  } & \cellcolor[rgb]{ 1.000,  1.000,  1.000}\textcolor[rgb]{ 0.000,  0.000,  0.000}{0.844 }\\
& $V_{fl}$                                 & \cellcolor[rgb]{ 0.964,  0.477,  0.286}\textcolor[rgb]{ 0.000,  0.000,  0.000}{6.54  } & \cellcolor[rgb]{ 1.000,  1.000,  1.000}\textcolor[rgb]{ 0.000,  0.000,  0.000}{0.854 } & \cellcolor[rgb]{ 0.964,  0.477,  0.286}\textcolor[rgb]{ 0.000,  0.000,  0.000}{5.66  } & \cellcolor[rgb]{ 1.000,  1.000,  1.000}\textcolor[rgb]{ 0.000,  0.000,  0.000}{0.794 } & \cellcolor[rgb]{ 0.939,  0.974,  0.665}\textcolor[rgb]{ 0.000,  0.000,  0.000}{2.14  } & \cellcolor[rgb]{ 1.000,  1.000,  1.000}\textcolor[rgb]{ 0.000,  0.000,  0.000}{0.64  } & \cellcolor[rgb]{ 0.964,  0.477,  0.286}\textcolor[rgb]{ 0.000,  0.000,  0.000}{5.9   } & \cellcolor[rgb]{ 1.000,  1.000,  1.000}\textcolor[rgb]{ 0.000,  0.000,  0.000}{0.734 } & \cellcolor[rgb]{ 0.999,  0.969,  0.697}\textcolor[rgb]{ 0.000,  0.000,  0.000}{2.74  } & \cellcolor[rgb]{ 1.000,  1.000,  1.000}\textcolor[rgb]{ 0.000,  0.000,  0.000}{0.662 }\\
& $\sigma\left(V_{fl}\right)$              & \cellcolor[rgb]{ 0.964,  0.477,  0.286}\textcolor[rgb]{ 0.000,  0.000,  0.000}{5.83  } & \cellcolor[rgb]{ 1.000,  1.000,  1.000}\textcolor[rgb]{ 0.000,  0.000,  0.000}{0.907 } & \cellcolor[rgb]{ 0.964,  0.477,  0.286}\textcolor[rgb]{ 0.000,  0.000,  0.000}{6.88  } & \cellcolor[rgb]{ 1.000,  1.000,  1.000}\textcolor[rgb]{ 0.000,  0.000,  0.000}{0.916 } & \cellcolor[rgb]{ 0.964,  0.477,  0.286}\textcolor[rgb]{ 0.000,  0.000,  0.000}{7.49  } & \cellcolor[rgb]{ 1.000,  1.000,  1.000}\textcolor[rgb]{ 0.000,  0.000,  0.000}{0.894 } & \cellcolor[rgb]{ 0.964,  0.477,  0.286}\textcolor[rgb]{ 0.000,  0.000,  0.000}{6.1   } & \cellcolor[rgb]{ 1.000,  1.000,  1.000}\textcolor[rgb]{ 0.000,  0.000,  0.000}{0.909 } & \cellcolor[rgb]{ 0.964,  0.477,  0.286}\textcolor[rgb]{ 0.000,  0.000,  0.000}{7.4   } & \cellcolor[rgb]{ 1.000,  1.000,  1.000}\textcolor[rgb]{ 0.000,  0.000,  0.000}{0.893 }\\
\cline{2-12}
& $\left(\chi; Q\right)$\textsubscript{LFS-LP} & \multicolumn{2}{c}{$ \textbf{(0.96; \ 4.83)} $ } & \multicolumn{2}{c}{$ \textbf{(1.0; \ 4.97)} $ } & \multicolumn{2}{c}{$ \textbf{(0.94; \ 4.82)} $ } & \multicolumn{2}{c}{$ \textbf{(0.98; \ 4.85)} $ } & \multicolumn{2}{c}{$ \textbf{(0.98; \ 4.66)} $ }\\
\midrule
\multirow{12}{*}{\makecell{Wall Langmuir\\probes for\\high-field-side\\target}}
& $n$                                      & \cellcolor[rgb]{ 0.985,  0.627,  0.355}\textcolor[rgb]{ 0.000,  0.000,  0.000}{4.49  } & \cellcolor[rgb]{ 1.000,  1.000,  1.000}\textcolor[rgb]{ 0.000,  0.000,  0.000}{0.879 } & \cellcolor[rgb]{ 0.995,  0.802,  0.481}\textcolor[rgb]{ 0.000,  0.000,  0.000}{3.75  } & \cellcolor[rgb]{ 1.000,  1.000,  1.000}\textcolor[rgb]{ 0.000,  0.000,  0.000}{0.909 } & \cellcolor[rgb]{ 0.993,  0.702,  0.397}\textcolor[rgb]{ 0.000,  0.000,  0.000}{4.22  } & \cellcolor[rgb]{ 1.000,  1.000,  1.000}\textcolor[rgb]{ 0.000,  0.000,  0.000}{0.884 } & \cellcolor[rgb]{ 0.851,  0.937,  0.545}\textcolor[rgb]{ 0.000,  0.000,  0.000}{1.62  } & \cellcolor[rgb]{ 1.000,  1.000,  1.000}\textcolor[rgb]{ 0.000,  0.000,  0.000}{0.884 } & \cellcolor[rgb]{ 0.965,  0.487,  0.290}\textcolor[rgb]{ 0.000,  0.000,  0.000}{4.97  } & \cellcolor[rgb]{ 1.000,  1.000,  1.000}\textcolor[rgb]{ 0.000,  0.000,  0.000}{0.924 }\\
& $T_e$                                    & \cellcolor[rgb]{ 0.964,  0.477,  0.286}\textcolor[rgb]{ 0.000,  0.000,  0.000}{13.7  } & \cellcolor[rgb]{ 1.000,  1.000,  1.000}\textcolor[rgb]{ 0.000,  0.000,  0.000}{0.958 } & \cellcolor[rgb]{ 0.964,  0.477,  0.286}\textcolor[rgb]{ 0.000,  0.000,  0.000}{6.21  } & \cellcolor[rgb]{ 1.000,  1.000,  1.000}\textcolor[rgb]{ 0.000,  0.000,  0.000}{0.926 } & \cellcolor[rgb]{ 0.999,  0.983,  0.721}\textcolor[rgb]{ 0.000,  0.000,  0.000}{2.61  } & \cellcolor[rgb]{ 1.000,  1.000,  1.000}\textcolor[rgb]{ 0.000,  0.000,  0.000}{0.923 } & \cellcolor[rgb]{ 0.991,  0.996,  0.737}\textcolor[rgb]{ 0.000,  0.000,  0.000}{2.43  } & \cellcolor[rgb]{ 1.000,  1.000,  1.000}\textcolor[rgb]{ 0.000,  0.000,  0.000}{0.888 } & \cellcolor[rgb]{ 0.997,  0.893,  0.569}\textcolor[rgb]{ 0.000,  0.000,  0.000}{3.29  } & \cellcolor[rgb]{ 1.000,  1.000,  1.000}\textcolor[rgb]{ 0.000,  0.000,  0.000}{0.938 }\\
& $V_{pl}$                                 & \cellcolor[rgb]{ 0.964,  0.477,  0.286}\textcolor[rgb]{ 0.000,  0.000,  0.000}{9.85  } & \cellcolor[rgb]{ 1.000,  1.000,  1.000}\textcolor[rgb]{ 0.000,  0.000,  0.000}{0.959 } & \cellcolor[rgb]{ 0.964,  0.477,  0.286}\textcolor[rgb]{ 0.000,  0.000,  0.000}{6.66  } & \cellcolor[rgb]{ 1.000,  1.000,  1.000}\textcolor[rgb]{ 0.000,  0.000,  0.000}{0.93  } & \cellcolor[rgb]{ 0.992,  0.686,  0.384}\textcolor[rgb]{ 0.000,  0.000,  0.000}{4.27  } & \cellcolor[rgb]{ 1.000,  1.000,  1.000}\textcolor[rgb]{ 0.000,  0.000,  0.000}{0.927 } & \cellcolor[rgb]{ 0.995,  0.825,  0.500}\textcolor[rgb]{ 0.000,  0.000,  0.000}{3.64  } & \cellcolor[rgb]{ 1.000,  1.000,  1.000}\textcolor[rgb]{ 0.000,  0.000,  0.000}{0.896 } & \cellcolor[rgb]{ 0.956,  0.982,  0.689}\textcolor[rgb]{ 0.000,  0.000,  0.000}{2.25  } & \cellcolor[rgb]{ 1.000,  1.000,  1.000}\textcolor[rgb]{ 0.000,  0.000,  0.000}{0.941 }\\
& $J_{sat}$                                & \cellcolor[rgb]{ 1.000,  0.988,  0.729}\textcolor[rgb]{ 0.000,  0.000,  0.000}{2.59  } & \cellcolor[rgb]{ 1.000,  1.000,  1.000}\textcolor[rgb]{ 0.000,  0.000,  0.000}{0.861 } & \cellcolor[rgb]{ 0.964,  0.477,  0.286}\textcolor[rgb]{ 0.000,  0.000,  0.000}{8.76  } & \cellcolor[rgb]{ 1.000,  1.000,  1.000}\textcolor[rgb]{ 0.000,  0.000,  0.000}{0.93  } & \cellcolor[rgb]{ 0.999,  0.964,  0.689}\textcolor[rgb]{ 0.000,  0.000,  0.000}{2.75  } & \cellcolor[rgb]{ 1.000,  1.000,  1.000}\textcolor[rgb]{ 0.000,  0.000,  0.000}{0.857 } & \cellcolor[rgb]{ 0.980,  0.991,  0.721}\textcolor[rgb]{ 0.000,  0.000,  0.000}{2.39  } & \cellcolor[rgb]{ 1.000,  1.000,  1.000}\textcolor[rgb]{ 0.000,  0.000,  0.000}{0.858 } & \cellcolor[rgb]{ 0.980,  0.597,  0.341}\textcolor[rgb]{ 0.000,  0.000,  0.000}{4.6   } & \cellcolor[rgb]{ 1.000,  1.000,  1.000}\textcolor[rgb]{ 0.000,  0.000,  0.000}{0.904 }\\
& $\sigma\left(J_{sat}\right)$             & \cellcolor[rgb]{ 0.983,  0.617,  0.350}\textcolor[rgb]{ 0.000,  0.000,  0.000}{4.51  } & \cellcolor[rgb]{ 1.000,  1.000,  1.000}\textcolor[rgb]{ 0.000,  0.000,  0.000}{0.797 } & \cellcolor[rgb]{ 0.993,  0.709,  0.403}\textcolor[rgb]{ 0.000,  0.000,  0.000}{4.16  } & \cellcolor[rgb]{ 1.000,  1.000,  1.000}\textcolor[rgb]{ 0.000,  0.000,  0.000}{0.874 } & \cellcolor[rgb]{ 0.978,  0.577,  0.332}\textcolor[rgb]{ 0.000,  0.000,  0.000}{4.65  } & \cellcolor[rgb]{ 1.000,  1.000,  1.000}\textcolor[rgb]{ 0.000,  0.000,  0.000}{0.8   } & \cellcolor[rgb]{ 0.964,  0.477,  0.286}\textcolor[rgb]{ 0.000,  0.000,  0.000}{5.08  } & \cellcolor[rgb]{ 1.000,  1.000,  1.000}\textcolor[rgb]{ 0.000,  0.000,  0.000}{0.856 } & \cellcolor[rgb]{ 0.993,  0.709,  0.403}\textcolor[rgb]{ 0.000,  0.000,  0.000}{4.16  } & \cellcolor[rgb]{ 1.000,  1.000,  1.000}\textcolor[rgb]{ 0.000,  0.000,  0.000}{0.806 }\\
& $\mathrm{skew}\left(J_{sat}\right)$      & \cellcolor[rgb]{ 0.997,  0.907,  0.593}\textcolor[rgb]{ 0.000,  0.000,  0.000}{3.2   } & \cellcolor[rgb]{ 1.000,  1.000,  1.000}\textcolor[rgb]{ 0.000,  0.000,  0.000}{0.796 } & \cellcolor[rgb]{ 0.993,  0.717,  0.409}\textcolor[rgb]{ 0.000,  0.000,  0.000}{4.14  } & \cellcolor[rgb]{ 1.000,  1.000,  1.000}\textcolor[rgb]{ 0.000,  0.000,  0.000}{0.767 } & \cellcolor[rgb]{ 0.964,  0.477,  0.286}\textcolor[rgb]{ 0.000,  0.000,  0.000}{5.6   } & \cellcolor[rgb]{ 1.000,  1.000,  1.000}\textcolor[rgb]{ 0.000,  0.000,  0.000}{0.835 } & \cellcolor[rgb]{ 0.964,  0.477,  0.286}\textcolor[rgb]{ 0.000,  0.000,  0.000}{7.62  } & \cellcolor[rgb]{ 1.000,  1.000,  1.000}\textcolor[rgb]{ 0.000,  0.000,  0.000}{0.903 } & \cellcolor[rgb]{ 0.964,  0.477,  0.286}\textcolor[rgb]{ 0.000,  0.000,  0.000}{5.7   } & \cellcolor[rgb]{ 1.000,  1.000,  1.000}\textcolor[rgb]{ 0.000,  0.000,  0.000}{0.711 }\\
& $\mathrm{kurt}\left(J_{sat}\right)$      & \cellcolor[rgb]{ 0.835,  0.930,  0.535}\textcolor[rgb]{ 0.000,  0.000,  0.000}{1.54  } & \cellcolor[rgb]{ 1.000,  1.000,  1.000}\textcolor[rgb]{ 0.000,  0.000,  0.000}{0.92  } & \cellcolor[rgb]{ 0.996,  0.878,  0.545}\textcolor[rgb]{ 0.000,  0.000,  0.000}{3.39  } & \cellcolor[rgb]{ 1.000,  1.000,  1.000}\textcolor[rgb]{ 0.000,  0.000,  0.000}{0.961 } & \cellcolor[rgb]{ 0.964,  0.477,  0.286}\textcolor[rgb]{ 0.000,  0.000,  0.000}{28.4  } & \cellcolor[rgb]{ 1.000,  1.000,  1.000}\textcolor[rgb]{ 0.000,  0.000,  0.000}{0.962 } & \cellcolor[rgb]{ 0.964,  0.477,  0.286}\textcolor[rgb]{ 0.000,  0.000,  0.000}{31.4  } & \cellcolor[rgb]{ 1.000,  1.000,  1.000}\textcolor[rgb]{ 0.000,  0.000,  0.000}{0.98  } & \cellcolor[rgb]{ 0.939,  0.974,  0.665}\textcolor[rgb]{ 0.000,  0.000,  0.000}{2.14  } & \cellcolor[rgb]{ 1.000,  1.000,  1.000}\textcolor[rgb]{ 0.000,  0.000,  0.000}{0.918 }\\
& $J_\parallel$                            & \cellcolor[rgb]{ 0.964,  0.477,  0.286}\textcolor[rgb]{ 0.000,  0.000,  0.000}{5.35  } & \cellcolor[rgb]{ 1.000,  1.000,  1.000}\textcolor[rgb]{ 0.000,  0.000,  0.000}{0.821 } & \cellcolor[rgb]{ 0.964,  0.477,  0.286}\textcolor[rgb]{ 0.000,  0.000,  0.000}{10.1  } & \cellcolor[rgb]{ 1.000,  1.000,  1.000}\textcolor[rgb]{ 0.000,  0.000,  0.000}{0.864 } & \cellcolor[rgb]{ 0.964,  0.477,  0.286}\textcolor[rgb]{ 0.000,  0.000,  0.000}{16.1  } & \cellcolor[rgb]{ 1.000,  1.000,  1.000}\textcolor[rgb]{ 0.000,  0.000,  0.000}{0.897 } & \cellcolor[rgb]{ 0.964,  0.477,  0.286}\textcolor[rgb]{ 0.000,  0.000,  0.000}{7.45  } & \cellcolor[rgb]{ 1.000,  1.000,  1.000}\textcolor[rgb]{ 0.000,  0.000,  0.000}{0.865 } & \cellcolor[rgb]{ 0.994,  0.771,  0.455}\textcolor[rgb]{ 0.000,  0.000,  0.000}{3.89  } & \cellcolor[rgb]{ 1.000,  1.000,  1.000}\textcolor[rgb]{ 0.000,  0.000,  0.000}{0.793 }\\
& $\sigma\left(J_\parallel\right)$         & \cellcolor[rgb]{ 0.993,  0.717,  0.409}\textcolor[rgb]{ 0.000,  0.000,  0.000}{4.15  } & \cellcolor[rgb]{ 1.000,  1.000,  1.000}\textcolor[rgb]{ 0.000,  0.000,  0.000}{0.84  } & \cellcolor[rgb]{ 0.997,  0.893,  0.569}\textcolor[rgb]{ 0.000,  0.000,  0.000}{3.31  } & \cellcolor[rgb]{ 1.000,  1.000,  1.000}\textcolor[rgb]{ 0.000,  0.000,  0.000}{0.876 } & \cellcolor[rgb]{ 0.999,  0.964,  0.689}\textcolor[rgb]{ 0.000,  0.000,  0.000}{2.77  } & \cellcolor[rgb]{ 1.000,  1.000,  1.000}\textcolor[rgb]{ 0.000,  0.000,  0.000}{0.906 } & \cellcolor[rgb]{ 0.997,  0.921,  0.617}\textcolor[rgb]{ 0.000,  0.000,  0.000}{3.07  } & \cellcolor[rgb]{ 1.000,  1.000,  1.000}\textcolor[rgb]{ 0.000,  0.000,  0.000}{0.906 } & \cellcolor[rgb]{ 0.993,  0.709,  0.403}\textcolor[rgb]{ 0.000,  0.000,  0.000}{4.16  } & \cellcolor[rgb]{ 1.000,  1.000,  1.000}\textcolor[rgb]{ 0.000,  0.000,  0.000}{0.843 }\\
& $V_{fl}$                                 & \cellcolor[rgb]{ 0.995,  0.809,  0.487}\textcolor[rgb]{ 0.000,  0.000,  0.000}{3.72  } & \cellcolor[rgb]{ 1.000,  1.000,  1.000}\textcolor[rgb]{ 0.000,  0.000,  0.000}{0.809 } & \cellcolor[rgb]{ 0.992,  0.686,  0.384}\textcolor[rgb]{ 0.000,  0.000,  0.000}{4.27  } & \cellcolor[rgb]{ 1.000,  1.000,  1.000}\textcolor[rgb]{ 0.000,  0.000,  0.000}{0.768 } & \cellcolor[rgb]{ 0.997,  0.921,  0.617}\textcolor[rgb]{ 0.000,  0.000,  0.000}{3.08  } & \cellcolor[rgb]{ 1.000,  1.000,  1.000}\textcolor[rgb]{ 0.000,  0.000,  0.000}{0.671 } & \cellcolor[rgb]{ 0.993,  0.740,  0.429}\textcolor[rgb]{ 0.000,  0.000,  0.000}{4.02  } & \cellcolor[rgb]{ 1.000,  1.000,  1.000}\textcolor[rgb]{ 0.000,  0.000,  0.000}{0.677 } & \cellcolor[rgb]{ 0.995,  0.817,  0.493}\textcolor[rgb]{ 0.000,  0.000,  0.000}{3.68  } & \cellcolor[rgb]{ 1.000,  1.000,  1.000}\textcolor[rgb]{ 0.000,  0.000,  0.000}{0.71  }\\
& $\sigma\left(V_{fl}\right)$              & \cellcolor[rgb]{ 0.964,  0.477,  0.286}\textcolor[rgb]{ 0.000,  0.000,  0.000}{5.77  } & \cellcolor[rgb]{ 1.000,  1.000,  1.000}\textcolor[rgb]{ 0.000,  0.000,  0.000}{0.852 } & \cellcolor[rgb]{ 0.964,  0.477,  0.286}\textcolor[rgb]{ 0.000,  0.000,  0.000}{5.6   } & \cellcolor[rgb]{ 1.000,  1.000,  1.000}\textcolor[rgb]{ 0.000,  0.000,  0.000}{0.877 } & \cellcolor[rgb]{ 0.964,  0.477,  0.286}\textcolor[rgb]{ 0.000,  0.000,  0.000}{6.79  } & \cellcolor[rgb]{ 1.000,  1.000,  1.000}\textcolor[rgb]{ 0.000,  0.000,  0.000}{0.837 } & \cellcolor[rgb]{ 0.964,  0.477,  0.286}\textcolor[rgb]{ 0.000,  0.000,  0.000}{7.04  } & \cellcolor[rgb]{ 1.000,  1.000,  1.000}\textcolor[rgb]{ 0.000,  0.000,  0.000}{0.865 } & \cellcolor[rgb]{ 0.964,  0.477,  0.286}\textcolor[rgb]{ 0.000,  0.000,  0.000}{6.38  } & \cellcolor[rgb]{ 1.000,  1.000,  1.000}\textcolor[rgb]{ 0.000,  0.000,  0.000}{0.834 }\\
\cline{2-12}
& $\left(\chi; Q\right)$\textsubscript{HFS-LP} & \multicolumn{2}{c}{$ \textbf{(0.94; \ 4.75)} $ } & \multicolumn{2}{c}{$ \textbf{(1.0; \ 4.84)} $ } & \multicolumn{2}{c}{$ \textbf{(1.0; \ 4.75)} $ } & \multicolumn{2}{c}{$ \textbf{(0.95; \ 4.79)} $ } & \multicolumn{2}{c}{$ \textbf{(0.99; \ 4.66)} $ }\\
\midrule
Overall
& $\chi$; $Q$                              & \multicolumn{2}{c}{$ \textbf{(0.87; \ 18.9)} $ } & \multicolumn{2}{c}{$ \textbf{(0.86; \ 19.4)} $ } & \multicolumn{2}{c}{$ \textbf{(0.83; \ 19.0)} $ } & \multicolumn{2}{c}{$ \textbf{(0.87; \ 19.2)} $ } & \multicolumn{2}{c}{$ \textbf{(0.92; \ 18.6)} $ }\\
\bottomrule
\end{tabular}
    }
    \includegraphics[width=0.5\textwidth]{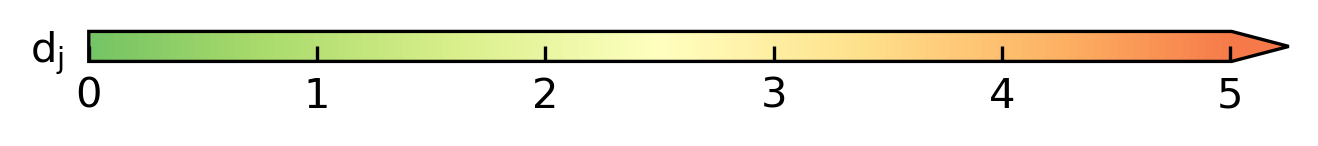}
    \caption{\textbf{Quantitative validation result for each observable}. For each code and field direction (indicated by $(+)$ for forward field and $(-)$ for reversed field), the $d_j$ (`normalised distance', Eq.\ref{eq:normalised_distance}) and $S_j$ (`sensitivity', Eq.\ref{eq:sensitivity}) terms are given. The normalised distance
    $d_j$ gives the root-mean-square Z-score of the difference between the experiment and simulation, with \textit{green} cells indicating good agreement ($d_j\to0$) and \textit{red} cells indicating poor agreement ($d_j \to \infty$, with the colour scale limited to $d_j \leq 5.0$).
    The sensitivity $S_j$ indicates the precision of each observable, with $S_j\to0$ for observables with high relative uncertainty and $S_j\to1$ for observables with low relative uncertainty. The combined level-of-agreement $\chi$ (Eq.\ref{eq:ricci_chi} for $d_0=1.0$ and $\lambda=0.5$) and the comparison quality $Q$ (Eq.\ref{eq:quality}) are given for each diagnostic individually as well as for the overall validation.}
    \label{tab:validation_results}
\end{table*}

For each simulation and each observable, the normalised distance $d_j$ (Eq.\ref{eq:normalised_distance}) and sensitivity $S_j$ (Eq.\ref{eq:sensitivity}) is computed. Points with a very low experimental uncertainty $\Delta e_j / e_j < 10^{-3}$ (typically due to a lack of repeat discharges to estimate the reproducibility error) are removed from the calculation of $d_j$. The values of $d_j$ and $S_j$ are used, together with the primacy hierarchies $H_j$ given in Tab.\ref{tab:summary_of_measurements_and_hierarchies}, to compute the overall composite metric $\chi$ (Eq.\ref{eq:ricci_chi}) and quality $Q$ (Eq.\ref{eq:quality}) for each simulation and including all observables. In addition, the effective composite metric and quality values $\chi_{diag}$ and $Q_{diag}$, are computed taking observables from a single diagnostic. The result is given in Tab.\ref{tab:validation_results}.
We find that the level of agreement computed from individual diagnostics varies significantly. Both the reciprocating midplane probe (FHRP) and divertor-entrance Thomson scattering (TS) show appreciable quantitative agreement, while the reciprocating divertor probe array (RDPA) and the divertor target profiles (HFS-LP/LFS-LP/LFS-IR) show poor agreement. This suggests that the change in agreement is due to the measurement location rather than the diagnostic itself: better agreement is found for diagnostics which are close to the confined region (where the TS values at the separatrix were used to tune the sources) than for diagnostics in the divertor volume or at the targets. To understand the quantitative result, we show comparisons of several observables, grouping the results by location.

\begin{figure*}
    \centering
    \includegraphics[width=\textwidth]{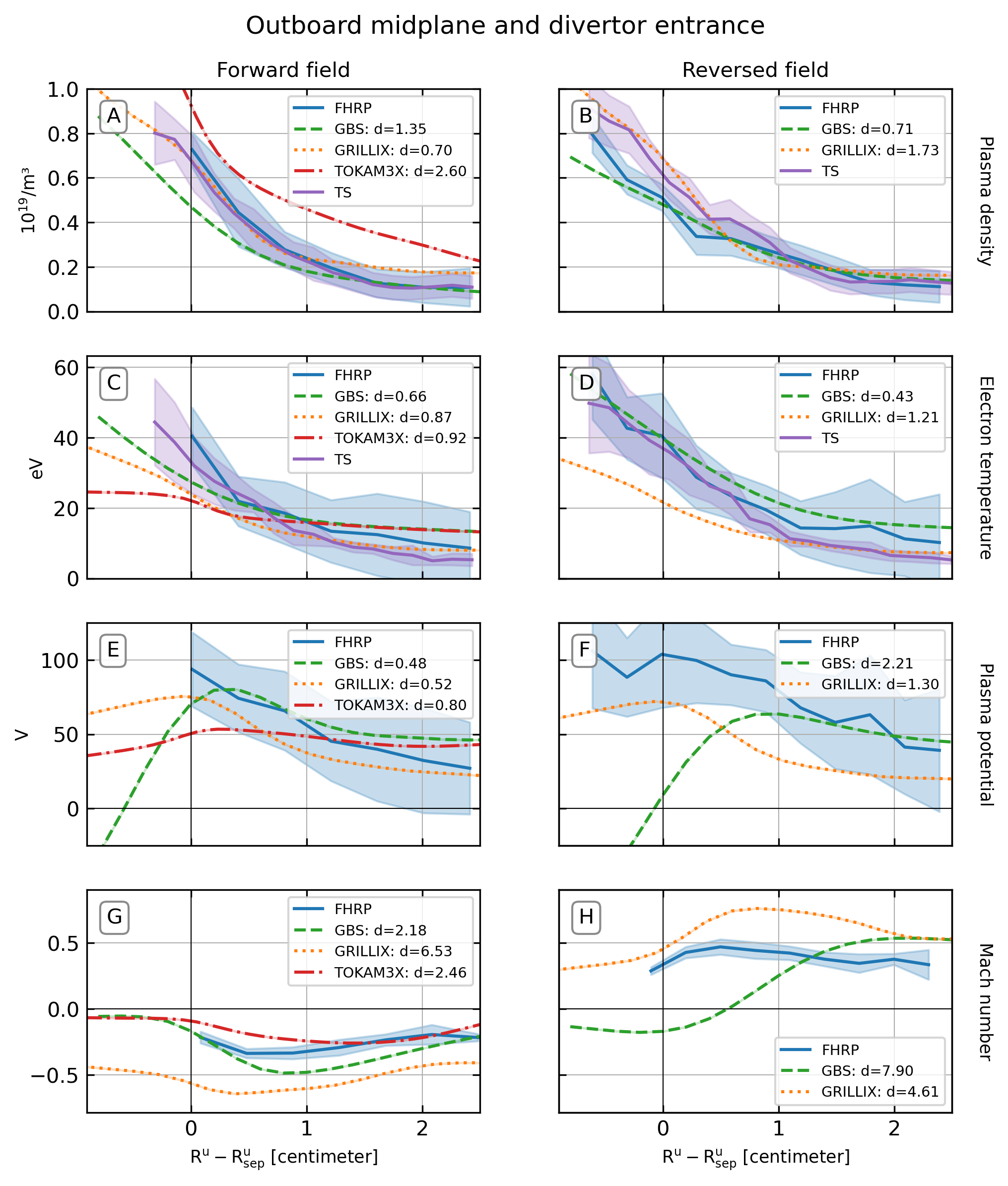}
    \caption{\textbf{Comparison of averaged profiles at the outboard midplane and divertor entrance}. \textit{Rows from top}: Profiles of the mean plasma density ($n_e$), electron temperature ($T_e$), plasma potential ($V_{pl}$) and the parallel Mach number ($M_\parallel$) in forward (\textit{left column}) and reversed field (\textit{right column}). The experimental data from the outboard-midplane reciprocating probe (FHRP) is indicated by the \textit{blue line}, with its uncertainty indicated by the \textit{blue shaded region}. The experimental data from the divertor entrance Thomson scattering (TS) diagnostic is indicated by the \textit{purple line}, with its uncertainty indicated by the \textit{purple shaded region}. The other lines give the mean simulated profiles \textit{at the outboard midplane} from the three codes (the simulated divertor entrance profiles are shown in \texttt{TCV-X21/3.results/summary\_fig/Divertor\_Thomson+density,electron\_temp.png}). The corresponding $d_j$ values (Eq.\ref{eq:normalised_distance}) at the outboard midplane are given in the legend for each code, while the $d_j$ values at the divertor entrance are given in Tab.\ref{tab:validation_results}. We note that $n$ and $T_e$ are not necessarily expected to agree between the FHRP and TS (since they are at different locations).}
    \label{fig:outboard_midplane_base_moments}
\end{figure*}
\begin{figure*}
    \centering
    \includegraphics[width=\textwidth]{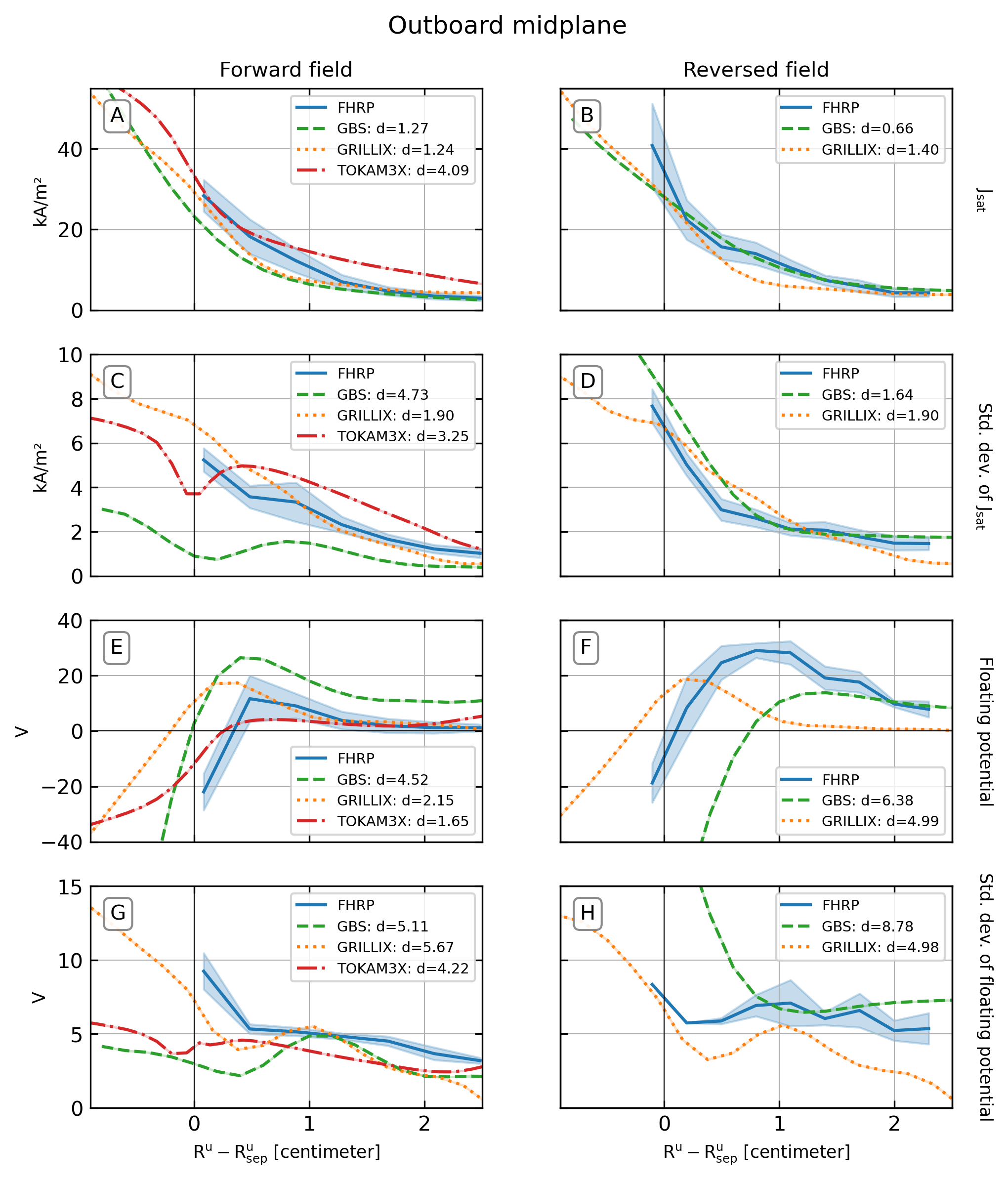}
    \caption{\textbf{Comparison of direct experimental measurements and their standard deviations at the outboard midplane}. \textit{Rows from top}: Profiles of the mean ion saturation current ($J_{sat}$) and its 
    standard deviation ($\sigma(J_{sat})$), mean floating potential ($V_{fl}$) and its 
    standard deviation ($\sigma(V_{fl})$) in forward (\textit{left column}) and reversed field (\textit{right column}).
    The experimental data from the outboard-midplane reciprocating probe (RPTCV) is indicated by the \textit{blue line}, with its uncertainty indicated by the \textit{blue shaded region}. The other lines give the corresponding simulated profiles from the three codes. The corresponding $d_j$ values (Eq.\ref{eq:normalised_distance}) are given in the legend for each code. We note that in Fig.\ref{fig:outboard_midplane_expt_moments}.H, points with very low experimental uncertainty are excluded from the calculation of $d_j$.}.
    \label{fig:outboard_midplane_expt_moments}
\end{figure*}
\subsection{Outboard midplane and divertor entrance profiles}\label{subsec:omp profiles}

In Fig.\ref{fig:outboard_midplane_base_moments} we show the outboard midplane profiles of the mean plasma density ($n$), electron temperature ($T_e$), plasma potential ($V_{pl}$) and parallel Mach number ($M_{||})$. In Fig.\ref{fig:outboard_midplane_expt_moments} we show the outboard midplane ion saturation current($J_{sat}$) and its standard deviation ($\sigma(J_{sat})$), and the floating potential ($V_{fl}$) and its standard deviation. The TCV divertor entrance profiles from TS are plotted together with the FHRP data in Fig.\ref{fig:outboard_midplane_base_moments}, and the comparison to simulation is shown in \path{TCV-X21/3.results/summary_fig/Divertor_Thomson+density,electron_temp.png}. The outboard midplane ion saturation current relative fluctuation, skew and kurtosis are shown in \path{TCV-X21/3.results/summary_fig/Outboard_midplane+jsat,}\linebreak\path{jsat_fluct,jsat_skew,jsat_kurtosis.png}

The overall agreement for the outboard midplane and divertor entrance profiles is typically very good, with the simulations matching both the amplitude and profile shape reasonably well for several observables. The experimental uncertainty of the FHRP $T_e$ and $V_{pl}$ profiles is very large, mostly due to the uncertainty in the four parameter fit of the IV curve, making quantitative agreement easier to achieve. Since the simulations tuned their sources to match both the separatrix $n$ and $T_e$ (GBS and TOKAM3X) or only the separatrix density (GRILLIX, with fixed power), the good agreement for these two observables at the separatrix is of course expected. As such, we are more interested in whether the profile shape is recovered for the two observables. This is addressed by fitting exponential decay functions of the form $A\exp{\left[-(R^u - R^u_{sep})/\lambda \right]}$ in the near-SOL (for $R^u-R^u_{sep}\in\left[0\si{\centi\meter}, 1.5\si{\centi\meter}\right]$) to the profiles (including the experimental error bars), to determine whether the codes are reproducing the observed fall-off lengths for the density and electron temperature. The results are given in Tab.\ref{tab:omp decay}.

\begin{table}[h]
    \centering
    \resizebox{1.0\columnwidth}{!}{
    \begin{tabular}{lcccc}
\toprule
{} &          TCV &          GBS &      GRILLIX &      TOKAM3X \\
\midrule
$\lambda_{n,OMP}$$^{+}$   &  0.9$\pm$0.2 &  1.1$\pm$0.1 &  1.0$\pm$0.1 &  1.7$\pm$0.1 \\
$\lambda_{n,DE}$$^{+}$    &  0.9$\pm$0.2 &  1.0$\pm$0.0 &  0.9$\pm$0.1 &  1.5$\pm$0.0 \\
$\lambda_{T_e,OMP}$$^{+}$ &  1.0$\pm$0.4 &  2.5$\pm$0.2 &  1.6$\pm$0.1 &  4.1$\pm$0.4 \\
$\lambda_{T_e,DE}$$^{+}$  &  1.0$\pm$0.1 &  2.3$\pm$0.1 &  1.5$\pm$0.1 &  6.3$\pm$0.2 \\
$\lambda_{n,OMP}$$^{-}$   &  2.0$\pm$1.1 &  1.4$\pm$0.0 &  0.9$\pm$0.1 &            - \\
$\lambda_{n,DE}$$^{-}$    &  1.1$\pm$0.2 &  1.1$\pm$0.0 &  0.9$\pm$0.1 &            - \\
$\lambda_{T_e,OMP}$$^{-}$ &  1.5$\pm$1.1 &  1.7$\pm$0.0 &  1.6$\pm$0.1 &            - \\
$\lambda_{T_e,DE}$$^{-}$  &  1.0$\pm$0.1 &  1.4$\pm$0.0 &  1.5$\pm$0.1 &            - \\
\bottomrule
\end{tabular}

    }
    \caption{\textbf{Near-SOL decay lengths (in cm)}, for the mean density ($n$) and electron temperature ($T_e$) measured at the outboard midplane (OMP) and divertor entrance (DE), in forward ($+$) and reversed ($-$) toroidal field direction. Profiles are fitted in the range $\left[0\si{\centi\meter}, 1.5\si{\centi\meter}\right]$. The observable uncertainty is included in the fitting uncertainty.}
    \label{tab:omp decay}
\end{table}

Experimentally, we see from Tab.\ref{tab:omp decay} that the OMP (FHRP) and divertor entrance (TS) give similar $\lambda_n$ and $\lambda_{T_e}$ fall-off lengths in forward-field, while in the reversed-field the fall-off lengths at the OMP are $1.5-2\times$ larger than at the divertor entrance, although with a much higher fit uncertainty. For the forward-field $\lambda_n$, TOKAM3X predicts broader n profiles, while GBS and GRILLIX match the experimental fall-off length within the uncertainty. All simulations predict too broad forward-field $T_e$ profiles, with GRILLIX matching the closest, then GBS, and then TOKAM3X. In the reversed-field case, GBS reproduces the narrowing of the $n$ and $T_e$ profiles between the OMP and divertor entrance (although to a smaller extent than in experiments), predicting $\lambda_n$ within the uncertainties. $\lambda_{T_e}$ is slightly too large, but still within uncertainty at the OMP. Conversely, GRILLIX predicts similar fall-off lengths as in the forward-field case. It does not show a narrowing of $\lambda_{n,OMP}$ between the OMP and the divertor-entrance, but still matches all fall-off-lengths except $\lambda_{T_e,DE}^-$ within uncertainty. We see that the higher fit uncertainty for the TCV reversed-field $\lambda_n$ is because the experimental $n$ profile (shown in Fig.\ref{fig:outboard_midplane_base_moments}.B) is not a simple exponential decay in the range $R^u-R^u_{sep}\in\left[0\si{\centi\meter}, 1.5\si{\centi\meter}\right]$. Instead, a flat region around around $R^u - R^u_{sep} = 0.25\si{\centi\metre}$ is seen in both the FHRP and OMP measurements. This is not, however, reproduced in the simulations.

The use of relaxed parameters is likely to be part of the cause of the broadened profiles for TOKAM3X and GBS. However, if relaxed parameters were the only cause for broadening, TOKAM3X (which uses values closer to the Braginskii values than GBS) should predict narrower $T_e$ profiles than GBS, while the opposite is found. Additionally, GRILLIX (which uses the Braginskii parameters directly) predicts broadened forward-field $T_e$ profiles -- indicating that relaxed parameters alone cannot explain the behaviour. It is likely that the lack of neutral dynamics is contributing to the broadened $T_e$ profiles, since neutral ionisation would add an additional sink of energy in the open field-line region. Another possible cause for the different profile widths is the different energy source rates (given in Sec.\ref{sec:free_parameters}).

For the $V_{pl}$ profiles (Figs.\ref{fig:outboard_midplane_base_moments}.E-F), we see in the experiment that the profiles are monotonically decreasing with a steeper slope in forward-field than in reversed-field. For all codes, $V_{pl}$ is positive in the SOL and of a similar amplitude as in the experiments, with a very good match in forward-field.
In GBS, $V_{pl}$ peaks near the separatrix in forward-field case, and further into the SOL in the reversed-field case. This is within uncertainty for the forward-field case, while in the reversed-field case a disagreement is found close to the separatrix. In GRILLIX, the peak of $V_{pl}$ is at the separatrix in both field directions, such that the agreement is within the error bars in forward-field, but not in reverse-field, where $V_{pl}$ is decreases too steeply into the near-SOL. In TOKAM3X (forward-field only), $V_{pl}$ is very flat, but still within uncertainty except in the vicinity of the separatrix. The reason for the worse agreement found for the reversed-field $V_{pl}$ profiles (compared to the reasonably good agreement in forward-field) is difficult to determine, since the potentials can be affected by both the confined region dynamics and the sheath boundary conditions \cite{rivaThreedimensionalSimulationsPlasma2018,zholobenkoElectricFieldTurbulence2021a}.

For the $M_\parallel$ profiles (Figs.\ref{fig:outboard_midplane_base_moments}.G-H), we see in the experiment that the parallel flow changes direction with the toroidal field reversal, increasing the flow speed in reversed-field. In forward-field, GBS and TOKAM3X predict similar profiles, which are reasonably close to the measured values. The agreement for GBS is considerably lower for the reversed-field case. The direction of the parallel flow is not matched near the separatrix, for $R^u - R^u_{sep}<\SI{0.5}{\centi\metre}$. For GRILLIX, the absolute values of the parallel flow are over-predicted, while the direction of the parallel flow is reproduced. The features in the simulated $M_\parallel$ appear to be consistent with the expected Pfirsch-Schlüter return flow -- where the parallel flows are determined by the radial electric field and radial ion pressure gradient \cite{pittsParallelSOLFlow2007}.

To determine whether the codes are capturing the time-dependent dynamics, we focus here on the statistical moments of $J_{sat}$ and $V_{fl}$, which are shown in (Fig.\ref{fig:outboard_midplane_expt_moments} and
\path{TCV-X21/3.results/summary_fig/Outboard_midplane+jsat,jsat_fluct,jsat_skew,}\linebreak\path{jsat_kurtosis.png}). We see that the mean $J_{sat}$ and $\sigma(J_{sat})$ profiles are recovered reasonably well by all codes, reflecting the agreement found for the mean n and Te. The only exception is $\sigma(J_{sat})$ for GBS in forward-field, which is low in the SOL, increasing inside the LFCS. The mean $V_{fl}$ is well-matched by TOKAM3X in forward-field, reproducing the drop across the separatrix into the confined region. GBS and GRILLIX also reproduce the qualitative behaviour observed in the experiment, but do not quantitatively match. For GBS, the profile is overestimated in forward-field, and underestimated in reversed-field. For GRILLIX, both profiles are shifted radially inwards. For the standard deviation of $\sigma(V_{fl})$, all codes are able to predict the magnitude of the profile in the SOL. GRILLIX also matches $\sigma(V_{fl})$ near the separatrix, while GBS and TOKAM3X underestimate the separatrix value in forward-field, and GBS overestimates the separatrix value in reversed-field.

\begin{figure*}
    \centering
    \includegraphics[width=\textwidth]{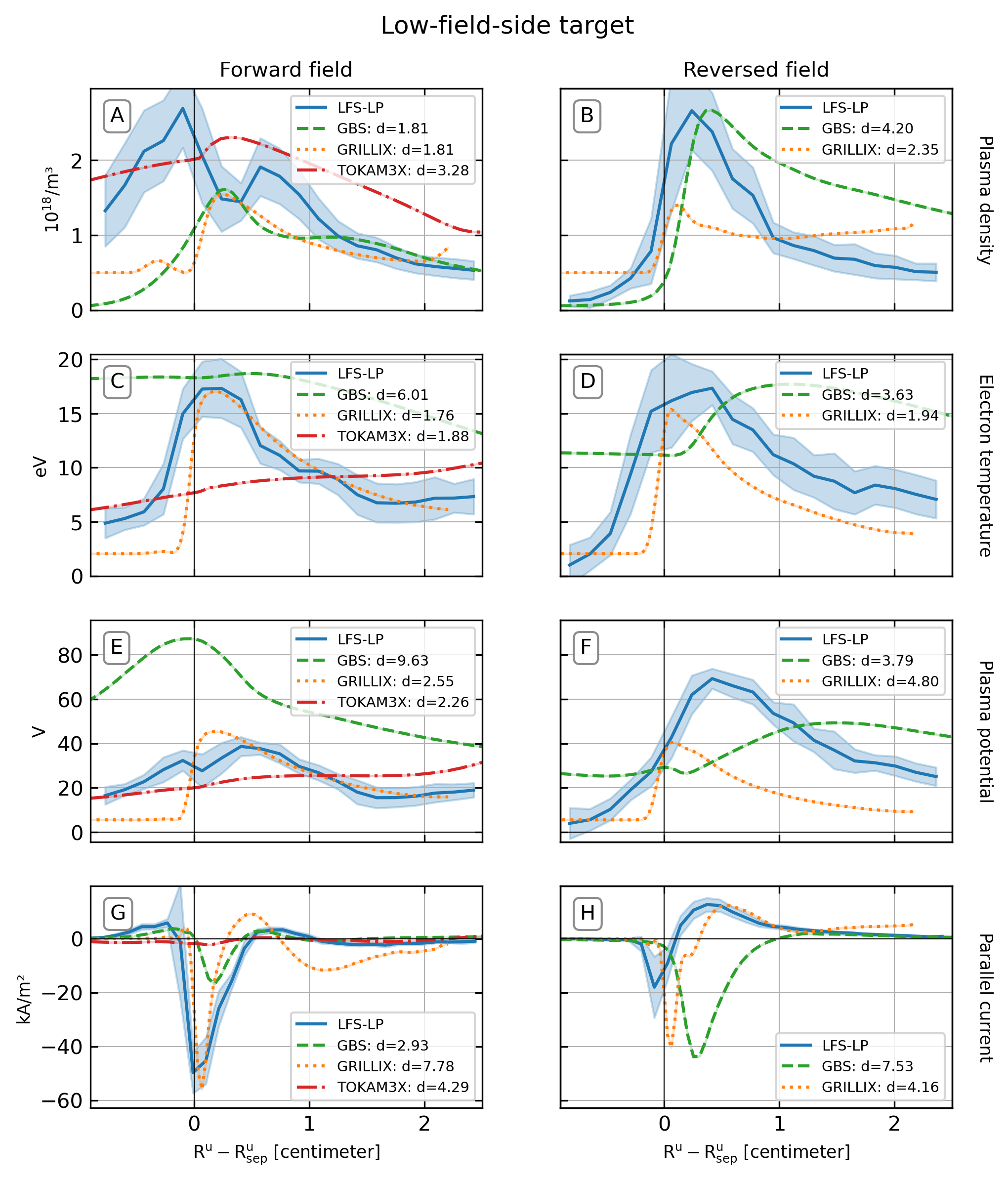}
    \caption{\textbf{Comparison of averaged profiles at the low-field-side divertor target}. \textit{Rows from top}: Profiles of the mean plasma density ($n_e$), electron temperature ($T_e$), plasma potential ($V_{pl}$) and the parallel current density ($j_\parallel$) in forward (\textit{left column}) and reversed field (\textit{right column}). The experimental data from the low-field-side Langmuir probe array (LFS-LP) is indicated by the \textit{blue line}, with its uncertainty indicated by the \textit{blue shaded region}. The other lines give the corresponding simulated profiles from the three codes. The corresponding $d_j$ values (Eq. \ref{eq:normalised_distance}) are given in the legend for each code.}
    \label{fig:lfs_base_moments}
\end{figure*}
\begin{figure*}
    \centering
    \includegraphics[width=\textwidth]{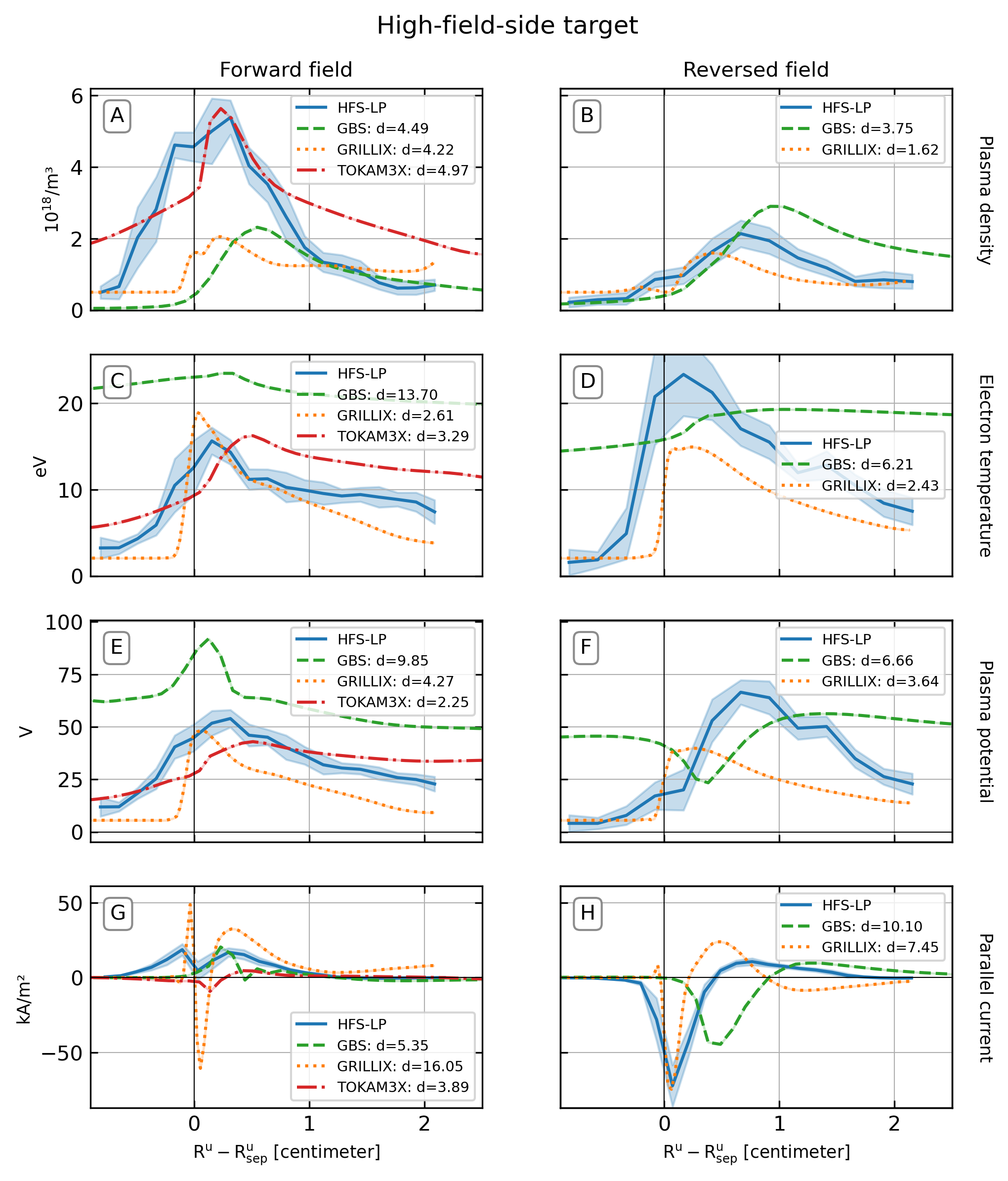}
    \caption{\textbf{Comparison of averaged profiles at the high-field-side divertor target}. \textit{Rows from top}: Profiles of the mean plasma density ($n_e$), electron temperature ($T_e$), plasma potential ($V_{pl}$) and the parallel current density ($j_\parallel$) in forward (\textit{left column}) and reversed field (\textit{right column}). The experimental data from the high-field-side Langmuir probe array (HFS-LP) is indicated by the \textit{blue line}, with its uncertainty indicated by the \textit{blue shaded region}. The other lines give the corresponding simulated profiles from the three codes. The corresponding $d_j$ values (Eq. \ref{eq:normalised_distance}) are given in the legend for each code.}
    \label{fig:hfs_base_moments}
\end{figure*}
\begin{figure*}
    \centering
    \includegraphics[width=\textwidth]{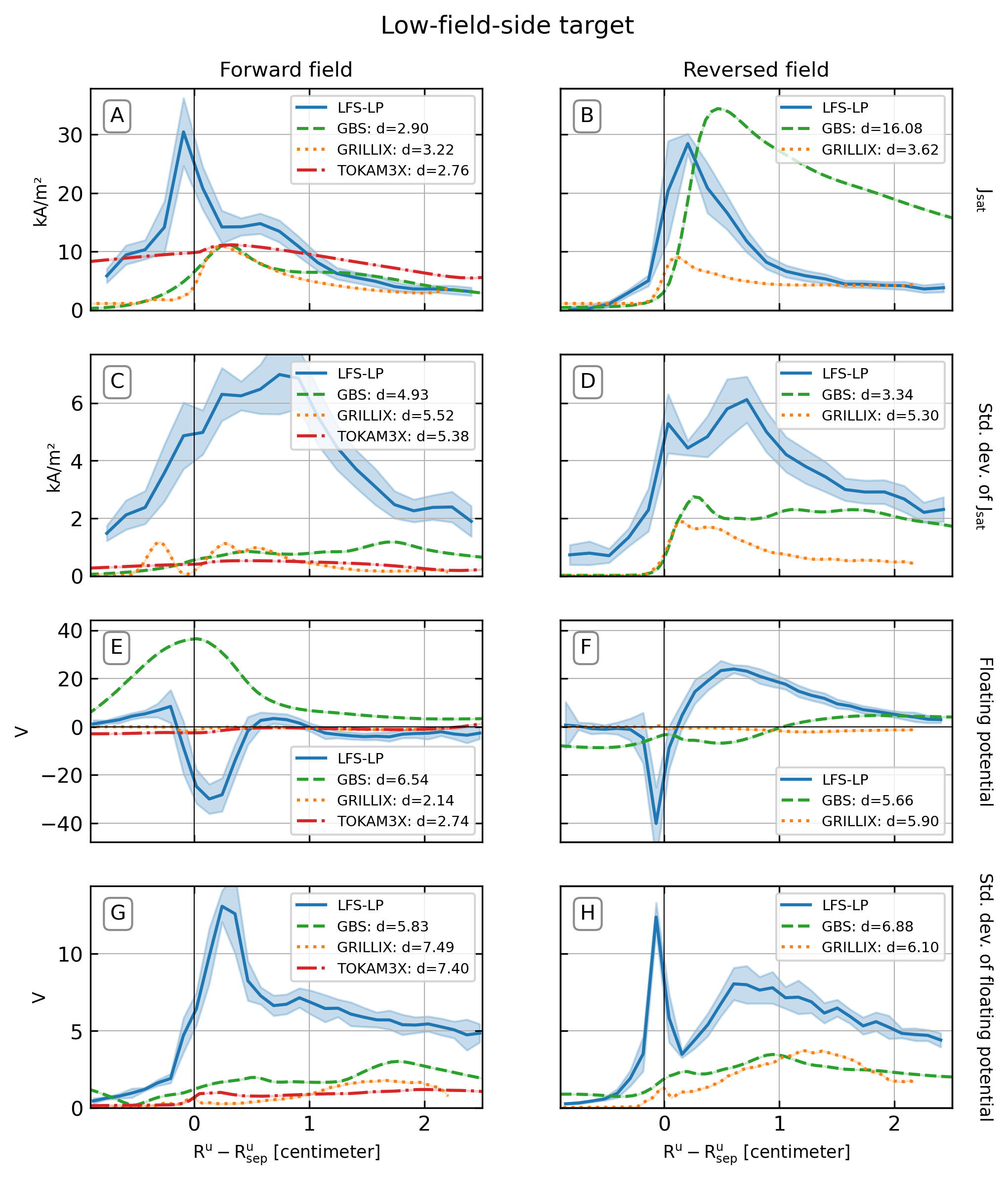}
    \caption{\textbf{Comparison of direct experimental measurements and their standard deviations at the low-field-side divertor target}. \textit{Rows from top}: Profiles of the mean ion saturation current ($J_{sat}$) and its 
    standard deviation ($\sigma(J_{sat})$), mean floating potential ($V_{fl}$) and its standard deviation ($\sigma(V_{fl})$) in forward (\textit{left column}) and reversed field (\textit{right column}).
    The experimental data from the low-field-side Langmuir probe array (LFS-LP) is indicated by the \textit{blue line}, with its uncertainty indicated by the \textit{blue shaded region}. The other lines give the corresponding simulated profiles from the three codes. The corresponding $d_j$ values (Eq. \ref{eq:normalised_distance}) are given in the legend for each code.}
    \label{fig:lfs_expt_moments}
\end{figure*}

\subsection{Low- and High-field-side target profiles}\label{subsec:divertor target profiles}

In Fig.\ref{fig:lfs_base_moments} and \ref{fig:hfs_base_moments} we show the profiles of the mean plasma density ($n$), electron temperature ($T_e$), plasma potential ($V_{pl}$) and parallel current density ($J_{||})$, at the low-field-side (LFS) target and high-field-side (HFS) target, respectively. In Fig.\ref{fig:lfs_expt_moments} we show the LFS ion saturation current ($J_{sat}$) and floating potential ($V_{fl}$), together with their standard deviations. The divertor target ion saturation current relative fluctuations, skewness and kurtosis are shown in \path{TCV-X21/3.results/summary_fig/TARGET+jsat,jsat_fluct,jsat_skew,jsat_kurtosis.png} and the standard deviation of the parallel current density is shown in \path{TCV-X21/3.results/summary_fig/TARGET+current,current_std} where \path{TARGET} is either \linebreak\path{Low-field-side_target} or \path{High-field-side_target}.

Overall, a worse match between simulation and experiment is found for the target profiles compared to the midplane profiles. Generally, for most observables, the codes capture the correct peak order-of-magnitude and the features visible in the experiment are (roughly) reproduced. However, the majority of observables are not matched within experimental uncertainty, and the broadness of the profiles is seen to vary significantly amongst the simulations. All simulations fail to accurately predict the $V_{fl}$ target profile, and under-predict the $\sigma(V_{fl})$ and $\sigma(J_{sat})$ by factors of $2$ or more. Furthermore, experimentally, a significant effect of the toroidal field reversal is seen for the $n$, $T_e$ and $V_{fl}$ and $J_\parallel$ profiles at the targets; at the HFS target the plasma is colder and denser in forward-field than in reversed-field, and on the LFS target a prominent private-flux peak in $n$ and $J_{sat}$ is observed in forward-field. The simulations are not able to reproduce the double peak $n$ profile at the LFS target in forward-field, missing the primary peak lying in the PFR. Generally, the codes provide very different predictions of the divertor target profiles. As such, we present the results from each code separately, highlighting general trends as well as observables with particularly good or poor agreement.\\

Starting with GBS, the broadness and peak values of the profiles vary depending on the toroidal field direction and between the targets. A significant effect of the toroidal field reversal is seen in the simulated $n$, $J_{sat}$ and $J_\parallel$ profiles at LFS and HFS targets. At the LFS target, the SOL $n$ peak value approximately matches the experiment in the reversed-field case, but the profile is broadened with respect to the experiment. Conversely, in the forward-field case, the width in the SOL matches more closely, but the PFR peak is missed. At the HFS target, the forward-field $n$ peak value is underestimated and shifted towards the SOL, while the reversed-field profile is again broadened with respect to the experiment.
Conversely, the $T_e$ and $V_{pl}$ profiles are much broader than the experimental profiles. This is likely due to the reduced heat conductivity, which changes the ratio between the parallel and perpendicular heat fluxes. Since the plasma potential is related to the electron temperature, it will also be broadened.
Of the three codes, GBS is the only code which predicts a significantly-non-zero $V_{fl}$, but the predicted profiles do not match the experiment. At least for the reversed-field case, this could be a consequence of the profile broadening.

For GRILLIX, the density at the targets is too low, while the shape is loosely recovered, except for the PFR peak observed at the LFS in forward-field. The SOL $T_e$ profile is matched well at both targets and in both field directions, while in the private-flux region the $T_e$ profile drops off too sharply. Due to the $V_{fl}=0 \implies V_{pl} = \Lambda T_e$ boundary condition, the sharp drop towards the PFR in $T_e$ leads to a corresponding drop in $V_{pl}$. This will in turn give a strong electric field across the separatrix. The $V_{fl}=0$ boundary condition is clearly incorrect, which causes the reversed-field SOL $V_{pl}$ to disagree regardless of the good match in $T_e$. Despite applying a potential boundary condition which corresponds to an insulating ($J_\parallel =0$) sheath, the $J_\parallel$ profile (which is set equal to the internal currents) is both significantly non-zero and also, surprisingly, a reasonably good match to the experiment (except for the HFS profile in forward-field). The current boundary conditions and the effect of the electric field along the target on $u_\parallel$ is discussed further in Sec.\ref{sec:discussion boundary conditions}.

For TOKAM3X, the HFS target is substantially better matched than the LFS target. Good matches are found for the HFS $n$, $T_e$ and $V_{pl}$, although the far-SOL profile is too broad. For the LFS target, it is seen that the $T_e$ and $V_{pl}$ profiles are \textit{increasing} into the far-SOL. Since there are no mechanisms to heat the electrons in the far-SOL in the TOKAM3X simulation, this appears to indicate a numerical issue such as pollution due to a buffer zone applied at the limiting flux surface. The LFS $n$ and $J_{sat}$ have the correct amplitude in the SOL, but both miss a prominent PFR peak. Both $V_{fl}$ and $J_{\parallel}$ at both targets are significantly lower than observed.\\

\begin{figure}[h]
    \centering
    \includegraphics[width=\columnwidth]{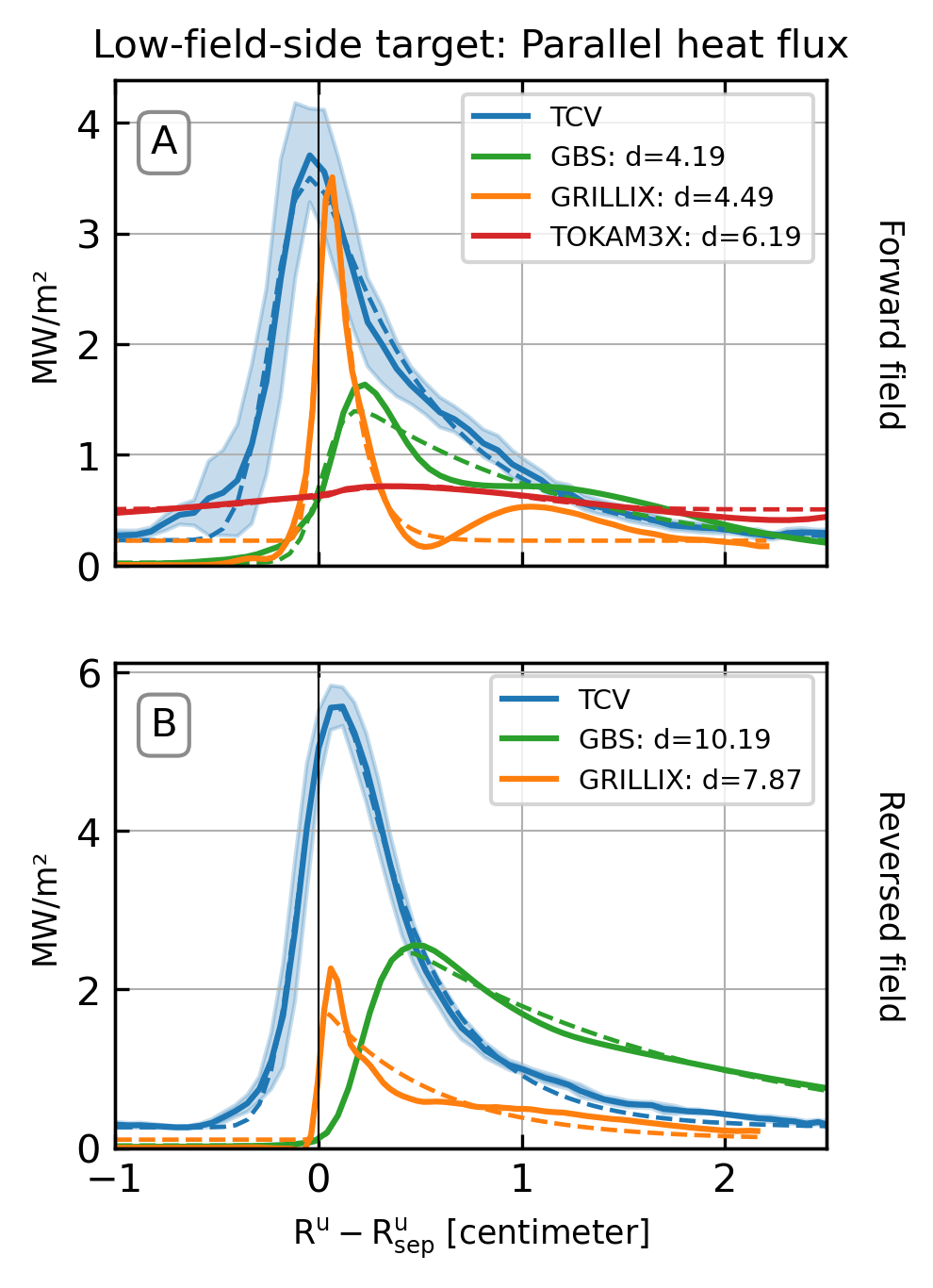}
    \caption{\textbf{Parallel heat flux profiles and Eich profile fits at the low-field-side target}. The experimental parallel heat flux data from the infrared camera for the low-field-side target is indicated by the \textit{blue line}, with its uncertainty indicated by the \textit{blue shaded region}. The other solid lines give the corresponding simulated profiles from the three codes, and the corresponding $d_j$ values (Eq.\ref{eq:normalised_distance}) is given in the legend for each code. For each profile shown, the corresponding Eich-type fit\cite{eichScalingTokamakScrapeoff2013a} is indicated by a dashed line of the same colour. The fitted parameters are given in Tab.\ref{tab:lambda q}. Profiles are fitted over the entire range of data.}
    \label{fig:q_parallel}
\end{figure}

We finally consider the heat flux and decay length of all codes. The parallel heat flux profiles measured at the LFS target by the infrared camera are shown in Fig.\ref{fig:q_parallel}. Since the simulations do not directly evolve the parallel heat flux, its value is calculated from the sum of the electron and ion contributions. In GBS, the parallel heat flux is computed from the parallel-convective heat flux $\frac{5}{2} n \left(v_\parallel T_e + u_\parallel T_i\right) +\frac{1}{2}m_i u_{\parallel}^3$, where $v_\parallel$ and $u_\parallel$ are the electron and ion velocities respectively, and the last term accounts for the kinetic energy associated with the bulk ion flow. In GRILLIX and TOKAM3X the heat flux to the sheath entrance is computed from the sum of a conductive heat flux $-\chi_{\parallel,e}T_e^{5/2} \nabla_\parallel T_e -\chi_{\parallel,i}T_i^{5/2} \nabla_\parallel T_i$ and the total convective heat flux $\frac{5}{2} n \left(\mathbf{v} T_e + \mathbf{u} T_i\right)\cdot\mathbf{\hat{b}}$, for $\mathbf{v}$ and $\mathbf{u}$ the total electron and ion velocities (the vector-sum of the parallel, $E\times B$ and -- for TOKAM3X only -- the $\nabla B$ components of the velocity). Note that the conductive component for GRILLIX and TOKAM3X is intended to mimic the electron cooling by the sheath, and that the $\chi_\parallel$ values used are the Braginskii values, which do not account for the modifications of the distribution function due to the wall \cite{loizu2011Sheath}. The calculation of the heat flux for GRILLIX is given in \path{TCV-X21/tcvx21/grillix_post/observables/heat_flux_m.py}.

We see that GRILLIX predict the peak heat flux in forward-field reasonably well while GBS under-predicts the peak heat flux. In reversed-field both GBS and GRILLIX predict a peak heat flux which is less than half the measured value. By comparison, in forward-field TOKAM3X predicts an extremely broad, low-amplitude heat flux profile, with a peak heat flux $\sim 25\%$ of the measured peak heat flux. We use a Levenberg-Marquardt algorithm \cite{2020SciPy-NMeth} to fit Eich-type profiles of the form \cite{eichScalingTokamakScrapeoff2013a}
\begin{eqnarray}
    q_\parallel(r) = &\frac{q_0}{2} \exp\left[\left(\frac{S}{2\lambda_q}\right)^2 - \frac{r-r_0}{\lambda_q} \right]\nonumber\\
    &\times \mathrm{erfc}\left(\frac{S}{2\lambda_q} - \frac{r-r_0}{S}\right) + q_{BG}
\end{eqnarray}
where $r=R^u-R^u_{sep}$ and $\lambda_q$, $S$, $q_0$, $q_{BG}$ and $r_0$ are fitted parameters corresponding to the (upstream mapped) heat flux decay length, the power spreading factor, the peak heat flux, the background heat flux, and a free radial shift. The parameter $r_0$ was introduced to account the uncertainty in the spatial calibration of the IR, which can change the strike-point position, affecting the fit of the Eich-profiles. We found that in forward and in reversed-field, respectively, $r_0=-2.1\si{\milli\meter}$ and $r_0=-0.6\si{\milli\meter}$. The fitted profiles are indicated by \textit{dashed lines} in Fig.\ref{fig:q_parallel}, and the fitted parameters for $\lambda_q$ and $S$ are given in Tab.\ref{tab:lambda q}. During fitting, the experimental uncertainty is neglected since this is found to improve the match between the raw and fitted profiles. The fitting procedure is provided in \path{TCV-X21/tcvx21/analysis/fit_eich_profile_m.py}.

\begin{table}[h]
    \centering
    \resizebox{1.0\columnwidth}{!}{
    \begin{tabular}{lcccc}
\toprule
{} &          TCV &           GBS &      GRILLIX &        TOKAM3X \\
\midrule
$\lambda_{q}^{+}$ &  5.5$\pm$0.2 &  11.6$\pm$0.5 &  1.1$\pm$0.1 &  0.1$\pm$929.3 \\
$S^{+}$           &  1.8$\pm$0.1 &   1.4$\pm$0.1 &  0.7$\pm$0.1 &   6.9$\pm$38.4 \\
$\lambda_{q}^{-}$ &  4.0$\pm$0.1 &  16.0$\pm$0.2 &  5.4$\pm$0.5 &              - \\
$S^{-}$           &  1.8$\pm$0.0 &   1.5$\pm$0.1 &  0.2$\pm$0.2 &              - \\
\bottomrule
\end{tabular}

    }
    \caption{\textbf{Eich profile fitted parameters (in mm)} for the mean parallel heat flux profile, for the heat flux decay length ($\lambda_q$) and power spreading factor ($S$), in forward ($+$) or reversed ($-$) toroidal field direction.}
    \label{tab:lambda q}
\end{table}

For the simulations, the simulated double-peak seen by GRILLIX for the forward-field clearly cannot be described by Eich-type fits. In the forward-field direction, GBS reproduces the experimental spreading factor $S$, but predicts a larger heat-flux decay length $\lambda_q$, while GRILLIX predicts smaller values for both parameters. For TOKAM3X, the fitted parameters have very high uncertainty, indicating a low quality fit. In the reversed-field direction, GBS again reproduces the experimental spreading factor and predicts a larger $\lambda_q$, while GRILLIX predicts a smaller spreading factor and a slightly larger $\lambda_q$. The low-amplitude TOKAM3X profile may be related to the very low injected power in that simulation (see Sec.\ref{sec:free_parameters}). The broadened GBS heat flux profiles are consistent with the profiles at the LFS target.
For GRILLIX, the smaller $S$ in both field directions is consistent with the interpretation of narrowed profiles due the fast parallel advection, discussed in Sec.\ref{sec:discussion boundary conditions}.

\begin{figure*}
    \centering
    \includegraphics[width=0.95\textwidth, trim=0cm 0.4cm 0cm 0cm] {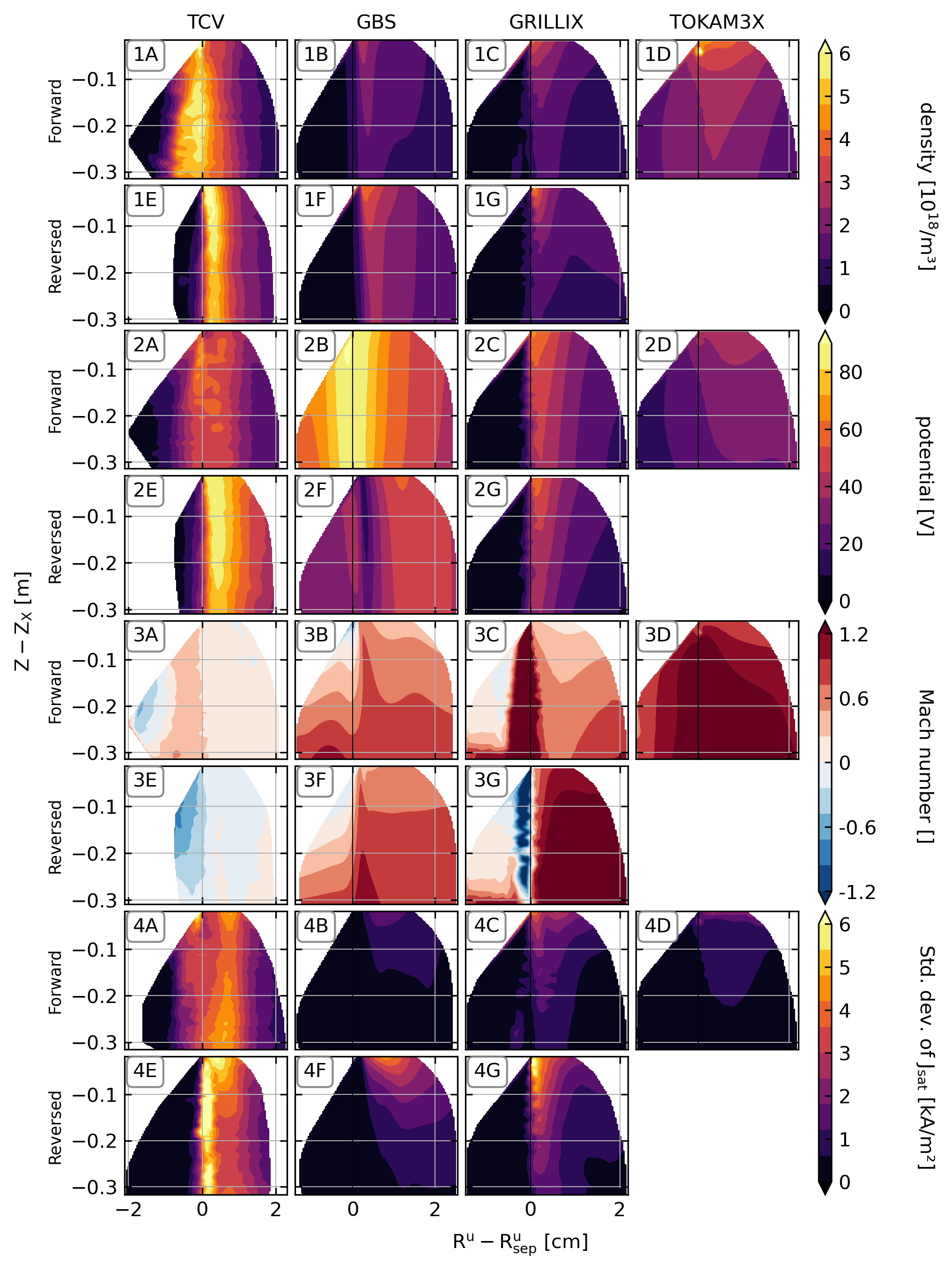}
    \caption{\textbf{Comparison of divertor volume measurements for selected profiles}. \textit{Rows from top}: Two-dimensional divertor volume profiles of the mean plasma density ($n_e$, Figs.\ref{fig:rdpa_plots}.1), potential ($V_{pl}$, Figs.\ref{fig:rdpa_plots}.2), parallel Mach number ($M_\parallel$, Figs.\ref{fig:rdpa_plots}.3) and the standard deviation of the ion saturation current ($\sigma(J_{sat})$, Figs.\ref{fig:rdpa_plots}.4). The experimental data from the reciprocating divertor probe array (RDPA) is given in the \textit{left column}, while the other columns give simulated profiles from the three codes. For each set of 7 subplots, forward-field data is given in the \textit{top row} and reversed-field data is given in the \textit{bottom row}. The parallel Mach number colour-bar is cropped to the range $M_\parallel \in [-1.2, 1.2]$ to better highlight the experimental results, while the range of the simulations is discussed in Sec.\ref{subsec:divertor volume measurements}.}
    \label{fig:rdpa_plots}
\end{figure*}

\subsection{Divertor volume profiles}\label{subsec:divertor volume measurements}

In Fig.\ref{fig:rdpa_plots} we show 2D profiles of the mean plasma density ($n$), plasma potential ($V_{pl}$), parallel Mach number ($M_{\parallel}$) and ion saturation current density ($J_{sat}$) fluctuations in the divertor volume, comparing the experiment and simulations. The mean, standard deviation, skew and kurtosis of $J_{sat}$ are shown in \path{TCV-X21/3.results/summary_fig/RDPA+jsat,jsat_std,jsat_skew,jsat_kurtosis.png}, and the electron temperature and the mean and fluctuations of floating potential are shown in figure
\path{TCV-X21/3.results/summary_fig/RDPA+electron_temp,}\linebreak\path{potential,vfloat,vfloat_std.png}. The RDPA measurements cover almost the entire low-field-side divertor leg, from just below the X-point at $Z - Z_X = 0\si{\meter}$ to just above the divertor targets at $Z-Z_X=-0.32\si{\meter}$.
In general, the quantitative agreement in the divertor volume is similar to the findings at the HFS and LFS targets, with $\chi_{RDPA}\sim0.87-1.00$. 

Experimentally, we see that the $n$ profile changes with the toroidal field direction. For the forward-field case in Fig.\ref{fig:rdpa_plots}.1A, the RDPA finds a broad $n$ profile extending from the SOL into the PFR, while in the reversed-field case shown in Fig.\ref{fig:rdpa_plots}.1E the $n$ profile is peaked close to the separatrix and limited to the SOL region, which is consistent with the LFS target measurements in Fig.\ref{fig:lfs_base_moments}.
The significant change in the density profile with the toroidal field reversal suggests a strong influence of the background drifts. This has previously been observed in diverted TCV studies -- where transport modelling suggested that $E\times B$ mean-field drifts were the dominant transport mechanism in the divertor \cite{christenExploringDriftEffects2017}. We see that, regardless of field direction, the $n$ profiles are approximately constant along flux-surfaces, with a drop of $\approx 30\%$  close to the divertor targets.
The simulations do not reproduce the effect of the toroidal field direction on the density profiles. Furthermore, all simulations predict that the peak value of $n$ should drop 
by approximately a factor $2$ along the divertor leg, in contrast with the experiment, where such a drop occurs only very close to the divertor target.

The shape of the experimental $V_{pl}$ profiles in Figs.\ref{fig:rdpa_plots}.2A and \ref{fig:rdpa_plots}.2E shows similar features to the $n$ profile, although with a broader profile in the radial direction. Again, the simulations do not correctly capture the toroidal field reversal, although in GBS we see that $V_{pl}$ does change with field direction. Both GBS and GRILLIX find that $V_{pl}$ is fairly constant along flux surfaces, which agrees with the experimental measurements, while TOKAM3X shows a less clear alignment to flux surfaces. For GBS, the $V_{pl}$ profiles are radially broad, which will result in reduced poloidal $E \times B$ flows in the divertor. Conversely, for GRILLIX, the $V_{pl}$ profile is very steep across the separatrix, leading to much stronger poloidal $E \times B$ flows near the separatrix.

In Figs.\ref{fig:rdpa_plots}.3A and \ref{fig:rdpa_plots}.3E we show the parallel ion velocity (relative to the local sound speed) in the divertor. Note that here $M_\parallel>0$ indicates a parallel flow towards the LFS target. Experimentally, we find that $M_\parallel$ is subsonic throughout the entire divertor leg, with $M_\parallel \in [-0.58,0.67]$ in forward-field and $M_\parallel \in [-0.78,0.3]$ in reversed-field. This may indicate the presence of a significant neutral ionisation source, which will reduce parallel flows \cite{Bufferand2014MachParticleSources}. In forward-field, the flow is directed towards the target, on the order of $\sim10-20\%$ of $c_s$ in the SOL and $\sim50\%$ of $c_s$ in the PFR. In reversed-field, a flow \textit{away from the target} is found for most of the divertor leg, which may indicate a returning flow for the strong poloidal $E \times B$ flow directed towards the target. Flow magnitudes $\sim <20\%$ of the $c_s$ are seen in the SOL (some weak flow towards the target is also observed), while in the PFR the flow reaches $80\%$ of $c_s$ directed away from the target near the X-point.
The weak parallel flows are not reproduced in the simulations, where the absolute values of $M_\parallel$ extend beyond the colour bar in Figs.\ref{fig:rdpa_plots}.3. For GBS, the parallel Mach number in the SOL is in the range $M_\parallel\in[-0.95, 1.05]$ in forward-field and $M_\parallel\in[-0.11, 1.12]$ in reversed-field, for GRILLIX, $M_\parallel\in[-0.13, 2.57]$ in forward-field and $M_\parallel\in[-0.23, 2.56]$ in reversed-field, and for TOKAM3X, $M_\parallel\in[0.57, 1.8]$ in forward-field.
In GBS (Figs.\ref{fig:rdpa_plots}.3B and \ref{fig:rdpa_plots}.3F), we see that $M_\parallel$ increases towards the divertor targets, reaching the local sound speed in the near-SOL close to the targets, consistent with the no-drift Bohm-Chodura boundary condition that imposes $u_\parallel=c_s$ at the targets. This boundary condition gives the correct sign for $M_\parallel$ in the forward-field case, but cannot reproduce the changes with toroidal field reversal.
For GRILLIX (Figs.\ref{fig:rdpa_plots}.3C and \ref{fig:rdpa_plots}.3G) the simulated parallel flows greatly exceed the local sound speed, while the direction of the parallel flow in the PFR appears to roughly correspond to the measurements. The fast parallel flows may be related to a feedback loop affecting the $E \times B$-drift correction $u_\parallel$ boundary condition used in GRILLIX, which is discussed more in Sec.\ref{sec:discussion boundary conditions}.
For TOKAM3X (Fig.\ref{fig:rdpa_plots}.3D), the effect of the $E \times B$ drift is weaker than in GRILLIX, consistent with the weak divertor $V_{pl}$ gradients (see Fig.\ref{fig:rdpa_plots}.2D), such that the values and direction of $M_\parallel$ are dominated by the drift-free component of the Bohm-Chodura condition.

The measured $J_{sat}$ fluctuations are approximately constant along the divertor leg (Figs.\ref{fig:rdpa_plots}.4A and 4E), with a moderate increase observed towards the target in forward-field and near the X-point in reversed-field. Radially, the fluctuation intensity shows the same pattern as the other fields, with a broader profile in forward-field and a marked peak near the separatrix in the reversed-field case. The relative fluctuation level $\sigma(J_{sat})/J_{sat}$ displays a marked hollowed profile in both field directions (not shown here), with a typical $\sigma(J_{sat})/J_{sat}\sim10\%$ around the separatrix, a $\sigma(J_{sat})/J_{sat}\sim50\%$ in the far-SOL, and a $\sigma(J_{sat})/J_{sat}\sim100\%$ deep in the PFR. We also remark that the region with low $\sigma(J_{sat})/J_{sat}$ is broad and extends from the SOL into the PFR in forward-field, while in reversed-field, the region of low relative fluctuations is only in the SOL and sits in a narrow region around $R^u-R^u_{sep}\approx0.1\si{\centi\meter}$. In contrast, the $J_{sat}$ fluctuations (the absolute fluctuations) in the simulations are significantly smaller, with the fluctuation profiles showing a strong drop when moving from the divertor entrance to the LFS target (see Figs\ref{fig:rdpa_plots}). We find that the lower simulated absolute fluctuation levels also translate to lower $\sigma(J_{sat})/J_{sat}$ levels (not shown here). In TOKAM3X, $\sigma(J_{sat})/J_{sat}<10\%$ throughout the divertor. In GBS, the simulations in both field directions recover the qualitative shape of the experimental $\sigma(J_{sat})/J_{sat}$ profiles in the SOL. The region of reduced relative fluctuations along the divertor leg is found with $\sigma(J_{sat})/J_{sat}\sim10\%$ as well, as the increase in the radial direction. In the reversed-field simulation, this is up to $\sim50\%$ in the region at $Z-Z_{X}\approx 0 \si{\meter}$ and $\sim15\%$ at $Z-Z_{X}\approx -0.35 \si{\meter}$. Conversely, the forward-field simulation shows a less steep poloidal gradient with $\sigma(J_{sat})/J_{sat}\sim15\%-25\%$, where the maximum and minimum values are found, respectively, at $Z-Z_{X}\approx 0 \si{\meter}$ and $Z-Z_{X}\approx -0.35 \si{\meter}$. In GRILLIX, in forward-field $\sigma(J_{sat})/J_{sat}\sim50\%$ is seen localised near the separatrix in the PFR, while $\sigma(J_{sat})/J_{sat}<10\%$ is seen in the SOL. In reversed-field, a peaked relative fluctuation profile is observed with a $\sigma(J_{sat})/J_{sat}\sim 25\%$ along the separatrix, while the rest of the divertor displays $\sigma(J_{sat})/J_{sat}<10\%$. We find that GBS simulations show a reasonably good quantitative agreement for the skewness and kurtosis of $J_{sat}$, as shown in Tab.\ref{tab:validation_results}, suggesting the blob dynamics are being reasonably well reproduced \cite{dippolitoConvectiveTransportIntermittent2011a}.

\section{Discussion}\label{sec:discussion}
We now consider the overall outcome from the quantitative and qualitative validations. The immediate result from both analyses is that outboard midplane and divertor entrance profiles are captured to a reasonable degree by all codes, whereas the divertor volume and target profiles show a worse agreement. The disagreement between simulation and experiment is particularly apparent for the parallel Mach number in the divertor volume (Figs.\ref{fig:rdpa_plots}.3), the standard deviation of $J_{sat}$ at the divertor targets (Figs.\ref{fig:lfs_expt_moments}.C and \ref{fig:lfs_expt_moments}.D) and the floating potential at the divertor targets (Figs.\ref{fig:lfs_expt_moments}.E and \ref{fig:lfs_expt_moments}.F). For observables where the simulations disagree with experiment, we find that the simulations also typically disagree with each other, which indicates that the simulations are sensitive to the differences between the models, model parameters, and numerical parameters.

In Sec.\ref{subsec:disc_quant_and_sensitivity} we discuss the result of the quantitative validation, and the observed sensitivity of the simulated profiles to the numerical resolution. Then, in Sec.\ref{sec:discussion reduced params}, we discuss how the choice of physical parameters was found to affect the results. In Sec.\ref{sec:discussion boundary conditions}, we discuss the sensitivity of the simulations to the choice of sheath boundary conditions -- particularly for the parallel ion velocity, parallel current and floating potential. In Sec.\ref{subsec:disc_toroidal_field}, we discuss the effect of toroidal field reversal, contrasting the results to previous transport modelling. We then close our discussion of the simulations in Sec.\ref{subsec:towards_a_match} by assessing what additional physics might be required to match the experimental results, such as neutral dynamics, and to provide predictions beyond \path{TCV-X21}. Finally, in Sec.\ref{subsec:disc_expt_extensions} we suggest minor extensions to the experimental reference dataset, which could further constrain the models and test additional capabilities of the simulations.

\subsection{Quantitative validation and sensitivity analysis}\label{subsec:disc_quant_and_sensitivity}

In this validation exercise, the overall level-of-agreement according to the composite metric is quite poor, with $\chi=0.83-0.94$ for all codes (as a reminder, $\chi=1$ indicates complete disagreement). This indicates that a majority of observables have $d_j$ values significantly above the agreement threshold $d_0=1$. However, the quality of the validation is very high, with values of $Q=18.6-19.4$, reflecting the large number of observables used in the validation. The results of the quantitative analysis correspond to a lower level of agreement and a significantly higher quality compared to previous works using the same methodology, such as Ricci et al., 2015 \cite{ricciApproachingInvestigationPlasma2015a}, which found a $\chi=0.5-0.75$ with a $Q\approx 4$ for TORPEX plasmas and Riva et al., 2020 \cite{rivaShapingEffectsScrapeoff2020} which found a value of $\chi=0.45$ and $Q\approx 4$ for limited TCV discharges. We note that a companion work to this paper, published as Galassi et al., 2021 \cite{Galassi2021} found $\chi=0.85-1.0$ and $Q\approx 4$ in TORPEX plasmas with an internal X-point.
The reduced level of agreement found in this work is partly due to our assumption of zero simulation uncertainty, in contrast to the previous studies which -- except for Ref.\cite{Galassi2021} -- estimated non-zero simulation uncertainties. We can explore the impact of non-zero $\Delta s_j$ by setting the simulation uncertainty $\Delta s_j$ to the experimental uncertainty $\Delta e_j$, which gives $\chi$ between $0.7-0.87$. This is closer to the range of the previous studies, although still at a lower level of agreement.

Beyond reducing our level of agreement, we see that our assumption of zero uncertainty also causes the metric to occasionally give non-intuitive results. This is particularly noticeable when the experimental uncertainty varies strongly across a profile, since low uncertainty points effectively dominate the quantitative level of agreement. To avoid this, we could introduce a finite $\Delta s_j$, or alternatively we could explore alternative metrics such as Gaussian process regression \cite{hoApplicationGaussianProcess2019}.\\

To estimate the simulation uncertainty, we could use error-estimation methods such as a sensitivity analysis \cite{terryValidationFusionResearch2008} or Richardson extrapolation \cite{Roy2019-ROYEAU}. However, such techniques require repeat simulations, and due to computational cost of the simulations, a rigorous error analysis was not performed within this work. Instead, GRILLIX was used to explore the effect of varying the poloidal resolution. We performed a resolution scan, comparing simulations at 2\si{\milli\meter}, 1\si{\milli\meter} and 0.5\si{\milli\meter} perpendicular resolutions. At the OMP, similar mean profiles were found across the resolution scan, while a phase-shift analysis showed an increase of drift-wave turbulence relative to ballooning turbulence with increasing resolution. In the open field-line region, higher fluctuation levels (particularly in the PFR) and more transport into the far-SOL was found with increasing resolution. This suggests that a resolution of a few \si{\milli\meter} is sufficient to resolve most of the primary turbulence-drive by confined-region instabilities, while higher resolutions are required to capture more of the secondary turbulence driven locally in open field-line regions with low temperature (or alternatively, small drift scales). We also investigated the statistical uncertainty of the GRILLIX results by using the bootstrap method \cite{Efron1979Jan}. This indicated that higher-order statistical moments, such as skewness and kurtosis, had significant statistical uncertainty, while the mean and standard deviation had negligible statistical uncertainty.

\subsection{Impact of relaxed parameters}\label{sec:discussion reduced params}

Whereas GRILLIX used the Braginskii values for resistivity and heat conductivities directly, relaxed parameters were used in GBS ($\eta_\parallel$ increased by factor $3$ and $\chi_{\parallel,e,i}$ reduced by factor between $4.8$ (at the targets) and $20$ (at the OMP)) and TOKAM3X ($\eta_\parallel$ increased by factor $1.8$ and $\chi_{\parallel,e,i}$ reduced by factor $1.8$) in order to reduce computational costs and improve the numerical stability. Previous studies have indicated that artificially increased collisionality leads to broadened SOL profiles \cite{rivaThreedimensionalPlasmaEdge2019, tataliImpactCollisionalityTurbulence2021}, larger blobs and higher fluctuation levels \cite{tataliImpactCollisionalityTurbulence2021}. Increased resistivity can directly affect the plasma potential (affecting the scale of potential fluctuations and the magnitude of the $E \times B$ transport) \cite{tataliImpactCollisionalityTurbulence2021}, extend the inertial regime of SOL blobs \cite{easyInvestigationEffectResistivity2016} and change the position where blobs are generated \cite{nespoliBlobPropertiesFullturbulence2017}. Reduced heat conductivity leads to a higher fraction of the heat flux being convected by parallel flows \cite{wersalComparisonRefinedTwopoint2017}, broader profiles \cite{zholobenkoThermalDynamicsFluxcoordinate2020} and allows larger parallel temperature gradients to develop.

In the poloidal density profile (Fig.\ref{fig:simulation_snaps}), we see that the codes predict noticeably different blob sizes -- large for GBS, intermediate for TOKAM3X, and small for GRILLIX -- which is the ordering expected if physical parameters were the dominant factor determining blob size. Comparing the profiles, the difference between the codes at the OMP is small and within experimental uncertainty, while at the divertor targets the differences between the codes are more pronounced. In general, GBS approximately matches the width of the $n$, $J_{sat}$, and $J_\parallel$ profiles in forward-field, while in reversed-field the profiles are slightly broadened. Conversely, the reversed-field simulations often predict the peak value more accurately than the forward-field simulations. For GRILLIX, the target profiles are noticeably narrower, which is likely due to the fast parallel advection (see Sec.\ref{sec:discussion boundary conditions}). For TOKAM3X, the LFS target profiles returns a low quantitative and qualitative agreement, while the HFS target gives appreciable agreement -- which may indicate that the effect of relaxed parameters is increased for longer leg lengths. However, we note that the electron temperature profile \textit{increasing into the far-SOL} for the TOKAM3X low-field-side target profile cannot be explained by relaxed parameters, and may indicate numerical pollution via the buffer zone applied on the limiting flux surface. It is also unclear whether the use of relaxed parameters can explain the reduced fluctuations in the divertor leg, since in Ref.\cite{tataliImpactCollisionalityTurbulence2021} fluctuations were found to \textit{increase} with increasing collisionality. Here, it is likely that grid resolution (and the numerical discretisation scheme) is having an impact, but a combined sensitivity analysis and resolution scan was not performed in this work. It is also unclear which of the increased resistivity or the reduced heat conductivity is having a more significant effect on the target profiles. A parameter scan would be useful to determine the effect of relaxed parameters on the simulated target profiles, and reveal whether different parametric dependencies are found in limiter and divertor geometries.

\subsection{Influence of sheath boundary conditions}\label{sec:discussion boundary conditions}

The codes use different sets of sheath boundary conditions, detailed in Sec.\ref{sec:codes} and Appendix \ref{sec:boundary_conditions}. One effect of the different sheath boundary conditions can be directly seen in the simulated parallel Mach number ($M_\parallel$) in the divertor volume, shown in Fig.\ref{fig:rdpa_plots}.3. Here, GBS shows the expected flow profiles for a sheath-limited regime without drifts or ionisation in the divertor \cite{Bufferand2014MachParticleSources}. The Mach number is large relative to the experiment throughout the divertor, approaching $M_{\parallel}=1$ at the target, and similar in both forward and reversed field. This is consistent with the $u_\parallel=c_s$ boundary condition, but it does not match the direction of the measured $M_\parallel$ profile, which is seen to be dependent on the toroidal field direction. The effect of the toroidal field reversal on $M_\parallel$ appears to be matched better by the $E\times B$-drift corrected $u_\parallel$ boundary conditions used by GRILLIX, which are described in Ref.\cite{stangebyIonVelocityBohm1995}. However, the GRILLIX simulations find excessively fast parallel flows with $|{M_\parallel}| \gg 1$). This might indicate a feedback mechanism, where the parallel flows affect the $T_e$ profile. Due to the $V_{pl} = \Lambda T_e$ boundary condition, this in turn affects the poloidal $E \times B$ drift and the $u_\parallel$ boundary condition. It appears that this leads to self-steepening mechanism, eventually leading to very fast parallel ion velocities. The increased parallel flow rate in GRILLIX may explain part of the disagreement in other profiles. Faster parallel advection will lead to a lower target $n$ for a given upstream $n$, and by modifying the balance of parallel and perpendicular transport it will also lead to narrower profiles (such as Fig.\ref{fig:q_parallel}).

The boundary conditions also set the potential and current at the sheath. For $V_{pl}$ particularly, since it is solved via an elliptic equation, the potential boundary condition can have a significant effect throughout the open-field line region \cite{loizuBoundaryConditionsPlasma2012,zholobenkoElectricFieldTurbulence2021a}. In GBS and TOKAM3X, the potential and current boundary conditions are coupled in the electrostatic limit, giving $J_\parallel = J_{sat} \left(1 - \exp\left[\Lambda - V_{pl}/T_e\right]\right)$ and a coupled $V_{pl}$ boundary condition. In GRILLIX, the current was allowed to freely flow and the potential was set to $\Lambda T_e$. Considering the target $V_{fl}$ profiles, the GBS potential boundary condition is seen to give a non-negligible $V_{fl}$ (with a reasonable magnitude, albeit a non-matching shape), while in GRILLIX and TOKAM3X $V_{fl}\approx 0$. The GBS current profile approximately matches the shape (but not the amplitude) in the forward-field, while in reversed-field it is broadened and shifted with respect to the experiment (likely due to the used of relaxed parameters). In contrast, the free-flowing boundary condition used by GRILLIX gives a reasonable match to the experiment for $J_\parallel$ (except at the forward-field HFS), which is surprising since this boundary condition does not have a theoretical basis. To investigate this further, a GRILLIX simulation with an insulating current ($J_\parallel=0$) boundary condition was performed. It was found that the interior currents were similar to the free-flowing current simulation, but a very strong heating near the boundaries was caused by the compression required to force $J_\parallel \to 0$. Therefore, it appears that in GRILLIX the currents observed at the boundary are driven internally, independently from the boundary conditions. This result warrants further investigation, to determine whether the match to the experiment is fortuitous and -- if not -- what is driving the current internally. Finally, TOKAM3X predicts very low currents at the boundary. This gives a reasonable match at the HFS target (where the other profiles also agree well), but not at the LFS target, potentially due to the disagreement in the other profiles.

\subsection{Toroidal field reversal and in-out asymmetry} \label{subsec:disc_toroidal_field}

The codes are seen to mostly under-predict the effect of the toroidal field reversal and the in-out asymmetry between the low- and high-field-side divertor targets. This could be due to an overall underestimation of large-scale plasma drifts, which reverse with toroidal field direction \cite{ROGNLIEN1999,wensingDriftrelatedTransportPlasmaneutral2021, christenExploringDriftEffects2017}, or due to a lack of symmetry-breaking terms such as ion-orbit-loss \cite{brzozowskiGeometricModelIon2019}. Here, it is interesting to compare to previous modelling of TCV with the UEDGE transport code \cite{christenExploringDriftEffects2017}, which found that the HFS target profile was $\sim 3\times$ colder and denser in forward-field than in reversed-field, primarily due to $E \times B$ drifts. This effect was also observed experimentally, although it appears that UEDGE was overestimating the effect of the background drifts, in contrast to this study where it appears that we are underestimating their effect. As such, it would be interesting to compare turbulence and transport simulations of the \path{TCV-X21} scenario, to compare the $E \times B$ drifts, as well as to provide additional information about the neutral dynamics and power and particle sinks in the divertor.

\subsection{Towards an improved match}\label{subsec:towards_a_match}

In future works, we expect that the match to experiment should improve when GBS and TOKAM3X use more realistic values for the resistivity and heat conductivity. For all simulations (and for GRILLIX especially), the overall match should improve if the parallel ion velocity can be reduced towards experimental values, which might be achieved by using other boundary conditions for $v_\parallel$ and/or $V_{pl}$. Theoretical work to identify a consistent coupled set of boundary conditions applicable to the full system of equations (including electromagnetic and ion thermal effects, potentially extending on Ref.\cite{loizuBoundaryConditionsPlasma2012}) could significantly improve the fidelity of the simulations in the divertor.

To continue improving the quantitative match between experiment and simulation will likely also require additional physics to be considered. Despite targeting a sheath-limited low-recycling regime in the \path{TCV-X21} scenario, it is nevertheless likely that the most significant missing physics term is the neutral dynamics. For instance, the Mach number measurements in the divertor volume (Fig.\ref{fig:rdpa_plots}.3) suggest that there is a non-negligible neutral ionisation source in the divertor leg \cite{Bufferand2014MachParticleSources}. By adding neutral dynamics, we expect that the simulated parallel advection in the divertor will be reduced. The neutrals will also introduce plasma cooling in the vicinity of the X-point and an asymmetry in the density source, which can change the profiles and, thus, the turbulence drive \cite{zholobenkoNeutrals2021}.

As such, new versions of the codes are currently targeting simulations of the TCV-X21 scenario. For GBS, a new version has recently been developed. This version does not use the Boussinesq approximation, includes electromagnetic effects and couples to a kinetic neutrals model \cite{giacomin2021}. For GRILLIX, a refactored version will be used for testing boundary conditions, including neutrals and extending the resolution scan \cite{zholobenkoNeutrals2021}. For TOKAM3X, or rather its successor SOLEDGE3X \cite{Bufferand2021}, the Boussinesq approximation will not be used, neutral and impurity physics will be provided via a coupling to EIRENE, and additional terms for the parallel viscosity and a complete Reynolds-stress tensor will be included.

Beyond \path{TCV-X21}, further extensions of the model beyond the drift-reduced fluid approach might be necessary, especially for future validations against larger experimental devices and more challenging plasma conditions. Kinetic corrections to the parallel heat flux \cite{tskhakayaKineticEffectsParallel2008}, ion orbit loss effects \cite{brzozowskiGeometricModelIon2019} and finite Larmor radius corrections (in the spirit of gyrofluid or velocity-space-decomposed-gyrokinetic \cite{freiGyrokineticModelPlasma2020} models) may become important.  In this context, comparisons to gyrofluid and gyrokinetic simulations of the \path{TCV-X21} reference case and future divertor validation cases are of great interest.

\subsection{Extensions to the experimental dataset}\label{subsec:disc_expt_extensions}

The reference experimental dataset provides unparalleled diagnostic coverage of the divertor and SOL. The \path{TCV-X21} repository further allows for the extension of the dataset, including additional experimental data and the results of future code validations. Future work will add data from the gas-puff imaging systems at the OMP and in the X-point region, providing information about the size and velocity of SOL and divertor blobs. We also intend to add parallel heat flux measurements at the HFS target from IR cameras, neutral pressure measurements from pressure gauges and spectroscopy measurements of $T_i$ and the neutral density in the SOL. In anticipation of future $T_i$ measurements, we have included the simulated $T_i$ profiles in the repository, which are shown in comparison to $T_e$ at \path{TCV-X21/3.results/analysis_fig/Ion_temperature.png}. These extensions will help to further establish the \path{TCV-X21} reference case as a standard test case for divertor and SOL simulations.

\section{Summary and conclusions}\label{sec:conclusions}
The predictive capabilities of divertor turbulence simulations was rigorously assessed via a validation against a series of dedicated diverted TCV discharges in both forward and reversed field direction. The discharges were performed at a lower toroidal magnetic field than typical for TCV ($0.95\si{\tesla}$ vs $1.45\si{\tesla}$) to decrease the computational cost, thus allowing direct, full-size simulations of the experimental scenario. Moreover, the discharges were carried out in low density, sheath-limited conditions, designed to reduce the effect of the neutrals, which are currently not included in the simulations. The discharges were repeated several times to improve the statistics of the experimental measurements and to investigate the effect of toroidal field reversal. An extensive experimental dataset, which we refer to as \path{TCV-X21}, was collected for the purpose of validating the codes, which includes a broad range of 1D and 2D measurements of fluctuation and time-averaged quantities at the outboard midplane, at the divertor entrance, throughout the divertor volume, and at the target plates. The validation dataset is provided in an open repository, as a reference for future validations and to allow benchmarking of boundary turbulence simulations.\\

Full-size simulations of the \path{TCV-X21} reference case, in both toroidal field directions, were performed by three separate edge turbulence modelling groups, using the GBS, GRILLIX and TOKAM3X codes. The simulations were flux-driven, meaning that the plasma profiles, turbulence and transport were evolved self-consistently. As such, the simulations were validated against the full set of experimental measurements, including both mean profiles and fluctuations. The only tunable physics parameters in these simulations were the position and strength of the density source and of the temperature or power source. The source positions were chosen to approximately match the positions of Ohmic power deposition and neutral ionisation, and the rates were adjusted to approximately match the experimental outboard midplane density and temperature (or power) at the position of the separatrix. To evaluate the quality of the match between the simulations and experiment, the different observables were compared graphically and via a quantitative validation metric.

An appreciable match between simulation and experiment was found, particularly at the outboard midplane. The simulations were able to match several of the outboard midplane profiles, including both mean profiles and fluctuations. By comparison, the simulation results lie mostly outside the experimental error bars at the divertor targets and in the divertor volume, while the order-of-magnitude and approximate shape of several profiles agree with the experiment. Additionally, although at the outboard midplane the three codes provided similar results, the difference between the simulations increased towards the targets. Since the simulations adjusted their sources for the outboard midplane separatrix values, the reduced agreement in the divertor was partially expected. However, it also appears that there are additional physical and numerical effects which are becoming important in the divertor.

By comparing the simulations against the experiment and each other, we find possible causes for the reduced agreement in the divertor, which may indicate how the simulations could be further improved. We see that simulations using relaxed parameters -- artificially increased resistivity or reduced heat conductivity -- found broadened divertor target profiles. The divertor profiles were also seen to be sensitive to the energy source rate, which is consistent with results from a two-point model. Simulations which resolved the local drift scale found increased fluctuations away from the confined region. Since computational cost is the main limit on the choice of parameters and resolution, continued improvement of the numerical efficiency and scalability of the simulations should help to improve the match to the experiment.

The simulated divertor profiles are found to be sensitive to the choice of sheath boundary conditions applied at the divertor targets. The use of drift-corrections in the parallel ion velocity boundary condition was found to give much faster velocities than the experiment, while simulations which neglected drift corrections found a lower flow speed but did not reproduce the flow pattern observed in the divertor volume. Additionally, the codes differed in their description of the current crossing the sheath. Here, a simple `free-flowing' boundary condition was found to give reasonable agreement when compared to the experiment or to simulations employing the usual Bohm-current boundary condition. In both cases, the tight coupling of the boundary conditions and profiles prevent generalisation of the results. Instead, we highlight that further work in investigating sheath effects and identifying an optimal, numerically-stable set of boundary conditions would be highly beneficial for divertor simulations.

Beyond this, it is expected that additional physics is required to achieve a quantitative match between simulation and the \path{TCV-X21} reference. The addition of neutral physics is expected to be particularly important -- since despite experimental efforts to reduce the divertor neutral pressure, our simple treatment of the neutrals as acting as only a confined-region density source neglects several important effects such as localised ionisation near the X-point and in the far-SOL. As such, the codes will target repeat simulations of the \path{TCV-X21} scenario including neutral physics, more realistic physical parameters and improved boundary conditions.\\

This validation shows that the \path{TCV-X21} reference case developed in this work allows for the rigorous validation and bench-marking of turbulence simulations. The results of this first validation indicate that turbulence simulations are already providing a promising match to the experiment, which can be further improved by targeted development of the codes. This validation methodology can be extended to other codes since the \path{TCV-X21} reference data has been publicly released. Additionally, similar validations could be performed at more challenging plasma parameters, with neutral physics included, in more complex magnetic geometries, at larger magnetic field strengths and on larger machines. Continued validation of turbulence simulations over a broad parameter space will accelerate the development of the codes, to improve their usefulness for interpreting results from existing machines, and eventually to enable predictive simulations of future fusion reactors.

\section{Author Information}

D. S. Oliveira performed and analysed the GBS forward-field simulations, was session leader for the majority of the TCV-X21 discharges, and post-processed the experimental results.
T. Body extended the GRILLIX code for this work, performed and post-processed the GRILLIX simulations, developed the TCV-X21 repository and performed the comparison of the validation results.
D. S. Oliveira and T. Body jointly wrote the paper with the assistance of coauthors, and as such are recognised as equal-first authors.
In addition to the first authors, D. Galassi, C. Theiler, P. Tamain and A. Stegmeir contributed extensively to the paper writing.
D. Galassi helped the simulation groups to set up their simulations, contributed to the development of GBS, and performed and analysed the GBS reversed-field simulations.
C. Theiler initiated and led the project and developed the TCV-X21 reference scenario.
E. Laribi performed the TOKAM3X simulations, which were post-processed by P. Tamain.
A. Stegmeir provided extensive support in the set up and analysis of the GRILLIX simulations.

This work was made possible by the contributions of several teams of collaborators. The TCV experimental team for this project was led by C. Theiler and consisted of D. S. Oliveira, D. Galassi, J. A. Boedo, C. Colandrea, H. de Oliveira, S. Gorno, N. Offeddu, H. Reimerdes, C. K. Tsui and C. Wüthrich.
The GBS team was led by P. Ricci and consisted of D. S. Oliveira, D. Galassi, C. Theiler, M. Giacomin, G. Fourestey and N. Varini.
The GRILLIX team was led by F. Jenko and consisted of T. Body, A. Stegmeir, W. Zholobenko and D. Coster.
The TOKAM3X team was led by P. Tamain and consisted of E. Laribi, H. Bufferand, G. Ciraolo and E. Serre.
The TCV-X21 repository was developed by T. Body, in collaboration with D. Coster, F. Imbeaux, F. Jenko, C. Theiler and the Fair4Fusion team.
V. Naulin and N. Vianello worked with C. Theiler to initiate this project.
M. Wiesenberger also participated in the modelling effort (with FELTOR), and contributed extensively to discussions regarding modelling the scenario.

\section{Acknowledgements}

This work has been carried out within the framework of the EUROfusion Consortium and has received funding from the Euratom research and training programme 2014-2018 and 2019-2020 under grant agreement number 633053. The views and opinions expressed herein do not necessarily reflect those of the European Commission. This work was supported in part by the Swiss National Science Foundation. This work was partly supported by the U.S. Department of Energy under Grant No. DE-SC0010529.

This work was completed within the framework of the EUROFusion Enabling Research ENR-MFE19.EPFL-02, led by Christian Theiler. It was supported by the EUROFusion Theory Simulation Verification and Validation project 3, led by Patrick Tamain.\\

This work was supported by a grant from the Swiss National Supercomputing Centre (CSCS) under project IDs s1126, s1129, s1028 and s882. The simulations in this work were performed within the 4th and 5th cycles of MARCONI-FUSION HPC.
The required computational resources for GBS were allocated under the projects \texttt{DIVturb} and \texttt{XTurb}. We acknowledge PRACE for awarding us access to SuperMUC-NG at GCS@LRZ, Germany. The required computational resources for GRILLIX were allocated under the projects \texttt{FUA34\_EST3D}, \texttt{FUA35\_EST3D} and \texttt{FUA35\_TSVV3}.
TOKAM3X simulation were additionally performed on HPC systems provided by GENCI.

The \path{TCV-X21} repository was made possible by several collaborators agreeing to publish their data openly. The data analysis relied on several open-source software packages, particularly the \texttt{scipy} \cite{2020SciPy-NMeth}, \texttt{numpy} \cite{harris2020array}, \texttt{matplotlib} \cite{Hunter:2007}, \texttt{xarray}, \cite{hoyer2017xarray} \texttt{Jupyter} \cite{Kluyver:2016aa}, \texttt{pint} and \texttt{netCDF4} Python packages.

\pagebreak
\FloatBarrier
\begin{appendices}
\section{Boundary conditions}\label{sec:boundary_conditions}

The model equations for each of the codes can be found in Giacomin and Ricci, 2020 \cite{giacominInvestigationTurbulentTransport2020} for GBS, in appendix A of Zholobenko et al., 2021\cite{zholobenkoElectricFieldTurbulence2021a} for GRILLIX and in Tatali et al., 2021 \cite{tataliImpactCollisionalityTurbulence2021} for TOKAM3X. The models equations require the use of boundary conditions at the edge of the computational domain. Of these, the choice of boundary conditions applied at divertor targets is particularly impactful, since these boundary conditions are used to include sheath effects in the simulations.

The sheath is known to have a significant effect on the plasma dynamics, affecting the plasma flows, currents and potential. However, the assumptions in the fluid approximation break down in the sheath (and magentic presheath \cite{loizuBoundaryConditionsPlasma2012}), and so the sheath cannot be self-consistently modelled by fluid codes (or by gyrokinetics -- self-consistent sheath modelling requires fully-kinetic 6D simulations \cite{tskhakayaOnedimensionalPlasmaSheath2017}). Instead, in this work, `sheath boundary conditions' aim to mimic the effect of the sheath. The choice of boundary conditions for each code is given here.

\subsection{GBS boundary conditions}

GBS uses a set of magnetic pre-sheath boundary conditions for $u_\parallel$, $v_\parallel$, $n$, $V_{pl}$, $T_e$ and for the vorticity $\omega = \nabla^2_\perp V_{pl}$ \cite{giacomin2021,loizuBoundaryConditionsPlasma2012} that have been rigorously derived via a first-principles analysis of the sheath dynamics and which are in agreement with kinetic simulations of this region.  In our simulations, we use the same boundary conditions as in Giacomin et al. \cite{giacominInvestigationTurbulentTransport2020}, which neglect corrections due to the gradients of density and plasma potential along the wall. 

\subsection{GRILLIX boundary condition}

For GRILLIX, the parallel velocity is set according to a flow-reversal-limited Bohm-Chodura boundary condition, written as
\begin{equation} \label{eq:weakBohmChoduraGRILLIX}
    u_\parallel \hat{b}\cdot\hat{n} = \mathrm{max}\left[c_s \hat{b}\cdot\hat{n} - \mathbf{u}_{E \times B} \cdot \hat{n}, u_\parallel^+, 0 \right]
\end{equation}
where $f^+$ means the nearest parallel-neighbouring point in the direction towards the main plasma volume, $\hat{b}$ is the parallel unit vector and $\hat{n}$ is the wall-normal unit vector. This boundary condition modifies the parallel velocity to prevent inwards $E \times B$-drifts across the boundary. The projection into the wall-normal direction $\hat{n}$ is to prevent cross-field flow oblique to the parallel direction, although we note that the $\hat{b}\cdot\hat{n}$ projection of the parallel velocity can lead to very fast flows. We additionally set free-flowing boundary conditions for the parallel current density $J_\parallel = J_\parallel^+$ and density $n = n^+$. The electrostatic potential is set to the floating potential $V_{pl} = \Lambda T_e$ for $\Lambda = 2.69$ the sheath potential drop. No boundary condition is used for the parallel electromagnetic potential $A_\parallel$. Finally, for the electron- and ion-temperatures we set sheath-heat-transmission boundary conditions of the form
\begin{equation}
    \nabla_\parallel \log(T_e) = \frac{-\tilde{\gamma_e}}{\chi_{\parallel,e}} n u_\parallel
\end{equation}
where $\tilde{\gamma_e} = 2.5$ is the anomalous electron sheath-heat-transmission factor. An equation of the same form is used for the ions replacing $T_e$ with $T_i$ and using $\tilde{\gamma_i} = 0.1$.

\subsection{TOKAM3X boundary conditions}

In TOKAM3X, the boundary conditions are developed in Tatali et al., 2021 \cite{tataliImpactCollisionalityTurbulence2021}. Bohm-Chodura boundary conditions are enforced for the parallel velocity of ions:
\begin{equation}
    \left(u_\parallel \hat{b} + \mathbf{u}_{E \times B} + \mathbf{u}_{\nabla B}^i \right) \cdot \hat{n} = \mathrm{max}\left[c_s, u_\parallel^+\right] \hat{b} \cdot \hat{n}
\end{equation}
where the total first order ion drift velocities (ExB plus curvature) are taken into account, $c_s=\sqrt{\frac{T_e+T_i}{m_i}}$ is the local acoustic velocity and $\hat{n}$ is the normal to the surface in the outgoing direction. Here $u_\parallel^+$ stands for the velocity at the nearest discretization point in the poloidal direction. The boundary condition on the parallel current $J_\parallel$ links it to the dimensionless plasma potential $V_{pl}$ and the floating sheath potential drop $\Lambda T_e = -0.5 \ln \left( 2 \pi \frac{m_e}{m_i} \left( 1 + \frac{T_i}{T_e} \right) \right) T_e$  according to the following relation\cite{stangeby2000plasma}:
\begin{equation}
    J_\parallel = \textrm{sign} \left( \hat{b} \cdot \hat{n} \right) n c_s \left( 1 - \exp \left( \Lambda - \frac{V_{pl}}{T_e} \right) \right)
\end{equation}
Finally, heat fluxes follow the standard sheath boundary conditions derived in \cite{stangeby2000plasma}, very similarly to what is used in GRILLIX:
\begin{equation}
    \hat{q}_{e/i} \cdot \hat{n} = \gamma_{e/i} T_{e/i} \hat{\Gamma}_{i} \cdot \hat{n}
\end{equation}
where $\hat{\Gamma}_i$ is the total particle flux and $\hat{q_{e/i}}$ is the total heat flux for electrons (e) or ions (i). The sheath heat transmission factors were set respectively to $\gamma_e = 4.5$ and $\gamma_i = 2.5$ for these simulations.

\section{Nomenclature, units and sign convection}\label{sec:nomenclature and units}

The sign convection of the parallel Mach number $M_\parallel$ and the parallel current is defined to match the experimental measurements. For the parallel current, measured at the divertor targets, positive values indicate current flowing into the targets and negative values indicate values flowing from the targets. For the parallel Mach number, measured via immersed probes, the sign convention is such that positive $M_\parallel$ indicates flows towards the low-field-side divertor target, and negative $M_\parallel$ indicate flows towards the high-field-side divertor target.

\paragraph{\textbf{Simulation codes}}
\begin{itemize}
    \item GBS: A fully-non-aligned 3D fluid turbulence code (developed at SPC, Lausanne)
    \item GRILLIX: A locally-field-aligned 3D fluid turbulence code (developed at IPP, Garching)
    \item TOKAM3X: A flux-aligned 3D fluid turbulence code (developed at CEA, Cadarche)
    \item LIUQE: An equilibrium reconstruction code for TCV
\end{itemize}

\paragraph{\textbf{Diagnostics and locations}}
\begin{itemize}
    \item FHRP: Horizontally-reciprocating probe at the outboard midplane
    \item RDPA: Reciprocating Divertor Probe array
    \item LP: Langmuir Probe (in TCV, wall embedded Langmuir Probes) 
    \item TS: Thomson Scattering system
    \item LFS-LP Low-field-side divertor target Langmuir probes
    \item HFS-LP: High-field-side divertor target Langmuir probes
    \item LFS-IR: Infrared camera measurements at the low-field-side divertor target
    \item OMP: outboard mid-plane
    \item DE: divertor entrance (TS measurement position)
    \item SOL: scrape-off-layer
    \item PFR: private flux region
    \item LFS: low-field-side, or low-field-side divertor target
    \item HFS: high-field-side, or high-field-side divertor target
\end{itemize}

\paragraph{\textbf{Plasma quantities (and base units)}}
\begin{itemize}
    \item $R$: radial displacement from the axis of rotational symmetry, in metres, or radial direction
    \item $Z$: vertical displacement from magnetic axis, in metres, or vertical direction
    \item $\phi$: toroidal direction
    \item $\parallel$: parallel-to-magnetic-field direction
    \item $R^u-R^u_{sep}$: flux surface label, giving the radial distance between the flux surface and the separatrix, in metres
    \item $Z - Z_X$: vertical coordinate relative to the X-point, in metres
    \item $\mathbf{B}$: magnetic field vector
    \item $B$: magnetic field strength, in tesla
    \item $I_p$: plasma current, in amperes
    \item $\rho_s$: sound drift scale, in metres
    \item $\beta$: the ratio of kinetic to magnetic pressure
    \item $\eta_\parallel$: parallel resistivity
    \item $\chi_{\parallel,e}$: parallel electron heat conductivity
    \item $\chi_{\parallel,i}$: parallel ion heat conductivity
    \item $n$: plasma density for electrons ($n_e$) and ions ($n_i$), in particles per cubic metre
    \item $\Lambda$: sheath potential drop, in units of per-coulomb
    \item $T_e$: electron temperature, in electron-volts
    \item $T_i$: ion temperature, in electron-volts
    \item $V_{pl}$: plasma electrostatic potential, in volts
    \item $J_{sat}$: ion saturation current density, in ampere-per-square-metre
    \item $V_{fl}$: floating potential, in volts
    \item $J_\parallel$: parallel current density, in ampere-per-square-metre
    \item $q_\parallel$: parallel heat flux, in watts-per-square-metre
    \item $\lambda_q$: parallel heat flux decay length, in metres
    \item $M_\parallel$: parallel advective velocity, normalised by the local sound speed
    \item $c_s$: sound speed, in metres-per-second
    \item $u_\parallel$: the parallel ion velocity, in metres-per-second
    \item $v_\parallel$: the parallel electron velocity, in metres-per-second
    \item $A_\parallel$: the parallel component of the electromagnetic vector potential, in tesla-metre
    \item $P$: power, in watts. $P_{sep}$ is the power crossing the separatrix, in watts
\end{itemize}

\paragraph{\textbf{Mathematical operators}}
\begin{itemize}
    \item $\langle f \rangle$ or $\overline{f}$: average of $f$. Note that if no operator is indicated on a function, the average is indicated unless stated otherwise
    \item $\sigma(f)$: standard deviation of $f$
    \item skew$(f)$: skewness of $f$
    \item kurt$(f)$: Pearson kurtosis of $f$. Note that \textit{this is not the same as the excess (or Fisher) kurtosis}
    \item $\nabla f$: vector-gradient of $f$
    \item $\mathbf{f} \times \mathbf{g}$: vector cross-product of $\mathbf{f}$ and $\mathbf{g}$
    \item $\mathbf{f} \cdot \mathbf{g}$: vector dot-product of $\mathbf{f}$ and $\mathbf{g}$
    \item $\hat{x}$: unit vector in the $x$ direction
\end{itemize}

\paragraph{\textbf{Quantitative validation factors}}
\begin{itemize}
    \item $\chi$: composite validation metric -- between 0 (agreement) and 1 (disagreement) [Eq.\ref{eq:ricci_chi}]
    \item $Q$: composite validation quality (higher values indicate validation has more high-precision observables) [Eq.\ref{eq:quality}]
    \item $d_j$: root-mean-square of inverse-error-weighted residual between experiment and simulation for some observable $j$ [Eq.\ref{eq:normalised_distance}]
    \item $R(d_j)$: level of agreement for some observable $j$ [Eq.\ref{fig:level_of_agreement_function}], parametrised by $d_0$ (agreement threshold) and $\lambda$ (transition sharpness)
    \item $H_j$: primacy hierarchy [Eq.\ref{eq:hierarchy}]
    \item $S_j$: sensitivity [Eq.\ref{eq:sensitivity}]
    \item $\Delta e_j$ and $\Delta s_j$: total uncertainty from the experiment and simulation respectively, for some observable $j$. $\Delta e_j$ consists of the uncertainty related to fitting experimental data to a model ($\Delta e_{fit}$), inherent diagnostic uncertainty ($\Delta e_{dia}$) and the uncertainty related to the reproducibility across repeat discharges ($\Delta e_{rep}$)
\end{itemize}
\end{appendices}

\printbibliography

\end{document}